\documentclass{article} 
 \usepackage{JOEconfig}

\usepackage[utf8]{inputenc}
\usepackage{textcomp}

\usepackage{graphicx}
\usepackage{amsmath}
\usepackage[version=4]{mhchem}
\usepackage{siunitx}
\usepackage{longtable,tabularx}
\usepackage{caption}
\usepackage{subcaption}

\usepackage{amsmath}
\usepackage{amssymb} 
\usepackage{amsthm} 
\usepackage{amsfonts} 
\usepackage{bm} 
\usepackage{xfrac}
\usepackage{xcolor}

\usepackage{upgreek}
\usepackage[artemisia]{textgreek} %

 \usepackage{hyperref}
\hypersetup{
    linktocpage,  
    colorlinks,
    citecolor=black,
    filecolor=blue,
    linkcolor=blue,
    urlcolor=blue
}

\DeclareMathSizes{10}{9.75}{7}{5.5} 
\DeclareMathSizes{9}{9.5}{7}{5}
 \DeclareMathSizes{8}{8}{6}{5}
 \DeclareMathSizes{7}{7}{5}{4}

\title{Connecting the Equinoctial Elements and Rodrigues Parameters: A New Set of Elements}

\author{Joseph T. A. Peterson\footnote{Graduate Research Assistant, Aerospace Engineering,   Texas A\& M University, College Station, Texas 77843-3141.} 
,
Vishala Arya\footnote{Postdoctoral Researcher, Aerospace Engineering, Texas A\&M University, College Station, Texas 77843-3141.}
,
John L. Junkins\footnote{Distinguished Professor of Aerospace Engineering and Director of The Hagler Institute for Advanced Study, Texas A\&M University, College Station, Texas 77843-3141.}
}

\begin{document}
\date{}
\maketitle

\newcommand*{\fnsz}[1]{\text{\footnotesize{$#1$}} }

\newcommand*{\nmsz}[1]{{\displaystyle{#1}}}
\newcommand*{\scrsz}[1]{{\scriptstyle{#1}}}
\newcommand*{\ssz}[1]{{\scriptstyle{#1}}}
\newcommand*{\ii}[1]{{\scriptscriptstyle{#1}}} 
\renewcommand*{\ss}[1]{ {\scalebox{0.7}{$#1$}} }

\renewcommand{\d}[0]{\text{d}}
\newcommand{\tx}[1]{\text{#1}}

\newcommand{\pderiv}[2]{\tfrac{\partial #1}{\partial #2}}
\newcommand{\ppderiv}[3]{\tfrac{\partial^2 #1}{\partial #2 \partial #3}}
\newcommand{\Pderiv}[2]{\frac{\partial #1}{\partial #2}}
\newcommand{\PPderiv}[3]{\frac{\partial^2 #1}{\partial #2 \partial #3}}

\newcommand{\diff}[2]{ \tfrac{\d\,#1}{\d #2} }
\newcommand{\ddiff}[2]{ \tfrac{\d^2{#1}}{\d{#2}^2} }
\newcommand{\Diff}[2]{ \frac{\d#1}{\d#2} }
\newcommand{\DDiff}[2]{ \frac{\d^2{#1}}{\d{#2}^2} }

\newcommand{\fdiff}[1]{ \,\tfrac{ {\!\!\!}^{ \text{\scriptsize{$#1$}} \!} \d }{\,\d t} }
\newcommand{\sfdiff}[1]{ \,\tfrac{ {\!\!\!\!\!}^{ \text{\scriptsize{$#1$}} }\!\d }{\,\d t} }
\newcommand{\fDiff}[3]{ \,\frac{ \!\!\!\!^{ \text{\footnotesize{$#3$}} }\d #1 }{\;\d\, #2} }
\newcommand{\fddiff}[1]{ \,\tfrac{ \!\!\!\!^{ \text{\scriptsize{$#1$}} }\d^2 }{\;\d\, t^2} }
\newcommand{\fDDiff}[3]{ \,\frac{ \!\!\!\!^{ \text{\footnotesize{$#3$}} }\d^2 #1 }{\;\d\, #2^2} }

\newcommand{\mbf}[1]{{\mathbf{#1}}}
\newcommand{\tbf}[1]{{\textbf{#1}}}

\newcommand{\gvec}[1]{{\bm{#1}}}
\newcommand{\gv}[1]{{\bm{#1}}}
\newcommand{\gvb}[1]{{\mbf{#1}}}

 \newcommand{\sigvec}[0]{\vec{\pmb{\upsigma}} }
 \newcommand{\sigtvec}[0]{\widetilde{\pmb{\upsigma}} }

\newcommand{\uvec}[1]{\hat{\bm{#1}}}
\newcommand{\ihat}[0]{\hat{\bm{\iota}}}
\newcommand{\ehat}[0]{\hat{\bm{e}}} 
\newcommand{\bhat}[0]{\hat{\bm{b}}}
\newcommand{\shat}[0]{\hat{\bm{s}}}
\newcommand{\nhat}[0]{\hat{\bm{n}}}
\newcommand{\ohat}[0]{\hat{\bm{o}}}
\newcommand{\uhat}[0]{\hat{\bm{u}}}

\newcommand{\uvecb}[1]{\hat{\mbf{#1}}}
\newcommand{\ihatb}[0]{\hat{\textbf{\textiota}}} 
\newcommand{\ehatb}[0]{\uvecb{e}} 
\newcommand{\bhatb}[0]{\!\!\uvecb{\;b}}
\newcommand{\hhatb}[0]{\!\uvecb{\;h}}
\newcommand{\shatb}[0]{\uvecb{s}}
\newcommand{\nhatb}[0]{\uvecb{n}}
\newcommand{\ohatb}[0]{\uvecb{o}}
\newcommand{\ahatb}[0]{\uvecb{a}}
\newcommand{\uhatb}[0]{\uvecb{u}}

\newcommand{\wt}[1]{ \widetilde{#1}  }
\newcommand{\wtb}[1]{ \widetilde{\bm{#1}}  }
\newcommand{\tv}[1]{ \widetilde{\bm{#1}}  }
\newcommand{\tvb}[1]{\widetilde{\mbf{#1}}}

\newcommand{\tvec}[1]{\tilde{\bm{#1}}}
\newcommand{\bv}[1]{\bar{\bm{#1}}}
\newcommand{\bvb}[1]{\bar{\mbf{#1}}}


\newcommand{\hdot}[2][.08ex]{\dot{\raisebox{0pt}[\dimexpr\height+#1][\depth]{$#2$}}}
\newcommand{\hddot}[2][.2ex]{\ddot{\raisebox{0pt}[\dimexpr\height+#1][\depth]{$#2$}}}

\newcommand{\ggvec}[2][-.2ex]{\vec{\raisebox{0pt}[\dimexpr\height+#1][\depth]{$\bm{#2}$}}}

\newcommand{\pbrak}[2]{\big\{#1\;;\;#2\big\}}

\newcommand{\eval}[2]{\left.{#1}\right\vert_{#2}}

\newcommand{\abs}[1]{ \big| #1 \big| }
\newcommand{\Abs}[1]{ \left| #1 \right| }
\renewcommand{\mag}[1]{ |\hspace{-0.1em}| #1 |\hspace{-0.1em}| }

\newcommand{\bwedge}[1]{ {\textstyle \bigwedge ^{\!#1}\, } }
\newcommand{\botimes}[1]{ {\textstyle \bigotimes ^{#1} } }

\newcommand{\scr}[1]{ \mathscr{#1} }

\newcommand{\tham}[1]{\tilde{\mathscr{#1}}}

\newcommand{\mfrak}[1]{ \mathfrak{#1} }
\newcommand{\mbb}[1]{ \mathbb{#1} }
\newcommand{\mrm}[1]{ \mathrm{#1} }
\newcommand{\mcal}[1]{ \mathcal{#1} }


\newcommand{\inv}[1]{ #1^{\scriptscriptstyle{-\!1}} }
\newcommand\trans[1]{ #1^{\scriptscriptstyle{-\!1}} }
\newcommand{\trn}[1]{ #1^{\ii{\textsf{\textup{T}}}}  }
\newcommand{\invtrn}[1]{ #1^{\ii{\textsf{\textup{--T}}}} }

\newcommand{\coord}[2]{[#1]_{_{\uvec{#2}}} }
\newcommand{\crd}[2]{#1^{\!\ii{\,\hat{#2}}}}

\newcommand{\del}[1]{ \nabla_{\!\!\!{\scriptsize {#1}}}  }

\newcommand{\hodge}[0]{ {\,^{ \text{\normalsize{${\star}$}} } \!} }

\newcommand{\ax}[0]{ {^{\times}} }
\newcommand{\axx}[1]{ {#1^{\times}} }

\let\oldcdot\cdot
\renewcommand*{\cdot}{{\mkern1.5mu\oldcdot\mkern1.5mu}}
\newcommand{\slot}[0]{{\oldcdot}} 
\newcommand{\cdt}[0]{ \pmb{\cdot} }
\newcommand{\bdt}[0]{ \bs{\cdot} }

\newcommand{\cddott}[0]{ \text{\normalsize{${\,\,\pmb{:}\,\,}$}} }

\newcommand{\cddot}[0]{ \text{\normalsize{${\,\,\underline{\pmb{:}}\,\,}$}} }

\newcommand{\bltt}[0]{ \text{\large{${\,\,\bullet\,\,}$}} }
\newcommand{\sblt}[0]{ \text{\scriptsize{${\,\,\bullet\,\,}$}} }

\newcommand{\blt}[0]{ \text{\large{${\,\,\underline{\bullet}\,\,}$}} }

\newcommand{\ring}[1]{\mathring{#1} }
\newcommand{\rring}[1]{ \accentset{\circ\circ}{#1} }
\newcommand{\strdot}[1]{\overset{_{*}}{#1}}


\newcommand{\nsp}[0]{\!\!\!\!}    
\newcommand{\nquad}[0]{\hspace{-1em}} 
\newcommand{\nqquad}[0]{\hspace{-2em}} 

\newcommand{\gvgr}[1]{{\tbf{#1}}}
\newcommand{\uvgr}[1]{\hat{\tbf{#1}}}
\newcommand{\tvgr}[1]{\widetilde{\tbf{#1}}}
\newcommand{\bvgr}[1]{\bar{\tbf{#1}}}

\newcommand{\gvbb}[1]{\vec{\pmb{#1}}}
\newcommand{\uvbb}[1]{\hat{\pmb{#1}}}
\newcommand{\tvbb}[1]{\widetilde{\pmb{#1}}}
\newcommand{\bvbb}[1]{\bar{\pmb{#1}}}

\renewcommand{\a}[0]{{\alpha}}
\renewcommand{\b}[0]{{\beta}}
\renewcommand{\g}[0]{{\gamma}}
\newcommand{\y}[0]{{\lambda}}
\renewcommand{\v}[0]{{\nu}}
\renewcommand{\u}[0]{{\mu}}
\renewcommand{\r}[0]{{\rho}}
\renewcommand{\s}[0]{{\sigma}}
\newcommand{\ep}[0]{\epsilon}
\newcommand{\w}[0]{\omega}
\renewcommand{\th}[0]{\theta}

\newcommand{\upa}[0]{{\upalpha}}
\newcommand{\upb}[0]{{\upbeta}}
\newcommand{\upg}[0]{{\upgamma}}
\newcommand{\upy}[0]{{\uplambda}}
\newcommand{\upd}[0]{{\updelta}}
\newcommand{\upv}[0]{{\upnu}}
\newcommand{\upu}[0]{{\upmu}}
\newcommand{\upr}[0]{{\uprho}}
\newcommand{\upsig}[0]{{\upsigma}}
\newcommand{\upw}[0]{{\upomega}}
\newcommand{\upe}[0]{\upepsilon}
\newcommand{\upth}[0]{\uptheta}

\newcommand{\txa}[0]{{\normalsize\text{\textalpha}}}
\newcommand{\txb}[0]{{\text{\textbeta}}}
\newcommand{\txg}[0]{{\text{\textgamma}}}
\newcommand{\txy}[0]{{\text{\textlambda}}}
\newcommand{\txv}[0]{{\text{\textnu}}}
\newcommand{\txu}[0]{{\text{\textmu}}}
\newcommand{\txr}[0]{{\text{\textrho}}}
\newcommand{\txsig}[0]{{\normalsize\text{\textsigma}}}
\newcommand{\txw}[0]{{\normalsize \text{\textomega}}}
\newcommand{\txe}[0]{{\normalsize \text{\textepsilon}}}
\newcommand{\txth}[0]{{\text{\texttheta}}}
\newcommand{\txphi}[0]{{\text{\textphi}}}
\newcommand{\txpsi}[0]{{\text{\textpsi}}}
\newcommand{\txi}[0]{{\text{\textiota}}}

\newcommand{\wtxt}[0]{{\textomega}}
\newcommand{\ytxt}[0]{{\textlambda}}
\newcommand{\thtxt}[0]{{\texttheta}}
\newcommand{\btxt}[0]{{\textbeta}}
\newcommand{\atxt}[0]{{\textalpha}}
\newcommand{\gtxt}[0]{{\textgamma}}

\newcommand{\eq}[1]{${#1}$}

\newcommand{\rnote}[1]{
{\small\textit{{\color{red}note:} {\color{darkgray}  #1}}} }

\newcommand{\bnote}[1]{
{\small\textit{{\color{blue}note:} {\color{darkgray}  #1}}} }

\newcommand{\gnote}[1]{
{\small\textit{{\color{darkgray}\textbf{note:}  #1}}} }



\newtheorem{theorem}{Theorem}
\newtheorem{remark}{Remark}
\newtheorem{defn}{Definition}

\newtheoremstyle{remrmsmall}
{8pt} 
{12pt} 
{\rmfamily\itshape}
{}
{\rmfamily\small\bfseries}
{.}
{.3em}
{}

\theoremstyle{remrmsmall}
\newtheorem{remrm}{Remark}[section]

\begin{abstract}
 A geometric interpretation of the equinoctial elements is given with a connection to orthogonal rotations and attitude dynamics in Euclidean 3-space. An identification is made between the equinoctial elements and classic Rodrigues parameters.  A new set of equinoctial elements are developed using the modified Rodrigues parameters, thereby removing the coordinate singularity for retrograde equatorial orbits present in previous versions of these elements. A low-thrust trajectory optimization problem is set up using the new elements to numerically verify convergence for the two-point boundary problem, as compared to their predecessors.\footnote[4]{A shorter version of this work appeared as a conference paper entitled \textit{On Equinoctial Elements and Rodrigues Parameters}, AAS 22-832, presented at the AAS/AIAA Astrodynamics Specialist Conference, Charlotte NC, August 2022. }
\end{abstract}

\renewcommand{\contentsname}{}
\tableofcontents
\newpage

\section{Introduction}


The classic orbit elements (COEs) are set of six state-space coordinates used to model and study the dynamics of an orbiting particle in the classic two-body problem (2BP). Five of these elements are integrals of motion for the unperturbed Kepler problem and are osculating elements (i.e., ``slow'' variables) in the presence of a non-Keplerian perturbing force (per unit reduced mass), \eq{\gvb{a}\in\mathbb{E}^3}. These five elements — along with their governing equations in the form of Gauss's variational equations — are given as follows \cite{brouwer2013methods,schaub2003analytical}:
\begin{small}
\begin{align}  
\label{COE_eom_0}
&\begin{array}{rllll}
  \fnsz{\text{semi-major axis, }} a \quad &\quad 
 \dot{a} \;=\; 2\tfrac{r}{h}a^2(e\sin \nu \,\uvecb{u}_1 + w\,\uvecb{u}_2) \cdot\gvb{a}
\\[4pt]
  \fnsz{\text{eccentricity, }} e \quad &\quad 
 \dot{e} 
    \;=\; \tfrac{r}{h}\big[ w \sin{\nu}\,\uvecb{u}_1 + \big( (w+1)\cos{\nu}+e\big)\uvecb{u}_2\big]\cdot\gvb{a}
\\[4pt]
 \fnsz{\text{inclination, }} i  \quad &\quad
 \diff{i}{t} \;=\; \tfrac{r}{h}\cos{\theta}\,\uvecb{u}_3\cdot\gvb{a} 
    &(\theta=\upomega+\nu)
\\[4pt]
     \text{RAAN, } \Omega  \quad &\quad 
    \hdot{\Omega}  \;=\; \tfrac{r}{h}\tfrac{\sin{\theta}}{\sin i}\uvecb{u}_3 \cdot\gvb{a}
\\[4pt]
  \fnsz{\text{argument of periapse, }} \upomega \quad &\quad 
 \dot{\upomega}  \;=\;  \tfrac{r}{h}\big[ \tfrac{1}{ e}\big( -w\cos{\nu}\,\uvecb{u}_1 + (w+1)\sin{\nu}\,\uvecb{u}_2 \big) -  \tfrac{\sin{\theta}}{\tan i}\uvecb{u}_3  \big] \cdot\gvb{a}
 \end{array}
\end{align}
\end{small}
where RAAN stands for \textit{right ascension of ascending node} and where all symbols appearing on the right-hand-side are described in the footnote.\footnote[5]{The unit vectors \eq{\uhatb_i\in\mathbb{E}^3} are the orthonormal local vertical, local horizontal (LVLH) basis vectors of the orbiting body, \eq{\mu} the gravitational parameter of the central body, \eq{r}  the radial distance from the central body, \eq{h}  the specific orbital angular momentum magnitude, \eq{p} the \textit{semilatus rectum}, \eq{\theta=\upomega+\nu} the \textit{argument of latitude}, \eq{b} the semi-minor axis, \eq{n} the mean motion, and \eq{w} unnamed.  In terms of the COEs:
\begin{align} \label{somethings}
    p = a(1-e^2) 
   &&,&&
      h^2  =   \mu p
   &&,&&
   r  =  \tfrac{a(1-e^2)}{1+e\cos{\nu}} = \tfrac{p}{1+e\cos{\nu}}
&&,&&
    w = \tfrac{p}{r} = 1+\cos\nu
&&,&& 
   n^2  =  \tfrac{\mu}{a^3}
   &&,&&
   b^2  =  ap  
\end{align}
}
The sixth COE is related to the phase angle and is typically chosen as one of the following anomalies:
\begin{small}
\begin{align} \label{d_angle}
\begin{array}{rllll}
    \fnsz{\text{true anomaly, }} \nu \quad &\quad 
    \dot{\nu}  \;=\;  \tfrac{r}{h e}\big[  w\cos{\nu}\uvecb{u}_1 - ( w+1)\sin{\nu}\uvecb{u}_2 \big] \cdot\gvb{a} \;+\; \tfrac{h}{r^2}
\\[4pt]
     \fnsz{\text{mean anomaly, }}  M \quad &\quad 
     \dot{M} \;=\; \tfrac{rb}{ha e}\big[ (w \cos{\nu}-2e)\uhatb_1 - (w+1)\sin{\nu}\uvecb{u}_2 \big] \cdot\gvb{a} \;+\;  n
\\[4pt]
     \fnsz{\text{eccentric anomaly, }} E \quad &\quad 
    \hdot{E} \;=\;   \tfrac{1}{nae}\big[ (\cos{\nu}-e)\uhatb_1 - (1+\tfrac{r}{a})\sin{\nu}\uvecb{u}_2 \big] \cdot\gvb{a} \;+\; \tfrac{a}{r}n
\end{array}
\end{align}
\end{small}
Unlike Eq.~\eqref{COE_eom_0}, the above equations do not reduce to zero for the Kepler problem (\eq{\gvb{a}=\gvb{0}}); \eq{\nu}, \textit{M}, and \textit{E}   are ``fast'' variables, cycling through  \eq{2\pi} every Kepler orbit.  There are also
``slow'' versions of these variables in use.\footnote{The common ``slow'' counterparts to Eq.~\eqref{d_angle} are: \eq{\tau_0} (time of periapse) instead of \eq{\nu}, \eq{M_0} (\eq{M} at periapse), or \eq{E_0} (\eq{E} at periapse).  Furthermore, although less common, rather than \eq{\nu}, \eq{M}, or \eq{E}, some sources instead use the argument of latitude, \eq{\theta=\upomega +\nu}, the mean argument of latitude, \eq{\upomega +M}, or the eccentric argument of latitude, \eq{\upomega +E}, as the sixth element. Or, alternatively, the ``slow'' versions, \eq{\upomega+M_0} or \eq{\upomega+E_0}.     }
The equations for  \eq{\hdot{\Omega}} and \eq{\dot{\upomega}} in Eq.~\eqref{COE_eom_0}  have singularities at \eq{e=0} (circular orbit)  and/or \eq{i=0\pm n\pi} for any integer \eq{n} (equatorial and retrograde equatorial orbits).  As seen in
Eq.~\eqref{d_angle}, the equations for \eq{\dot{\nu}}, \eq{\dot{M}}, and \eq{\hdot{E}} also all have a singularity at \eq{e=0} (this is not resolved by using their slow counterparts).  
Furthermore, as is well known,  the elements \eq{\Omega} and \eq{\upomega}  become undefined as \eq{i\to 0} and \eq{e\to 0}, respectively. 
Several alternative sets of elements have been developed to
avoid these troublesome issues, one of the most well-known being the \textit{equinoctial elements}.

\textit{The Equinoctial Elements. }
There are several closely related sets of elements referred to as some form of the equinoctial elements \cite{arsenault1970orbit,broucke1972equinoctial,cefola1972equinoctial,walker1985set,bau2021generalization,biria2018equinoctial}. 
Although their origins can be traced back to Lagrange \cite{lagrange1783theorie}, it was not until centuries later that the term \textit{equinoctial} elements came into use when Arsenault introduced the so-called equinoctial reference frame and defined a set of elements in relation to this frame.  Shortly after, Broucke and  Cefola presented a slightly different set of equinoctial elements along with their Lagrange and Poisson brackets for the state transition matrix \cite{broucke1972equinoctial,cefola1972equinoctial}. A little over a decade later, another small but noteworthy modification to the equinoctial elements was made by Walker et al  \cite{walker1985set}. These three sets of orbit elements, all referred to as some form of the equinoctial elements, are defined  in terms of the classic orbit elements as follows (using the original labeling of the cited authors):\footnote{The element \eq{h} appearing in Eq.~\eqref{EE_defs} is \textit{not} the specific angular momentum magnitude, as is the case everywhere else that \eq{h} appears in this paper.  }
\begin{small}
\begin{align} \label{EE_defs}
\begin{array}{ll}
    \text{\underline{Arsenault et al, 1970 \cite{arsenault1970orbit}}} 
\\
     n \,=\, \sqrt{\tfrac{\mu}{a^3}} 
\\
     a_f  \,=\, e\cos(\Omega+\upomega)
 \\
     a_g  \,=\,  e\sin(\Omega+\upomega)
 \\
      \psi  \,=\, \tan\tfrac{i}{2}\cos\Omega
  \\
      \chi  \,=\, \tan\tfrac{i}{2}\sin\Omega
\\
     L  \,=\, \Omega\,+\, \upomega \,+\, M
\end{array}
\quad\;\;
\left| \quad\;\;
\begin{array}{ll}
 \text{\underline{Broucke \& Cefola, 1972 \cite{broucke1972equinoctial,cefola1972equinoctial}}} 
\\
     a
\\
     k  \,=\, e\cos(\Omega+\upomega)
\\
     h  \,=\,  e\sin(\Omega+\upomega)
\\
      q  \,=\, \tan\tfrac{i}{2}\cos\Omega
\\
      p  \,=\, \tan\tfrac{i}{2}\sin\Omega
\\
     \lambda_0  \,=\, \Omega\,+\, \upomega \,+\, M_0
\end{array} \right.
\quad\;\;
\left| \quad\;\;
\begin{array}{ll}
\text{\underline{Walker et al, 1985 \cite{walker1985set}}}
& \quad  \text{\underline{Our Notation}}
\\
     p  \,=\, a(1-e^2) 
     &\qquad= p
\\
     f  \,=\, e\cos(\Omega+\upomega)  &\qquad = e_1
     \\
     g  \,=\,  e\sin(\Omega+\upomega) &\qquad = e_2
     \\
     h  \,=\, \tan\tfrac{i}{2}\cos\Omega &\qquad = q_1
      \\
      k  \,=\, \tan\tfrac{i}{2}\sin\Omega &\qquad =  q_2
    \\
     L  \,=\, \Omega\,+\, \upomega \,+\, \nu &\qquad = l
\end{array} \right.
\end{align}
\end{small}
The term \textit{equinoctial elements}, with no further descriptors, usually refers to the elements in the second column above.
The elements in the third column are typically referred to as the \textit{modified equinoctial elements} (MEEs) and have seen significant use in the decades since their introduction, particularly in spacecraft GNC \cite{kechichian1996optimal,peterson2020lyapunov} and trajectory optimization, where they have demonstrated efficient convergence for the two point boundary value problem \cite{junkins2019exploration}. Unlike the COEs, the above elements, as well as their governing equations (given in Eq.~\eqref{MEE_eom_tru} of this paper) are free of singularities at \eq{e=0} and \eq{i=0} \cite{broucke1972equinoctial,battin1999introduction,schaub2003analytical}. The MEEs (third column) in particular possess only a single coordinate singularity at \eq{i=\pm \pi} (retrograde equatorial orbit). 
While the notation  differed between the above choices for equinoctial elements, the second through fifth elements remained the same and yet, to the authors' knowledge, have never been named. We change the notation once again and, for reasons that will become apparent, denote these  as \eq{e_1}, \eq{e_2}, \eq{q_1}, and \eq{q_2}.



The element sets listed in \eqref{EE_defs} are
perhaps the the three most well-known stages in the evolution of the equinoctial elements. However, earlier iterations exist. The elements we label \eq{e_1} and \eq{e_2} can be   found in the 1961 edition of the text from Brouwer and Clemence (pp.~287) under the labels \eq{k} and \eq{h} \cite{brouwer2013methods}. A similar version of the elements we label \eq{q_1} and \eq{q_2} can be found in this same text (pp.~288), albeit in a different form that uses \eq{\tan{i}} (singular at  \eq{i=\pm \tfrac{\pi}{2}}) rather than \eq{\tan\tfrac{i}{2}}. In fact, this  \eq{\tan{i}} version of \eq{q_1} and \eq{q_2} can be traced back centuries earlier to Lagrange's theory of planetary motion, where they are found along with low-inclination approximations of \eq{e_1} and \eq{e_2}  \cite{lagrange1783theorie}. 
On the other side of the timeline, in the last decade, subsequent versions of the MEEs have been developed which generalize them for other geometries and perturbations. Biria and Russell used spheroidal coordinates to  develop the \textit{oblate spheroidal equinoctial elements} in \cite{biria2018equinoctial} and used them to solve Vinti's problem in \cite{biria2020analytical}. Baù et al further generalized the MEEs, developing a version which embeds any conservative perturbing forces into the definitions of the elements themselves \cite{bau2021generalization}.  

\textit{Outline of this Work. }
Although the various versions of equinoctial elements resolve most of the singularity issues associated with the COEs, their connection to the geometry of the 2BP is incomplete and seldom discussed. When encountered in standard texts on the subject such as \cite{brouwer2013methods,battin1999introduction}, the equinoctial elements  are typically introduced simply by  defining them algebraically in relation to the COEs as in Eq.~\eqref{EE_defs}, with little discussion of their origins or meaning (with the exception of the semilatus rectum, \textit{p}).
This paper sheds some light on the matter.  We show that the equinoctial elements are closely related to concepts commonly encountered in the study of orthogonal rotations and attitude kinematics in Euclidean 3-space. It is shown that the elements \eq{e_1,e_2,q_1,q_2,l} arise organically when examining kinematic relations between the local vertical local horizontal (LVLH), perifocal, and \textit{equinoctial} bases (discussed in section \ref{sec:frames}).  In particular,  it is shown that two of the equinoctial elements (\eq{q_1} and \eq{q_2}) are in fact attitude coordinates known as the \textit{Classic Rodrigues Parameters} (CRPs) — a detail that, it seems, has not yet been  identified in the literature. 
The CRPs are known to have a coordinate singularity for principal rotations of \eq{\pm \pi}, corresponding to the coordinate singularity in the MEEs for retrograde equatorial orbits.

Upon recognizing that the MEEs contain two CRP attitude coordinates,  we facilitate the next  step in their evolution by replacing these coordinates with the closely related \textit{Modified Rodrigues Parameters} (MRPs), thereby shifting the coordinate singularity to \eq{i=\pm 2\pi}. 
This new set of equinoctial elements is creatively dubbed as \textit{MRP modified equinoctial elements} (MRP MEEs) and  we will occasionally refer to the usual MEEs in Eq.~\eqref{EE_defs} as the CRP MEEs. 
Although we give a short overview of the CRPs and MRPs that is sufficient for the purpose of this work, readers interested in a more thorough exposition of these attitude coordinates are referred to \cite{schaub1996stereographic,marandi1987preferred} and chapter 3 of \cite{schaub2003analytical}.

The remainder of this paper is structured as follows.  We begin in section \ref{sec:att} by reviewing four attitude representations in Euclidean 3-space ($\mbb{E}^3$): the principal axis and angle, the CRPs, the MRPs, and Euler angles. Then, in section \ref{sec:frames}, 
we review three orthonormal bases encountered in celestial mechanics: the local vertical, local horizontal (LVLH) basis; the perifocal basis; and the lesser-known \textit{equinoctial basis}.
In section \ref{sec:MEE}
we show the aforementioned connection between the equinoctial elements and the geometry of the 2BP and, in section \ref{sec:MRP MEE}, develop the new MRP equinoctial elements. 
Lastly, in section \ref{sec:traj}, a low-thrust trajectory optimization problem is formulated and solved for demonstrating numerical utility and performance of the new set. A review of orthogonal rotations and angular velocity in Euclidean 3-space can be found in Appendix \ref{app:rot}.  A collection of partial derivatives needed for the the costate equations of the optimal control problem are found in Appendix \ref{app:costate}.

\paragraph{Conventions \& Notation.}
The body of this paper, particularly sections \ref{sec:att} and \ref{sec:frames}, assumes familiarity with orthogonal rotations and angular velocity in Euclidean 3-space (denoted \eq{\mbb{E}^3}), a review of which may be found in Appendix \ref{app:rot}. For the reasons given at the beginning of said appendix, no distinction will be made between \eq{\mbb{E}^3} and its dual space; all bases are assumed orthonormal and all indices will appear as subscripts. As such, tensors are treated as ``cartesian tensors'' wherein no distinction is made between contravariant and covariant components; 
all tensors of order \eq{0\leq k\leq 3} are referred to as \textit{k}-tensors and considered elements of \eq{\botimes{k}\mbb{E}^3} (the \eq{k^{\tx{th}}} tensor power).  Likewise, all alternating (skew-symmetric) \textit{k}-tensors may be regarded as elements of \eq{\bwedge{k}\mbb{E}^3\subset \botimes{k}\mbb{E}^3} (the \eq{k^{\text{th}}} exterior power) such that \eq{\bwedge{2}\mbb{E}^3} is regarded the space of skew-symmetric 2-tensors, and \eq{\bwedge{3}\mbb{E}^3} is identified with the 1-dimensional space of volume forms on \eq{\mbb{E}^3}.  The symbol \eq{\wedge} denotes the exterior product (``wedge'' product) which, for vectors, is regarded as the atisymmetrized tensor product (e.g., \eq{\gvb{u}\wedge \gvb{v}=\gvb{u} \otimes \gvb{v} -\gvb{v}\otimes\gvb{u} = -\gvb{v} \wedge\gvb{u}} for any \eq{\mbf{u},\mbf{v}\in\mbb{E}^3}).  
\eq{\delta_{ij}} and \eq{\epsilon_{ijk}} will, respectively, denote the three-dimensional Kronecker and Levi-Civita symbols defined in Eqs.~\eqref{LC_3d} and \eqref{LCdel} of Appendix \ref{app:rot}.\footnote{That is, \eq{\delta_{ij}} and \eq{\epsilon_{ijk}} are the components of the euclidean metric and volume form (Levi-Civita pseudotensor) in any right-handed orthonormal basis.}
Einstein summation notation is used throughout with summation from 1 to 3 over repeated indices (e.g., \eq{\delta_{ii} =3} and \eq{\mbf{u}\cdot\mbf{v}=u_i v_i := u_1v_1 + u_2v_2+u_3v_3}, where \eq{u_i=\mbf{u}\cdot\ehatb_i} and \eq{v_i=\mbf{v}\cdot\ehatb_i} are the vectors' components in any orthonormal basis, \eq{\ehatb_i}, for \eq{i=1,2,3}).  When used in the optimal control problem formulation of section \ref{sec:traj}, the indices instead range from \eq{1} to \eq{6}. 

\section{Review: Attitude Representations} \label{sec:att} 

The tools of attitude dynamics typically encountered in the context of rigid body dynamics are also applicable in the context of point-mass orbital dynamics.
The transformation of orthonormal bases is conceptually congruent with the rotation of rigid bodies. Accordingly, there is a strong analogy between rigid body dynamics and the motion of bases commonly used in orbital mechanics (see \cite{junkins1979analogy} for a more in-depth discussion).
For the entirety of this section, we consider two sets of right-handed  orthonormal — but otherwise arbitrary — basis vectors, \eq{\ehatb_i,\ehatb'_i\in\mbb{E}^3} (for \eq{i=1,2,3}). 
As discussed in Appendix \ref{app:rot}, there exists some  \eq{\mbf{R}\in \mrm{SO}(\mbb{E}^3)}, such that 
\begin{align} \label{ib_0}
    \ehatb'_i \,=\, \mbf{R}\cdot\ehatb_i  \,=\, R_{ji}\ehatb_j
     &&,&&
    \ehatb_i \,=\, \trn{\mbf{R}}\cdot\ehatb'_i \,=\, R_{ij} \ehatb'_j
    &&,&&
         \mbf{R}\,=\,\ehatb_i'\otimes\ehatb_i 
         \,=\, \invtrn{\mbf{R}}
     &&,&&
     |\mbf{R}| \,\dot{=}\, \det(\mbf{R})  =  1 
\end{align}
where \eq{R_{ij}=\ehatb_i\cdot\mbf{R}\cdot\ehatb_j \equiv \ehatb_i'\cdot\mbf{R}\cdot\ehatb_j'=R_{ij}' = \ehatb_i\cdot\ehatb'_j }.  We assume both  bases are rotating arbitrarily 
and use a dot to denote the inertial time-derivative and a ring to denote the \eq{\ehatb_i}-corotational time-derivative (Appendix \ref{sec:ang_vel}, Eq.~\eqref{frame_diff}):
\begin{small}
\begin{align}
    \dot{(\cdot)} \;\;\dot{=}\;\; \diff{}{t}(\cdot)
&&,&&
    \ring{(\cdot)} \;\;\dot{=}\;\; \fdiff{\hat{e}}(\cdot)
&&,&&
    \ring{\ehatb}_i = \gvb{0}
    \quad \qquad 
    \ring{\ehatb}'_i \,=\, \tvgr{\txw}\cdot\ehatb'_i
\end{align}
\end{small}
where \eq{\tvgr{\txw}\in\bwedge{2}\mbb{E}^3} is the angular velocity tensor of the  \eq{\ehatb'_i} basis (relative to \eq{\ehatb_i}), with corresponding axial vector \eq{\gvgr{\txw}} 
such that \eq{ \tvgr{\txw}\cdot\gvb{u} = \gvgr{\txw}\times\gvb{u}, \,\; \forall \gvb{u}\,\in\mathbb{E}^3} (see Appendix \ref{sec:ang_vel}, Eqs.~\eqref{hodge} - \eqref{ux_props}).\footnote{In $\mbb{E}^3$, the axial operator (Appendix \ref{app:axial}) is a self-inverse isomorphism between $\mbb{E}^3$ and $\bwedge{2}\mbb{E}^3$, such that, for any $\mbf{W}\in\bwedge{2}\mbb{E}^3$:
\begin{align} \label{hodge_3d_00}
    \mathbb{E}^3  \ni 
   \;\;
     \gvb{w} \,=\, \tvb{W} \,=\, -\tfrac{1}{2}\epsilon_{ijk}(\ehatb_j\cdot\mbf{W}\cdot\ehatb_k)\ehatb_i
\quad\;\;\leftrightarrow \quad\;\;
       \mbf{W}  \,=\, \tvb{w} \,=\,  
      -\tfrac{1}{2}\epsilon_{ijk}(\gvb{w}\cdot\ehatb_k)\ehatb_i\wedge\ehatb_j  
     \;\;  \in  \bwedge{2} \mathbb{E}^3 
\end{align}
} 
The skew-symmetric 2-tensor \eq{\tvgr{\txw}} and the (pseudo)vector \eq{\gvgr{\txw}}  — which are the ``axial duals'' of one-another 
— are given by
\begin{align} \label{wrel_0}
    \tvgr{\txw} \,=\, \ring{\mbf{R}}\cdot\trn{\mbf{R}}  \,=\, \tfrac{1}{2}(\ehatb'_i\cdot \ring{\ehatb}'_j)\ehatb'_i\wedge\ehatb'_j 
    \;=\; \ring{\ehatb}'_i\otimes \ehatb'_i
    && \leftrightarrow &&
    \gvgr{\txw} 
    \,=\,  \tfrac{1}{2}(\ehatb'_i\cdot \ring{\ehatb}'_j)\ehatb'_j\times\ehatb'_i  \,=\,  \tfrac{1}{2}\ehatb'_i\times \ring{\ehatb}'_i 
\end{align}
As seen in appendix \ref{app:orth}, \eq{\text{dim}(\mrm{SO}(\mbb{E}^n))=\tfrac{n}{2}(n-1)}.  Thus, any \eq{\mbf{R}\in \mrm{SO}(\mbb{E}^3)}, has three independent components in any basis; only three parameters are required to define the nine \eq{R_{ij}}, which encode the attitude of some orthonormal basis with respect to another.  
We now present four such attitude representations which will later be connected to the classical and equinoctial elements:  the principal axis and angle (four rather than three parameters), classic Rodrigues parameters (CRPs),   modified Rodrigues parameters (MRPs), and Euler angles.

\subsection{The Principal Axis and Angle} \label{sec:euler axis/angle}



An intuitive way of conceptualizing rotations in \eq{\mbb{E}^3} follows from Euler's rotation theorem, summarized as follows: \textit{Any rotation, or sequence of rotations, in Euclidean 3-space of a rigid-body with one point fixed is equivalent to a single rotation about an axis through this point.}
The axis of rotation, called the \textit{principal axis}, will be denoted by a unit vector, \eq{\uvgr{\txa}\in\mbb{E}^3}. The angle through which the body (basis) is rotated about \eq{\uvgr{\txa}} is called the \textit{principal angle} and will be denoted by \eq{\varphi\in\mbb{R}}.
Thus, for any orthonormal basis transformation (rigid rotation) as in Eq.~\eqref{ib_0}, the rotation tensor is fully defined by some \eq{\uvgr{\txa}} and \eq{\varphi} such that \eq{\ehatb'_i = \mbf{R}(\uvgr{\txa},\varphi)\cdot\ehatb_i}
where the left (resp., right) action of \eq{\mbf{R}(\uvgr{\txa},\varphi)} on any vector rotates said vector about axis \eq{\uvgr{\txa}} through angle \eq{\varphi} (resp., \eq{-\varphi}).  
\eq{\mbf{R}(\uvgr{\txa},\varphi)} is given by Rodrigues' rotation formula  \cite{shuster1993survey}:
\begin{align} \label{Rephi_00}
    \mbf{R}(\uvgr{\txa},\varphi) \,=\, 
   \cos\varphi\, \mbf{I} \,+\, \sin\varphi\,\tvgr{\txa} \,+\, (1-\cos\varphi)\uvgr{\txa}\otimes\uvgr{\txa} 
    \,\;=\;\,
    \mbf{I} \,+\, \sin{\varphi}\,\tvgr{\txa} \,+\, (1-\cos{\varphi})\tvgr{\txa}\cdot\tvgr{\txa}
    &&,&&
     \mbf{R}(\uvgr{\txa},\varphi)\cdot\uvgr{\txa}=\uvgr{\txa}
\end{align}
where \eq{\uvgr{\txa}} is the axial vector of \eq{\tvgr{\txa}\in \bwedge{2} \mathbb{E}^3}. 
It can be shown,  and is also rather intuitive, that the axis, \eq{\uvgr{\txa}}, is an eigenvector of \eq{\mbf{R}} with \eq{+1} eigenvalue, as indicated above. \footnote{The other two eigenvalues are expressed in terms of the angle, \eq{\varphi}, as a complex exponential, \eq{\text{e}^{\pm \text{i}\varphi}}.} 
The three components of \eq{\uvgr{\txa}} are thus the same in both bases: \eq{\alpha_i' = \uvgr{\txa}\cdot\ehatb_i' = \uvgr{\txa}\cdot\ehatb_i = \alpha_i}.
Given any \eq{\mbf{R}\in\mrm{SO}(\mbb{E}^3)}, we may extract it's effective axis and angle by taking the trace and antisymmetric part of Eq.~\eqref{Rephi_00}:\footnote{ \eq{\text{tr}(\mbf{I})=3},  \eq{\;\text{tr}(\tvb{u})=0}, \eq{\;\text{tr}(\gvb{u}\otimes\gvb{u})=u^2}, and  \eq{\;\text{tr}(\tvb{u}\cdot\tvb{u})=-2u^2}. The trace of a 2-tensor is invariant under proper orthogonal rotations.}
\begin{small}
\begin{align} \label{axis_Rax} 
\begin{array}{lllll}
      \text{tr}(\mbf{R}) \,=\,  R_{ii} 
    \,=\, 2\cos\varphi + 1 
\\[3pt]
     \tfrac{1}{2}(\mbf{R}-\trn{\mbf{R}}) \,=\, \sin\varphi \tvgr{\txa} 
\end{array}
 && \Rightarrow  &&
 \begin{array}{llll}
       \cos\varphi \,=\, \tfrac{1}{2}(R_{ii}-1)  
   \\[3pt]
      \uvgr{\txa} \sin\varphi  \;=\; \wt{\tfrac{1}{2} ( \mbf{R}-\trn{\mbf{R}}) } \;=\; \tvb{R} 
    \;=\; -\tfrac{1}{2}\epsilon_{ijk}R_{jk}\ehatb_i 
\end{array}
\end{align}
\end{small}
Although \eq{\uvgr{\txa}} and  \eq{\varphi} fully define the rotation tensor through Eq.~\eqref{Rephi_00}, the relation is not unique: 
\begin{small}
\begin{align}
    \mbf{R}(\uvgr{\txa},\varphi) \,=\, \mbf{R}(\text{–}\uvgr{\txa},\text{–}\varphi)  \,=\, \mbf{R}(\uvgr{\txa},\varphi\pm 2n\pi)
    \qquad, &&
    \inv{\mbf{R}}(\uvgr{\txa},\varphi) \,=\, \trn{\mbf{R}}(\uvgr{\txa},\varphi) \,=\,
    \mbf{R}(\text{–}\uvgr{\txa},\varphi)\,=\, \mbf{R}(\uvgr{\txa},\text{–}\varphi)
\end{align}
\end{small}

\paragraph{{Axis and Angle Attitude Kinematics.} } 
The angular velocity of \eq{\ehatb'_i} relative to \eq{\ehatb_i} is defined as in equation \eqref{wrel_0}. Thus, \eq{\tvgr{\txw}} may be expressed in terms of \eq{\uvgr{\txa}} and \eq{\varphi} by substitution of \eq{\mbf{R}(\uvgr{\txa},\varphi)} from Eq.~\eqref{Rephi_00} into \eq{\tvgr{\txw}=\ring{\mbf{R}}\cdot\trn{\mbf{R}}}. 
Alternatively, we may obtain \eq{\gvgr{\txw}} by differentiating both sides of Eq.~\eqref{axis_Rax} and then using \eq{\ring{\mbf{R}}=\tvgr{\txw}\cdot\mbf{R}} to obtain  \eq{\ring{\uvgr{\txa}}} and \eq{\dot{\varphi}} in terms of \eq{\gvgr{\txw}}. Taking either approach, one obtains the following \cite{gelman1971note,casey2009remark}:
\begin{align}   \label{wx_EE}
     &\gvgr{\txw} \,=\, \hdot{\varphi}\uvgr{\txa}  \,+\, \big[\sin\varphi\mbf{I} \,+\, (1-\cos\varphi) \tvgr{\txa} \, \big] \cdot\ring{\uvgr{\txa}} 
     &&,&& 
     \tvgr{\txw} \,=\, \hdot{\varphi}\tvgr{\txa}  \,+\, \sin\varphi \ring{\tvgr{\txa}} \,-\, (1-\cos\varphi) \uvgr{\txa}\wedge\ring{\uvgr{\txa}}
 \\  \label{EEdot_w} 
     &\ring{\uvgr{\txa}} 
    \,=\, -\tfrac{1}{2} (\, \tvgr{\txa} \,+\, \cot\tfrac{\varphi}{2}\tvgr{\txa}\cdot\tvgr{\txa} \, ) \cdot\gvgr{\txw}
    &&,&&
     \ring{\tvgr{\txa}} \,=\, 
      \tfrac{1}{2}\big( \uvgr{\txa}\wedge\gvgr{\txw} \,+\, \cot\tfrac{\varphi}{2} \big[ \,\tvgr{\txw} \,-\, (\gvgr{\txw}\cdot\uvgr{\txa}) \tvgr{\txa}\, \big] \big)
      &&,&&
      \dot{\varphi} \,=\, \gvgr{\txw} \cdot \uvgr{\txa} 
\end{align}
We are typically interested in the coordinate representations of the above tensor equations in either the \eq{\ehatb_i} or \eq{\ehatb'_i} basis. Since \eq{\alpha_i'=\alpha_i},  clearly \eq{\dot{\varphi}= \omega_i\alpha_i = \omega'_i\alpha_i}.  Writing the above expressions for  \eq{\gvgr{\txw}} and \eq{\ring{\uvgr{\txa}}} in both bases leads to \cite{shuster1993survey}:
\begin{align}   \label{EEdot_w_mat}  
\begin{array}{lllll}
      \omega_i \;=\;
    \dot{\varphi} \alpha_i \,+\, \big[ \delta_{ij}\sin\varphi  \,+\, (1-\cos\varphi) \wt{\alpha}_{ij} \big] \dot{\alpha}_j
 &\quad,&\quad
     \omega'_i \;=\;
    \dot{\varphi} \alpha_i \,+\, \big[ \delta_{ij}\sin\varphi  \,-\, (1-\cos\varphi) \wt{\alpha}_{ij} \big] \dot{\alpha}_j 
\\[3pt]
     \dot{\alpha}_i \,=\,  -\tfrac{1}{2}\big[\, \wt{\alpha}_{ij} \,+\, \cot(\tfrac{\varphi}{2})\wt{\alpha}_{in}\wt{\alpha}_{nj} \, \big] \omega_j
     &\quad\equiv&\quad
     \dot{\alpha}'_i 
          \;=\; \tfrac{1}{2}\big[\, \wt{\alpha}_{ij} \,-\, \cot(\tfrac{\varphi}{2})\wt{\alpha}_{in}\wt{\alpha}_{nj} \, \big] \omega'_j
\end{array}
\end{align}
where \eq{\alpha_i=\alpha'_i} and where sign differences follow from kinematic transport, \eq{\dot{\alpha}_i\ehatb_i=\dot{\alpha}_i\ehatb'_i+\tvgr{\txw}\cdot\uvgr{\txa}}.


\subsection{Classic Rodrigues Parameters (CRPs)} \label{sec:CRPdyn}
The Rodrigues vector (or Gibbs vector), \eq{\gvb{q}}, is an often-used attitude description which may be defined in terms of the principal axis and angle  as \eq{\gvb{ q} \;\dot{=}\; \tan{\tfrac{ \varphi}{2}}\uvgr{\txa} }.
It can be shown that the rotation tensor is given in terms of this vector as \cite{schaub2003analytical}
\begin{align} \label{CRPdef}  
    \gvb{q} \;\dot{=}\; \tan{\tfrac{ \varphi}{2}}\uvgr{\txa} 
    &&,&&
    \mbf{R}(\gvb{ q}) 
     \;=\; \tfrac{2}{1+q^2}\Big( \tfrac{1}{2}(1-q^2)\mbf{I} \;+\; \tvb{q} \;+\; \gvb{q}\otimes\gvb{q} \Big)
    \;\;=\;\;
    \mbf{I} \;+\; \tfrac{2}{1+ q^2}\tvb{q} \;+\; \tfrac{2}{1+ q^2}\tvb{q}\cdot\tvb{q}
\end{align}
where \eq{q^2=\gvb{q}\cdot\gvb{q}=\tan^2\tfrac{\varphi}{2}}.
Taking the trace and axial vector of the above \eq{\mbf{R}}, one finds \eq{  R_{ii} + 1 = \tfrac{4}{1+q^2} } and \eq{\tvb{R} = \tfrac{2}{1+q^2}\gvb{q} } such that  the Rodrigues vector may be extracted from \eq{\mbf{R}} as
\begin{align}
     \tvb{q} \;=\; \tfrac{1}{1+R_{ii}}(\mbf{R}-\trn{\mbf{R}}) 
&&
     \gvb{q} \;=\;
    \tfrac{2}{1+R_{ii}}\tvb{R} 
\end{align}
Note that \eq{\mbf{R}(\gvb{q})\cdot\gvb{q} =\gvb{q}} such that, for two bases related by \eq{\ehatb'_i=\mbf{R}\cdot\ehatb_i}, 
then \eq{\mbf{R}(\gvb{q})\cdot\gvb{q} =\gvb{q}} and \eq{ q_i = \gvb{q}\cdot\ehatb_i = \gvb{q}\cdot\ehatb'_i =  q_i'}.
These three components, \eq{q_i}, are called the classic Rodrigues parameters (CRPs). The  matrix representation  of the above \eq{\mbf{R}} (in the \eq{\ehatb_i} and \eq{\ehatb'_i} bases) is then given in terms of the CRPs by
\begin{small}
\begin{align} \label{Rij_crp}
  & R(q_i) = \fnsz{
     \frac{1}{1+ q^2}
 \begin{pmatrix}  
    1 +   q_1^2 - q_2^2 - q_3^2    &    2(  q_1  q_2 -q_3 ) &   2( q_1  q_3 +  q_2 )
 \\
  2( q_1  q_2 +  q_3 )  &  1+ q_2^2 -  q_1^2 -   q_3^2  &   2( q_2  q_3 - q_1 ) 
\\
    2( q_1  q_3 - q_2 )  &   2(  q_2  q_3 +  q_1 ) &  1+  q_3^2 -  q_1^2 -   q_2^2 
\end{pmatrix}  }
\end{align}
\end{small}



\paragraph{{CRP Attitude Kinematics.}  } 
By differentiating Eq.~\eqref{CRPdef}, and making use of the expressions for \eq{\dot{\varphi}} and \eq{\ring{\uvgr{\txa} }} given by Eq.~\eqref{EEdot_w}, it can be shown that \eq{\gvgr{\txw}} (angular velocity of the \eq{\ehatb'_i} basis relative to the \eq{\ehatb_i} basis)  and  \eq{\ring{\gvb{q}}} are related by
\begin{small}
\begin{align} \label{CRPdot_vec}
      \ring{\gvb{q}}  \,=\, \tfrac{1}{2}\big( \mbf{I} \,-\,  \tvb{q}   \,+\, \gvb{q}\otimes\gvb{q} \big) \cdot\gvgr{\txw}
&&\leftrightarrow&&
      \gvgr{\txw} \;=\;  \tfrac{2}{1+q^2}( \mbf{I} \;+\; \tvb{q})\cdot\ring{\gvb{q}}
      \qquad \Rightarrow \qquad
       \tvgr{\txw} \;=\; \tfrac{2}{1+q^2}( \ring{\tvb{q}} \,-\, \gvb{q}\wedge\ring{\gvb{q}} )
\end{align}
\end{small}
The above are general tensor equations valid in any basis. 
By writing the first two relations  in the \eq{\ehatb_i} and \eq{\ehatb'_i} bases, and using \eq{q'_i=q_i} along with the kinematic transport theorem to write \eq{\dot{q}_i\ehatb_i=\dot{q}_i\ehatb'_i+\tvgr{\txw}\cdot\gvb{q}}, we obtain\cite{shuster1993survey}
\begin{small}
\begin{align}  \label{w_crp}  
 \begin{array}{rl}
      \dot{q}_i &\!\!=\; 
       \tfrac{1}{2}(\delta_{ij}  \;-\; \wt{q}_{ij} \;+\; q_iq_j ) \omega_j 
\\[3pt]
    &\!\!=\;\;
     \tfrac{1}{2}(\delta_{ij}  \;+\; \wt{q}_{ij} \;+\; q_iq_j ) \omega'_j
\end{array}
&&\leftrightarrow&&  
\begin{array}{ll}
    \omega_i \;=\;  
    \tfrac{2}{1+q^2}(\delta_{ij} + \wt{q}_{ij})\dot{q}_j
   \\[3pt]
    \omega'_i \;=\;  
    \tfrac{2}{1+q^2}(\delta_{ij} - \wt{q}_{ij})\dot{q}_j
\end{array}   
\end{align}
\end{small}
which may also be collected into matrices. For instance, the above relations for \eq{\dot{q}_i} and \eq{\omega'_i}  may be written as 
 \begin{align}  \label{crp_ode_gen}
     [\dot{\bm{q}}]  = \frac{1}{2}
     \fnsz{\begin{pmatrix}
            q_1^2 + 1 & q_1 q_2 - q_3 & q_2 + q_1 q_3 \\
     q_3 + q_1 q_2 &   q_2^2 + 1 & q_2 q_3 - q_1 \\
     q_1 q_3 - q_2 & q_1 + q_2 q_3 &   q_3^2 + 1 \\
   \end{pmatrix} }
   [\bm{\omega}]'
&&, &&
      [\bm{\omega}]'= \ \frac{2}{1+q^2}
      \fnsz{  \begin{pmatrix}
        1 & q_3 & -q_2 \\
        -q_3 & 1 & q_1 \\
        q_2 & -q_1 & 1
     \end{pmatrix}   }
     [\dot{\bm{q}}] 
\end{align}

\begin{remrm}[CRP Singularities] \label{rem:crp sing}
Although there is no visual singularity  in the kinematics of Eq.~\eqref{crp_ode_gen}, the CRPs themselves, \eq{ q_i \;\dot{=}\; \tan{\tfrac{ \varphi}{2}}\alpha_i }, have a \emph{coordinate singularity} at principal angles of \eq{\varphi = \pm\pi}; \eq{\tan{\tfrac{ \varphi}{2}}\rightarrow \pm\infty} as \eq{\varphi\rightarrow\pm\pi}.
\end{remrm}

\subsection{Modified Rodrigues Parameters (MRPs)} \label{sec:MRPdyn}

The \textit{modified} Rodigues vector, \eq{\gvgr{\txsig}}, 
is defined in terms of the principal axis and angle in a similar manner as the classic Rodrigues vector, but with a magnitude of \eq{\tan{\tfrac{\varphi}{4}}} \cite{schaub2003analytical,schaub1996stereographic}:\footnote{The reason for this seemingly minor difference of \eq{\tan{\tfrac{\varphi}{4}}} and  \eq{\tan{\tfrac{\varphi}{2}}} arises from the interpretation of the  MRPs and CRPs as stereographic projection coordinates. We do not discuss this relation but refer readers to \cite{schaub1996stereographic,marandi1987preferred}.      }
\begin{align} \label{MRPdef}
     \gvgr{\txsig}\;\dot{=}\; \tan{\tfrac{ \varphi}{4}}\uvgr{\txa}  \;\;=\; \tfrac{1}{1+\sqrt{1+q^2}}\gvb{q}
     &&,&&
      \mbf{R}(\gvgr{\txsig}) 
    \;\;=\;\;
    \mbf{I} \;+\; 4\tfrac{1-\sigma^2}{(1+\sigma^2)^2} \tvgr{\txsig}\;+\; \tfrac{8}{(1+\sigma^2)^2} \tvgr{\txsig}\cdot \tvgr{\txsig}
\end{align}
where \eq{\sigma^2=\tan^2\tfrac{\varphi}{4}}.  
 Taking the trace and axial vector of the above \eq{\mbf{R}}, one finds \eq{R_{ii}+1 = 4\big(\tfrac{1-\sigma^2}{1+\sigma^2}\big)^2} and \eq{\tvb{R} = 4\tfrac{1-\sigma^2}{(1+\sigma^2)^2} \gvgr{\txsig} } such that \eq{\gvgr{\txsig}} is given in terms of \eq{\mbf{R}} as
\begin{small}
\begin{align}
    \tvgr{\txsig} \;=\; \tfrac{2}{(2+\sqrt{R_{ii}+1})(\sqrt{R_{ii}+1})}\tfrac{1}{2}(\mbf{R}-\trn{\mbf{R}}) 
\qquad,&&
    \gvgr{\txsig} \;=\; \tfrac{2}{(2+\sqrt{R_{ii}+1})(\sqrt{R_{ii}+1})}\tvb{R} 
\end{align}
\end{small}
We again note \eq{\mbf{R}(\gvb{q})\cdot\gvb{q} =\gvb{q}} such that, for two bases related by \eq{\ehatb'_i=\mbf{R}\cdot\ehatb_i}, 
then \eq{\mbf{R}(\gvgr{\txsig})\cdot\gvgr{\txsig} = \gvgr{\txsig}  }  and \eq{\sigma_i = \gvgr{\txsig}\cdot\ehatb_i = \gvgr{\txsig}\cdot\ehatb'_i =  \sigma'_i}.
These three components, \eq{\sigma_i}, are called the modified Rodrigues parameters (MRPs). 
The matrix representation of \eq{\mbf{R}} (in the \eq{\ehatb_i} and \eq{\ehatb'_i} bases) from Eq.~\eqref{MRPdef} is then given in terms of the MRPs by
\begin{small}
\begin{align} \label{Rij_mrp}
 &R \;= \tfrac{4}{(1+\sigma^2)^2}
    \fnsz{\begin{pmatrix}
     k^2 +  \sigma_1^2 -\sigma_2^2 - \sigma_3^2  &    2( \sigma_1 \sigma_2- k\sigma_3) &   2(\sigma_1 \sigma_3 + k\sigma_2 )
 \\
  2(\sigma_1 \sigma_2 + k \sigma_3 )  &  k^2+\sigma_2^2 - \sigma_1^2 -  \sigma_3^2  &   2(\sigma_2 \sigma_3 -k\sigma_1 ) 
\\
    2(\sigma_1 \sigma_3 - k\sigma_2 )  &   2(\sigma_2 \sigma_3 + k\sigma_1 ) &  k^2+ \sigma_3^2 - \sigma_1^2 -  \sigma_2^2 
    \end{pmatrix} }
    \;\;,&&
    k\;\dot{=}\; \tfrac{1}{2}(1-\sigma^2)
\end{align}
\end{small}

\paragraph{{MRP Attitude Kinematics.} } 
As before, we may use \eq{\tvgr{\txw}=\ring{\mbf{R}}\cdot\trn{\mbf{R}}} with \eq{\mbf{R}} given by Eq.~\eqref{MRPdef} to obtain an expression for \eq{\tvgr{\txw}} in terms of \eq{\ring{\gvgr{\txsig}}}. We may also differentiate Eq.~\eqref{MRPdef} and use equation  \eqref{CRPdot_vec} for \eq{\dot{\varphi}} and \eq{\ring{\uvec{\alpha}}}, or equation  \eqref{EEdot_w} for \eq{\ring{\gvb{q}}}, to obtain \eq{\ring{\gvgr{\txsig}}} in terms of \eq{\gvgr{\txw}}. Whatever method is used, the result may be written as
\begin{align} \label{MRPdot_vec}
 \begin{array}{rlll}
    &\ring{\gvgr{\txsig}} \;=\;  \tfrac{1}{2} \big[ \tfrac{1}{2}(1+\sigma^2)\mbf{I} \;-\; \tvgr{\txsig} \;+\; \tvgr{\txsig}\cdot\tvgr{\txsig} \big] \cdot \gvgr{\txw} 
    &=\;\;  \tfrac{1}{2} \big[ \tfrac{1}{2}(1-\sigma^2)\mbf{I} \;-\; \tvgr{\txsig} \;+\; \gvgr{\txsig}\otimes\gvgr{\txsig} \big] \cdot \gvgr{\txw} 
\\[4pt]
   & \gvgr{\txw} \;=\;  \tfrac{8}{(1+\sigma^2)^2}  \big[ \tfrac{1}{2}(1+\sigma^2)\mbf{I} \;+\; \tvgr{\txsig} \;+\; \tvgr{\txsig}\cdot\tvgr{\txsig} \big] \cdot \ring{\gvgr{\txsig}}
    &=\;\; \tfrac{8}{(1+\sigma^2)^2}  \big[ \tfrac{1}{2}(1-\sigma^2)\mbf{I} \;+\; \tvgr{\txsig} \;+\; \gvgr{\txsig}\otimes\gvgr{\txsig} \big] \cdot \ring{\gvgr{\txsig}}
\\[4pt]
  \fnsz{\text{also:}}  &\tvgr{\txw} \;=\; \tfrac{8}{(1+\sigma^2)^2}  \big[ \tfrac{1}{2}(1-\sigma^2) \ring{\tvgr{\txsig}} \;-\; \gvgr{\txsig}\wedge \ring{\gvgr{\txsig}} \;+\; (\gvgr{\txsig}\cdot\ring{\gvgr{\txsig}})\tvgr{\txsig} \big] 
\end{array}  
\end{align}
 We write Eq.~\eqref{MRPdot_vec} in the \eq{\ehatb_i} and \eq{\ehatb'_i} bases and, using \eq{\sigma'_i=\sigma_i} along with  \eq{\dot{\sigma}_i\ehatb_i=\dot{\sigma}_i\ehatb'_i+\tvgr{\txw}\cdot\gvgr{\txsig}}, we obtain the expressions for \eq{\dot{\sigma}_i} in terms of \eq{\omega_i} and \eq{\omega'_i}, as well as the inverse relations \cite{shuster1993survey}:
\begin{small}
\begin{align}  \label{mrp_dot}
      &\dot{\sigma}_i \;=\; 
       \tfrac{1}{2} \big[ \tfrac{1}{2}(1-\sigma^2)\delta_{ij} \;-\; \wt{\sigma}_{ij} \;+\; \sigma_{i} \sigma_{j} \big] \omega_j 
    \;\;=\;\;
     \tfrac{1}{2} \big[  \tfrac{1}{2}(1-\sigma^2)\delta_{ij} \;+\; \wt{\sigma}_{ij} \;+\; \sigma_{i} \sigma_{j} \big] \omega'_j
\\[2pt]  \label{w_of_mrp}
    &\omega_i \;=\;  
   \tfrac{8}{(1+\sigma^2)^2} \big[ \tfrac{1}{2}(1-\sigma^2)\delta_{ij} \;+\; \wt{\sigma}_{ij} \;+\; \sigma_{i} \sigma_{j} \big] \dot{\sigma}_j
\qquad,\qquad 
    \omega'_i \;=\;  
   \tfrac{8}{(1+\sigma^2)^2} \big[ \tfrac{1}{2}(1-\sigma^2)\delta_{ij} \;-\; \wt{\sigma}_{ij} \;+\; \sigma_{i} \sigma_{j} \big] \dot{\sigma}_j
\end{align}
\end{small}
Collecting  the terms into matrices, the above relations for \eq{\dot{\sigma}_i} and \eq{\omega'_i} may be written as 
\begin{small}
\begin{align}  \label{MRPdyn_gen}
 & [\dot{\bm{\sigma}}] \;= \frac{1}{2}
     \fnsz{\begin{pmatrix}
      \tfrac{1}{2}(1+ \sigma_1^2 -  \sigma_2^2 -  \sigma_3^2)  &   -( \sigma_3- \sigma_1  \sigma_2)    &     \sigma_2 +   \sigma_1  \sigma_3
    \\
      \sigma_3 +   \sigma_1  \sigma_2 &   \tfrac{1}{2}(1- \sigma_1^2 +  \sigma_2^2 -  \sigma_3^2)  &     -(\sigma_1-  \sigma_2  \sigma_3)
    \\
     -(  \sigma_2- \sigma_1  \sigma_3)  & \sigma_1 +  \sigma_2  \sigma_3 & \tfrac{1}{2}(1-  \sigma_1^2 -  \sigma_2^2 +  \sigma_3^2) 
\end{pmatrix}  }  [ \bm{\omega}]'
\\[2pt] \label{MRPdyn2_gen} 
    & [\bm{\omega}]' \;=
  \frac{8}{(1+\sigma^2)^2}
   \fnsz{ \begin{pmatrix}
      \tfrac{1}{2}(1+ \sigma_1^2 - \sigma_2^2 - \sigma_3^2)  &    \sigma_3 +  \sigma_1 \sigma_2 &   -( \sigma_2 - \sigma_1 \sigma_3)
    \\
     -(\sigma_3- \sigma_1 \sigma_2)  & \tfrac{1}{2}(1 - \sigma_1^2 + \sigma_2^2 - \sigma_3^2 )  &   \sigma_1 +  \sigma_2 \sigma_3
    \\
      \sigma_2 +  \sigma_1 \sigma_3 &   -(\sigma_1 - \sigma_2 \sigma_3) & \tfrac{1}{2}( 1- \sigma_1^2 - \sigma_2^2 + \sigma_3^2 )
\end{pmatrix}  }  [\dot{\bm{\sigma}}]
\end{align}
\end{small}

\begin{remrm}[MRP singularities] \label{rem:mrp sing}
From the definition given by Eq.~\eqref{MRPdef}, it is seen that \eq{\sigma_i\rightarrow\infty} as \eq{\varphi\rightarrow \pm2\pi}. 
Thus,  an  advantage of the MRPs over the CRPs is that the coordinate singularity is shifted from \eq{\varphi= \pm \pi} (for the CRPs) to \eq{\varphi= \pm2\pi} (for the MRPs). 
\end{remrm}

\begin{remrm}[``Shadow'' MRPs]
It is worthwhile to note that, in contrast to the CRPs, the MRPs are not unique; it can be shown that a numerically distinct ``shadow'' set of MRPs, \eq{\sigma_i^*}, describe the same rotation as the above \eq{\sigma_i} and are related   by \eq{\sigma_i^*\,\dot{=}\, -\tfrac{1}{\sigma^2}\sigma_i} and \eq{\sigma_i= -\tfrac{1}{{\sigma^*}^2}\sigma_i^* } such that, as one set blows up to infinity, the other goes to zero.
Furthermore, if \eq{\mag{\gvgr{\txsig}}\leq 1 }, then \eq{\mag{\gvgr{\txsig}^*}\geq 1 }, and vice versa, such that one set always lies within the closed unit 3-ball; as \eq{\mag{\gvgr{\txsig}}} leaves, \eq{\mag{\gvgr{\txsig}^*}} enters, and vice versa (with \eq{\mag{\gvgr{\txsig}}=\mag{\gvgr{\txsig}^*}=1 } at \eq{\varphi=\pm\pi}). Either \eq{\sigma_i}  or \eq{\sigma_i^*} may be used as attitude parameters, with \eq{\sigma_i^*} satisfying the same kinematic equations as \eq{\sigma_i} given above. By switching from \eq{\sigma_i} to \eq{\sigma_i^*} as the \eq{\sigma_i} leave the closed 3-ball, one can circumvent the coordinate singularity at \eq{\varphi=\pm 2\pi} without changing the governing differential equation given by Eq.~\eqref{MRPdyn_gen} \cite{schaub2003analytical,schaub1996stereographic}. If the switch from \eq{\sigma_i} to \eq{\sigma^*_i} is not made exactly as the two sets pass through the unit 2-sphere, then one uses the simple transformation \eq{\sigma_i^*= -\tfrac{1}{\sigma^2}\sigma_i}. 
\end{remrm}

\subsection{Simple Rotations \& Euler Angles}

\paragraph{{Simple Rotations.} }
For two bases related by \eq{\ehatb'_i=\mbf{R}\cdot\ehatb_i}, the rotation is said to be a simple 1-rotation, 2-rotation, or 3-rotation if \eq{\ehatb'_1=\ehatb_1}, \eq{\ehatb'_2=\ehatb_2}, or \eq{\ehatb'_3=\ehatb_3}, respectively. That is, a simple \textit{p}-rotation of any basis, \eq{\ehatb_i},  is a rotation about \eq{\ehatb_p}. In other words, \eq{\ehatb'_p=\mbf{R}\cdot\ehatb_p = \ehatb_p} (\textit{p} is not a free index here) is the principle axis of rotation; \eq{\mbf{R}=\mbf{R}(\ehatb_p,\theta)} for some angle \eq{\theta}. For any basis, there are three such simple rotations and we define three \textit{simple Eulerian rotation matrices}, \eq{M_p(\theta)}, as the matrix representation of \eq{\mbf{R}(\ehatb_p,\theta)} in the \eq{\ehatb_i} basis itself (the basis is arbitrary);  \eq{M_p(\theta) \;\dot{=}\; \crd{R}{e}(\ehatb_p,\theta)}.  These matrices are found using the matrix representation of Eq.~\eqref{Rephi_00}:\footnote{Although the matrix representation of a rotation tensor in a certain basis may take the form of one of the simple rotation matrices, these matrices themselves are defined without reference to any basis and depend only on the angle.  }
\begin{small}
\begin{align} \label{M_k}
    M_1(\theta)  \;\dot{=}\;  { \begin{bmatrix}
    1 & 0 & 0 \\
   0 &  \cos\theta & -\sin\theta \\
   0 & \sin\theta & \cos\theta
    \end{bmatrix} }
    &&, &&
    M_2(\theta)  \;\dot{=}\;   { \begin{bmatrix}
   \cos\theta & 0 & \sin\theta \\
   0 & 1 & 0 \\
   -\sin\theta & 0 & \cos\theta
    \end{bmatrix} }
    &&, &&
     M_3(\theta)  \;\dot{=}\;   { \begin{bmatrix}
   \cos\theta & -\sin\theta & 0\\
   \sin\theta & \cos\theta & 0\\
   0 & 0 & 1\\
    \end{bmatrix} }
\end{align}
\end{small}
where we have used the fact that the coordinate representation of any basis vector in its own basis is simply 
 \begin{small}
\begin{align} \label{Ihats}
  \crd{[\uvec{e}_1]}{e} \,=\,  \hat{1} \;\dot{=}\;  
  \trn{\big[ \, 1 \;\; 0 \;\; 0 \, \big]}
   &&
   \crd{[\uvec{e}_2]}{e} \,=\, \hat{2} \;\dot{=}\;  
  \trn{\big[ \, 0\;\; 1 \;\; 0 \, \big]}
   &&
    \crd{[\uvec{e}_3]}{e} \,=\,  \hat{3}  \;\dot{=}\;  
  \trn{\big[ \, 0 \;\; 0 \;\; 1 \, \big]}
\end{align}
\end{small}

\paragraph{{Euler Angles.} }
Any arbitrary orthonormal basis transformation can be decomposed into a sequence of three simple rotations. The angle of rotation for each of the individual rotations constitutes a set of three attitude coordinates called the \textit{Euler angles}. 
Denote by \eq{\mbf{R}_p\cdot\ehatb_i} the simple \textit{p}-rotation of basis  \eq{\ehatb_i}. We then define  \eq{\mbf{R}_{lmn}(\phi,\theta,\psi)} as a \textit{l-m-n} simple rotation \textit{sequence} through the Euler angles \eq{\phi}-\eq{\theta}-\eq{\psi}. That is, \eq{\ehatb'_i = \mbf{R}_{lmn}(\phi,\theta,\psi)\cdot\ehatb_i } is interpreted as
 three consecutive basis transformations:
a simple \textit{l}-rotation from  \eq{\ehatb_i} to some intermediate basis, \eq{\uvecb{a}_i}; followed by a simple \textit{m}-rotation from \eq{\uvecb{a}_i} to a second intermediate basis, \eq{\uvecb{b}_i}; and lastly a simple \eq{n}-rotation from \eq{\uvecb{b}_i} to the basis \eq{\ehatb'_i}:
\begin{small}
\begin{align} \label{Rdecomp}
\ehatb'_i  =  \mbf{R}_{lmn}(\phi,\theta,\psi) \cdot \ehatb_i  =  
&\left\{ \!\begin{array}{lll}
    1:   \;\uvecb{a}_i  =  \mbf{R}_l(\phi)\cdot\ehatb_i 
 \\
      2:  \; \uvecb{b}_i = \mbf{R}_m(\theta)\cdot\uvecb{a}_i
       =  \mbf{R}_{lm}(\phi,\theta)\cdot\ehatb_i
 \\
    3:  \; \ehatb'_i = \mbf{R}_n(\psi)\cdot\uvecb{b}_i  =  \mbf{R}_{mn}(\theta,\psi)\cdot\uvecb{a}_i
     =  \mbf{R}_{lmn}(\phi,\theta,\psi)\cdot \ehatb_i
\end{array}\!\right\} 
\,= 
\mbf{R}_n(\psi)\cdot\mbf{R}_m(\theta)\cdot\mbf{R}_l(\phi)\cdot\ehatb_i
\end{align}
\end{small}
\sloppy Or, in terms of principle axes and angles, \eq{\ehatb'_i  =\mbf{R}(\bhatb_n,\psi)\cdot\mbf{R}(\uvecb{a}_m,\theta)\cdot\mbf{R}(\ehatb_l,\phi)\cdot\ehatb_i   }.  The subscripts \eq{l},  \eq{m}, and  \eq{n} in Eq.~\eqref{Rdecomp} are \textit{not} free indices.  They simply denote the type of simple rotation (1, 2, or 3). We may choose \textit{l-m-n} to be any combination of 1, 2, and 3, 
so long as no two neighboring integers are the same (e.g.,  \textit{1-2-3, 3-1-2, 1-2-1,} etc.).\footnote{Two consecutive rotations about the same axis is no different than a single rotation about that axis;  \eq{\mbf{R}_p(\phi)\mbf{R}_p(\theta)=\mbf{R}_p(\phi+\theta)}. Although viable, this is simply redundant.}
There are therefore a total of twelve possible simple rotation sequences available for describing the same overall rotation.  
Let \eq{R_{lmn}(\phi,\theta,\psi)\in\mrm{M}_{3,3}(\mbb{R})} denote the matrix representation of \eq{\mbf{R}_{lmn}(\phi,\theta,\psi)\in\mbb{E}^3\otimes\mbb{E}^3} in the \eq{\ehatb_i} and \eq{\ehatb'_i}  bases.
Although the right-hand-side of Eq.~\eqref{Rdecomp} may seem to imply this would be  given in terms of the simple rotation matrices as \eq{ M_n(\psi) M_m(\theta) M_l(\phi)}, 
the sequence is actually reversed:\footnote{The matrix in \eq{\ehatb_i} is \eq{R_{lmn}(\phi,\theta,\psi)= \crd{{R}}{e}(\bhatb_n,\psi) \crd{{R}}{e}(\uvecb{a}_m,\theta) \crd{{R}}{e}(\ehatb_l,\phi)}. But, note that \eq{\crd{R}{e}(\bhatb_n,\psi) \neq M_n(\psi)= \crd{R}{b}(\bhatb_n,\psi)} and \eq{\crd{{R}}{e}(\uvecb{a}_m,\theta) \neq M_m(\theta)= \crd{{R}}{a}(\uvecb{a}_m,\theta)}.  We must use Eq.~\eqref{basisXform} to get each \eq{R} matrix into the form of one of the simple rotation matrices. Doing so leads to Eq.~\eqref{Rlmn_mat}  }
\begin{small}
\begin{align} \label{Rlmn_mat}
\text{for } \;\; \ehatb'_i=\mbf{R}_{lmn}(\phi,\theta,\psi)\cdot\ehatb_i  \quad
\left\{\; \begin{array}{ll}
       R_{lmn}(\phi,\theta,\psi) \;=\; M_{nml}(\psi,\theta,\phi) \;=\; M_l(\phi)M_m(\theta)M_n(\psi)
\\[4pt]
      \inv{\big[R_{lmn}(\phi,\theta,\psi)\big]} \;=\; \trn{\big[M_{nml}(\psi,\theta,\phi)\big]} \;=\; \trn{M}_n(\psi)\trn{M}_m(\theta)\trn{M}_l\!(\phi)
\end{array}\right.
\end{align}
\end{small}  
For a \textit{3-1-3} rotation sequence, the above leads to
\begin{small}
\begin{align} \label{R313} 
     R_{313}(\phi,\theta,\psi) &=
   M_3(\phi)M_1(\theta)M_3(\psi) =
   \fnsz{\begin{pmatrix}
       \cos{\phi}\cos{\psi} - \cos{\theta}\sin{\phi}\sin{\psi} & - \cos{\phi}\sin{\psi} - \cos{\psi}\cos{\theta}\sin{\phi} & \sin{\phi}\sin{\theta}
\\
 \cos{\psi}\sin{\phi} + \cos{\phi}\cos{\theta}\sin{\psi} & \cos{\phi}\cos{\psi}\cos{\theta} - \sin{\phi}\sin{\psi} & -\cos{\phi}\sin{\theta}
\\
\sin{\psi}\sin{\theta} &   \cos{\psi}\sin{\theta} &         \cos{\theta}
    \end{pmatrix} }
\end{align}
\end{small}

\paragraph{{Euler Angle Attitude Kinematics (\textit{3-1-3} sequence).}  }
For the relation given by equation  \eqref{Rdecomp}, one may find the angular velocity of  the \eq{\ehatb'_i} basis with respect to the \eq{\ehatb_i} basis in terms of Euler angles from Eq.~\eqref{wx_R} or \eqref{wx_R2} with \eq{R} given by Eq.~\eqref{Rlmn_mat}.  Alternatively, one may use the fact that the angular velocity is given by the sum of the angular velocities of each intermediate rotation in Eq.~\eqref{Rdecomp}.  E.g., for a \textit{3-1-3} rotation sequence, \eq{\ehatb'_i \;=\; \mbf{R}_{313}(\phi,\theta,\psi)\cdot \ehatb_i}, the angular velocity is \eq{\gvgr{\textomega} = \hdot{\phi}\ehatb_3 + \dot{\theta}\uvecb{a}_1 + \hdot{\psi}\ehatb'_3}.  
where \eq{\uvecb{a}_i=\mbf{R}_3(\phi)\cdot\ehatb_i} is an intermediate basis.
However, we would like this expressed in a single orthonormal basis.  Through inspection of Eq.~\eqref{Rdecomp}, we may re-write this as
\begin{small}
\begin{align}
    \gvgr{\textomega} \;=\; 
    \hdot{\phi}\ehatb_3 \,+\, \dot{\theta}\uvecb{a}_1 \,+\, \hdot{\psi}\ehatb'_3 \;=\;
    \hdot{\phi}\ehatb_3 \,+\, \dot{\theta}\mbf{R}_{3}(\phi)\cdot\ehatb_1 \,+\,
     \hdot{\psi}\mbf{R}_{313}(\phi,\theta,\psi)\cdot\ehatb_3
     \;=\;\hdot{\phi}\inv{\mbf{R}_{313}}(\phi,\theta,\psi)\cdot\ehatb'_3
      \,+\, \dot{\theta}\inv{\mbf{R}}_{3}(\psi)\cdot\ehatb'_1
      \,+\, \hdot{\psi}\ehatb'_3
\end{align}
\end{small}
Making use of equations \eqref{Rlmn_mat} and \eqref{Ihats}, this expressed in matrix form, in either the \eq{\ehatb_i} or \eq{\ehatb'_i} basis, as
\begin{small}
\begin{align} \nonumber
\begin{array}{ll}
      [\bm{\omega}] 
    \, =\,   \hdot{\phi}\hat{3} \, +\,  \dot{\theta} R_3(\phi)  \hat{1} \, +\,
    \hdot{\psi} R_{313}(\phi,\theta,\psi) \hat{3}
    &\!\!=\, \,   \hdot{\phi} \hat{3}  \, +\,  \dot{\theta} M_3(\phi) \hat{1} \, +\,   \hdot{\psi} M_3(\phi)M_1(\theta)M_3(\psi) \hat{3}
\\[4pt] 
    [\bm{\omega}]' 
     \, =\,   \hdot{\phi}\inv{R_{313}}(\phi,\theta,\psi)  \hat{3}
     \, +\,  \dot{\theta}\inv{R}_3(\psi) \hat{1} \, +\,  \hdot{\psi}\hat{3}
     &\!\! =\, \, 
     \hdot{\phi} \trn{M}_3(\psi)\trn{M}_1(\theta)\trn{M}_3(\phi) \hat{3} \, +\,  \dot{\theta} \trn{M}_3(\psi) \hat{1} \, +\,  \hdot{\psi}\hat{3}
\end{array}
\end{align}
\end{small}
where  \eq{[\bm{\omega}]} and \eq{[\bm{\omega}]'}  are the coordinate column vectors of \eq{\gvgr{\txw}} in the \eq{\ehatb_i} and \eq{\ehatb'_i} bases, respectively.  
Using Eq.~\eqref{M_k} the above equations, after considerable simplification and mapping back to \eq{\mathbb{E}^3}, give \eq{\gvgr{\textomega}} as
\begin{align} \label{w_313}
\begin{array}{rl}
     \gvgr{\textomega}  & \!\!\!=\; (\dot{\theta}\cos{\phi} +\hdot{\psi}\sin{\phi}\sin{\theta})\ehatb_1
    \;+\;
    (\dot{\theta}\sin{\phi} - \hdot{\psi}\cos{\phi}\sin{\theta}  )\ehatb_2
    \;+\; (\hdot{\phi}+\hdot{\psi}\cos{\theta})\ehatb_3
 \\[4pt]
     & \!\!\!=\; (\dot{\theta}\cos{\psi} +\hdot{\phi}\sin{\psi}\sin{\theta})\ehatb'_1
    \;+\;
    (\hdot{\phi}\cos{\psi}\sin{\theta} - \dot{\theta}\sin{\psi} )\ehatb'_2
    \;+\;(\hdot{\psi}+\hdot{\phi}\cos{\theta})\ehatb'_3
\end{array}
\end{align}
Or, keeping the equations in matrix form, these relations may be inverted to obtain 
\begin{small}
\begin{align} \label{Eang_dot_313}
    \begin{pmatrix}
       \omega'_1 \\
       \omega'_2 \\
       \omega'_3
    \end{pmatrix} =
    \begin{pmatrix}
       \sin\psi\sin\theta & \cos\psi & 0 \\
       \cos\psi\sin\theta & -\sin\psi & 0 \\
       \cos\theta &  0 & 1
    \end{pmatrix}
    \begin{pmatrix}
       \hdot{\phi} \\
       \dot{\theta} \\
       \hdot{\psi}
    \end{pmatrix}  
&&\Rightarrow&&
     \begin{pmatrix}
       \hdot{\phi} \\
       \dot{\theta} \\
       \hdot{\psi}
    \end{pmatrix} 
    =
    \begin{pmatrix}
       \tfrac{\sin\psi}{\sin\theta} & \tfrac{\cos\psi}{\sin\theta} & 0 \\
       \cos\psi & -\sin\psi & 0 \\
       -\cot\theta\sin\psi &  -\cot\theta\cos\psi & 1
    \end{pmatrix}  \begin{pmatrix}
       \omega'_1 \\
       \omega'_2 \\
       \omega'_3
    \end{pmatrix}
\end{align}
\end{small}

\begin{remrm}[Euler angle singularities]
For any given rotation, all twelve possible Euler angle representations will contain a singularity in their dynamics. For the 3-1-3 example given above, this singularity is at \eq{\theta=0\pm\pi}; the rightmost equation in Eq.~\eqref{Eang_dot_313} contains terms of the form  \eq{\tfrac{1}{\sin\theta}}.
\end{remrm}
\section{The LVLH, Perifocal, and Equinoctial Bases} \label{sec:frames}

\begin{figure}
    \centering    
    \includegraphics[width = 0.6\linewidth]{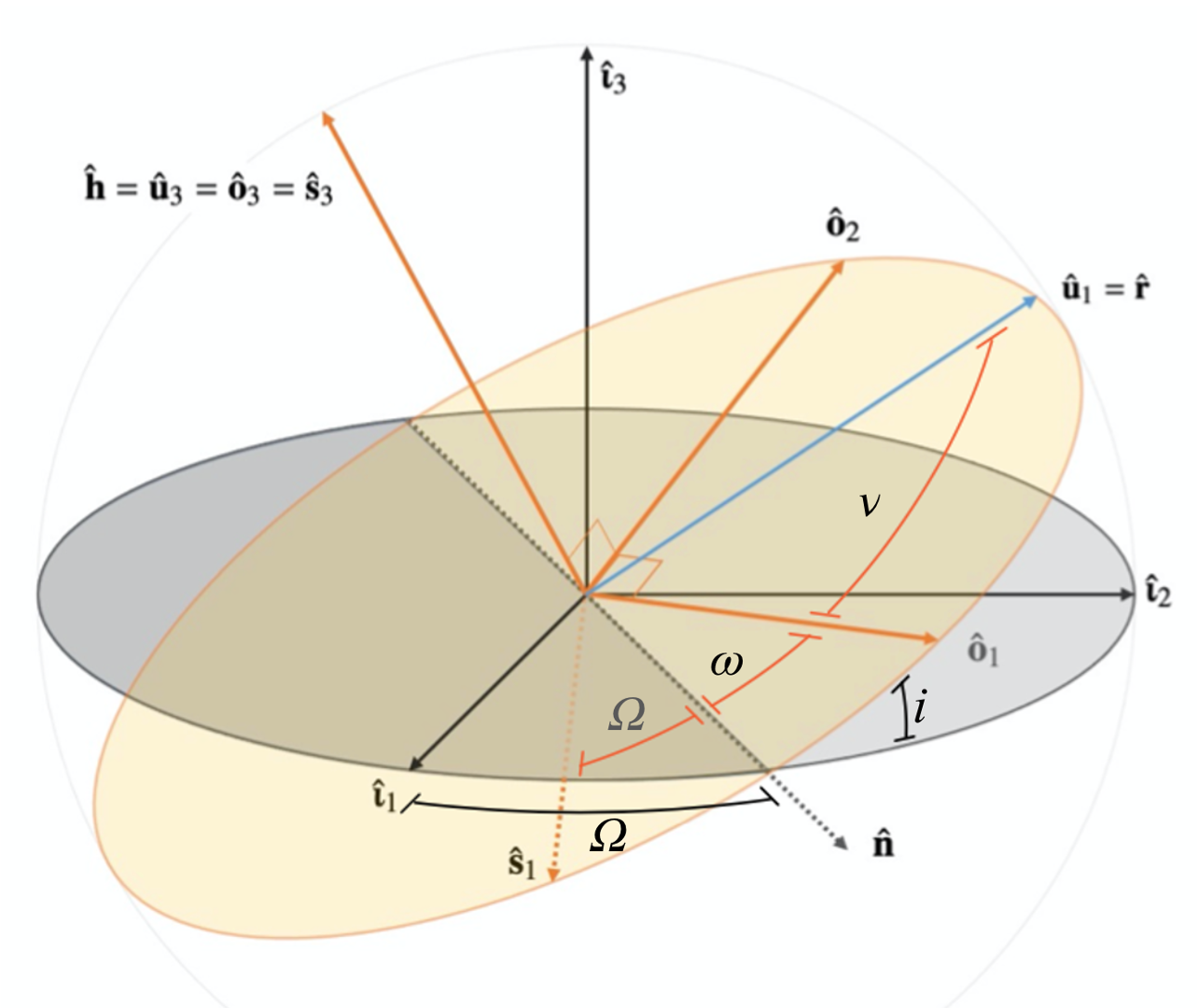}
    \caption{Orientation of the perifocal basis (\eq{\ohatb_i}), LVLH basis (\eq{\uhatb_i}), and equinoctial basis (\eq{\shatb_i}), relative to the inertial reference basis (\eq{\ihatb_i}). Orbital plane in orange.   }
    \label{fig:bases}
\end{figure}

In order to better understand the orbit elements introduced in subsequent sections, we will first define and briefly discuss some commonly used reference frames in celestial mechanics. Rather than use the general notation of \eq{\ehatb_i} and \eq{\ehatb'_i} as we have been doing, we will now denote specific bases as follows
\begin{small}
\begin{align}
\begin{array}{llll}
   \ihatb_i  \;=\; \text{inertial basis} 
   &\quad,&\quad \uvecb{u}_i  \;=\; \text{local vertical, local horizontal (LVLH) basis}
\\
    \ohatb_i \;=\; \text{perifocal basis}
 &\quad,&\quad
    \shatb_i \;=\; \text{equinoctial basis}
\end{array}
\end{align}
\end{small}
where \eq{i} ranges from \eq{1} to \eq{3} and where \eq{\ihatb_i} can be any inertial basis convenient for the problem at hand; it is simply a reference basis we use to define the other bases.\footnote{For instance \eq{\ihatb_i} may be the basis  corresponding to an earth-centered inertial (ECI) frame such as the the MJ2000 Earth Equator Frame, or it may the basis corresponding to the International Celestial Reference Frame (ICRF).  } 
The relative orientations of the above bases is depicted in Figure \ref{fig:bases}, which should serve as a useful reference for the following developments.\footnote{A few remarks on Figure \ref{fig:bases}:  all bases are depicted with their origin at the location of the central body (at \eq{\gvb{r}=\gvb{0}}) and the indicated unit vectors denote direction only; they are \textit{not} drawn to scale. \eq{\nhatb} denotes the line of nodes and \eq{\hhatb} the angular momentum direction. Note  \eq{\ohatb_1,\ohatb_2}, \eq{\uhatb_1,\uhatb_2}, and \eq{\shatb_1,\shatb_2} all lie in the osculating orbital plane (orange), with \eq{\ohatb_1} corresponding to the position of periapse (as well as the direction of the eccentricity vector), and \eq{\uhatb_1} to the orbiting body's instantaneous position.  \eq{\Omega}, \eq{\upw}, and \eq{\nu} denote the right ascension of the ascending node, argument of periapse, and true anomaly, respectively. \eq{i} denotes the inclination. The argument of latitude is \eq{\theta=\upw+\nu} and the true longitude is \eq{l = \Omega+\upw + \nu}.}

Several of the classic elements, \eq{e}, \eq{i}, \eq{\Omega}, \eq{\upomega}, \eq{\theta}, and \eq{\nu} will arise organically in defining the LVLH and perifocal bases. However, these relations are rather common textbook knowledge and we will not spend undue time rehashing them.
Later, it will also be shown that the less-common equinoctial basis, and the corresponding CRPs, provide a connection between the classical elements and the equinoctial elements; these relations are \textit{not} common textbook knowledge.

\subsection{The LVLH Basis} \label{sec:LVLH}
The instantaneous specific angular momentum (pseudo)vector of an orbiting body is given by \eq{\gvb{h}=\gvb{r}\times\dot{\gvb{r}}} (with \eq{\gvb{r}} defined relative to the central body). By definition, this vector is always orthogonal to the instantaneous orbital plane spanned by \eq{\gvb{r}} and \eq{\dot{\gvb{r}}}.   As such, we may define a rotating orthonormal basis — typically called the LVLH basis —  that rotates with the orbiting body and which is often denoted by \eq{\uvecb{r}=\tfrac{\gvb{r}}{r}},  \eq{\hhatb =\tfrac{\gvb{h}}{h}}, and \eq{\hat{\bm{\uptheta}}=\hhatb \times\uvecb{r}}.  
To facilitate the use of index notation, we denote these basis vectors as \eq{\{\uvecb{u}_1,\uvecb{u}_2,\uvecb{u}_3\}\,\dot{=}\, \{\uvecb{r},\hat{\bm{\uptheta}},\hhatb \}  },
\begin{small}
\begin{align} \label{LVLHdef}
     \uvecb{u}_1 \;\dot{=}\; \uvecb{r}  \quad, &&
    \uvecb{u}_2 \;\dot{=}\; \hhatb \times\uvecb{r} \;=\; \uvecb{u}_3 \times\uvecb{u}_1
    \quad,&&
    \uvecb{u}_3 \;\dot{=}\; \hhatb  \;=\; \frac{\gvb{r}\times\dot{\gvb{r}}}{\Vert \gvb{r}\times\dot{\gvb{r}} \Vert  }
\end{align}
\end{small}
Since this basis is orthonormal, there is some \eq{\mbf{R}\in\mrm{SO}(\mbb{E}^3)}, such that \eq{\uhatb_i=\mbf{R}\cdot\ihatb_i}. 
This rotation can be expressed as a sequence of three simple rotations, as in Eq.~\eqref{Rdecomp}.  It is well known that the classic elements \eq{\Omega} and \eq{i}, and the argument of latitude, \eq{\theta}, discussed in the introduction are precisely the Euler angles for \eq{\uhatb_i=\mbf{R}\cdot\ihatb_i} using a 3-1-3 rotation sequence:
\begin{small}
\begin{align} \label{nhat_def_R}
  \uhatb_i \;=\; \mbf{R}_{313}(\Omega,i,\theta)\cdot\ihatb_i 
    \;\;=\;
    \mbf{R}(\hhatb ,\theta)\cdot\mbf{R}(\uvecb{n},i)\cdot\mbf{R}(\ihatb_3,\Omega)
    \cdot\ihatb_i 
\end{align}
\end{small}
where the intermediate axis, \eq{\uvecb{n}}, is called the \textit{line of nodes} and is given by
\begin{small}
\begin{align} \label{nodes_1}
    \uvecb{n} 
    \;=\; \mbf{R}_3(\Omega)\cdot\ihatb_1 \;=\; \mbf{R}_3(-\theta)\cdot\uhatb_1
    \;=\; \cos\Omega\ihatb_1 \;+\; \sin\Omega\ihatb_2
    \;=\; \cos\theta\uhatb_1 \;-\; \sin\theta\uhatb_2
    \;\;=\; \frac{\ihatb_3 \times \mbf{h}}{\Vert  \ihatb_3  \times \mbf{h} \Vert }
\end{align}
\end{small}
From Eq.~\eqref{Rlmn_mat}, it is seen that the matrix representation (in the \eq{\uhatb_i} and \eq{\ihatb_i} bases) of the rotation tensor  in Eq.~\eqref{nhat_def_R} is given in terms of the simple rotation matrices as \eq{R_{313}(\Omega,i,\theta) = M_3(\Omega)M_1(i)M_3(\theta)}.
We can then note from Eq.~\eqref{w_313} that the angular velocity of the LVLH basis with respect to the inertial basis is given by
\begin{small}
\begin{align} \label{w_n/i}
    \gvgr{\txw}_{\hat{u}/\hat{\iota}} \;=\ \dot{\Omega}\,\ihatb_3 \;+\; \diff{i}{t}\uvecb{n} \;+\; \dot{\theta}\,\uhatb_3
    &\;=\;
    (\diff{i}{t}\cos{\theta}+\dot{\Omega}\sin{\theta}\sin{i})\uhatb_1
    \,+\,
    (\dot{\Omega}\cos{\theta}\sin{i} - \diff{i}{t}\sin{\theta} )\uhatb_2
    \,+\,(\dot{\theta}+\dot{\Omega}\cos{i})\uhatb_3
\\ \nonumber 
    &\;=\;  \tfrac{r}{h}\,(\uhatb_3\cdot\gvb{a}) \,\uhatb_1 \,+\,
    \tfrac{h}{r^2}\uhatb_3
\end{align}
\end{small}
where the second line follows from substitution \eq{\diff{i}{t}}, \eq{\dot{\Omega}},  \eq{\dot{\upomega}}, and \eq{\dot{\nu}} given in Eqs. \eqref{COE_eom_0} and \eqref{d_angle}:
\begin{small}
\begin{align} \label{diW}
    \diff{i}{t} \,=\, \tfrac{r}{h}\cos\theta\uvecb{u}_3\cdot\gvb{a} 
&&,&&\;\;
    \dot{\Omega}  \,=\, \tfrac{r}{h}\tfrac{\sin\theta}{\sin i}\uvecb{u}_3 \cdot\gvb{a}
    &&,&&\;\;
     \dot{\theta} \,=\, \dot{\upomega}+\dot{\nu} \,=\, -\tfrac{r}{ h}\tfrac{\sin\theta}{\tan{i}} \uhatb_3\cdot\gvb{a} \,+\, \tfrac{h}{r^2}
\end{align}
\end{small}
where \eq{\gvb{a}} encompasses any non-Keplerian perturbing forces (per unit reduced mass).  Note the \eq{\uhatb_2} component of \eq{\gvgr{\txw}_{\hat{u}/\hat{\iota}}}  vanishes.  The orbiting body's instantaneous position and velocity vectors are given in the LVLH basis by
\begin{small}
\begin{align} \label{v_lvlh}
    \gvb{r} \,=\, r \uhatb_1 
    &&,&&
    \dot{\gvb{r}} \,=\, \dot{r} \uhatb_1 \,+\, \gvgr{\txw}_{\hat{u}/\hat{\iota}} \times r\uhatb_1
    \;=\; \dot{r} \uhatb_1 \,+\, \tfrac{h}{r}\uhatb_2 
    \;=\;
    \tfrac{\mu}{h}e\sin{\nu} \uhatb_1 \,+\, \tfrac{h}{r}\uhatb_2
    \;=\;
    \tfrac{\mu}{h}e\sin{\nu} \uhatb_1 \,+\,  \tfrac{\mu}{h}(1+e\cos{\nu})\uhatb_2
\end{align}
\end{small}
where, for the second-to-last expression for \eq{\dot{\gvb{r}}}, we have substituted  \eq{\dot{r}=\tfrac{\mu}{h}e\sin{\nu}}, which follows from Eq.~\eqref{v_peri} in the next section. For the last expression for \eq{\dot{\gvb{r}}}, we have substituted \eq{r=\tfrac{h^2/\mu}{1+e\cos{\nu}}}.


\subsection{The Perifocal Basis} \label{sec:perifocal} 
The perifocal basis is another basis commonly encountered in celestial mechanics and is defined in connection with three conserved vectors of the classic two-body problem: the specific angular momentum, \eq{\gvb{h}}, the Laplace-Runge-Lenz (LRL) vector or eccentricity vector, \eq{\gvb{e}}, and the Hamilton vector, \eq{\gvb{k}}, which are defined from \eq{\gvb{r}} and \eq{\dot{\gvb{r}}} as
\footnote{The LRL and the Hamilton vectors are often defined (per unit reduced mass) as \eq{\gvb{A}=\dot{\gvb{r}}\times\gvb{h}-\mu \uvecb{r}} and \eq{\gvb{K}=\tfrac{1}{h^2}\gvb{h}\times\gvb{A}}, respectively.   The eccentricity vector, \eq{\gvb{e}}, and the vector \eq{\gvb{k}} seen in Eq.~\eqref{LRL_e} are simply a scaling of \eq{\gvb{A}} and \eq{\gvb{K}} by \eq{\mu}.  }
\begin{align} \label{LRL_e} 
    \gvb{e} \,=\,  \tfrac{1}{\mu}\dot{\gvb{r}}\times\gvb{h} \,-\,\uvecb{r} 
    \;=\; -(\tfrac{1}{\mu}
   \wt{\dot{\mbf{r}}}\cdot \wt{\dot{\mbf{r}}}\cdot\gvb{r} + \uvecb{r})
    \quad,
&&
    \gvb{k}  \,=\, \tfrac{1}{h^2}(\gvb{h}\times\gvb{e})
     \,=\, \tfrac{1}{\mu}(\dot{\gvb{r}} \,-\, \tfrac{\mu}{h^2}\gvb{h}\times\uvecb{r})  \quad,
&&
    \gvb{h} \,=\, \gvb{r}\times\dot{\gvb{r}}
\end{align}
where the magnitude \eq{e=\mag{\gvb{e}}}
is the familiar dimensionless eccentricity. 
The above vectors — which are all conserved for Keplerian motion\cite{goldstein2002classical, leach2003generalisations} —  are mutually orthogonal. Thus, the  normalization of any two of them defines an orthonormal basis, called the perifocal basis, which we denote as \eq{\ohatb_i}: 
\begin{small}
\begin{align}
    \ohatb_1 \;=\; \uvecb{e} \;=\; \tfrac{1}{\mu e}(\dot{\gvb{r}}\times\gvb{h} \;-\; \mu\uvecb{r}) \quad,&&
    \ohatb_2  \;=\; \uvecb{k} \;=\; \hhatb \times\uvecb{e} 
    \quad,&&
    \ohatb_3 \;=\; \hhatb  
\end{align}
\end{small}
For Keplerian motion, the above is an inertial basis with the \eq{\ohatb_i} directed along the three axes of symmetry of the  orbital ellipse; \eq{\ohatb_1} is directed parallel to the major axis towards periapse, \eq{\ohatb_2} is directed parallel to the minor axis, and \eq{\ohatb_3} is directed normal to orbital plane. 
Note that this basis shares an axis with the LVLH basis: \eq{\hhatb =\uvecb{u}_3=\ohatb_3}. As such, the perifocal and LVLH bases are related by a simple-3 rotation about this orbit normal axis. The principal angle for this rotation is likely familiar; it is the  true anomaly, \eq{\nu},  defined such that:
\begin{small}
\begin{align} \label{o_i_rel}
    \uhatb_i \;=\; \mbf{R}_3(\nu) \cdot \ohatb_i
&&,&&
    \ohatb_i \;=\;   \mbf{R}_3(-\nu) \cdot \uhatb_i
    &&\;&&
 \text{ \eq{ (\,\hhatb =\uvecb{u}_3=\ohatb_3\,)} }
\end{align}
\end{small}
Combining the above with Eq.~\eqref{nhat_def_R}, we see that \eq{\ohatb_i} is related to the inertial basis by
\begin{align}  \label{o_n_rel}
    \ohatb_i 
     \;=\; \mbf{R}_3(-\nu)\cdot \mbf{R}_{313}(\Omega,i,\theta)\cdot\ihatb_i 
    \;=\; \mbf{R}_{313}(\Omega,i,\upomega)\cdot\ihatb_i 
    &&
    \upomega\;\dot{=}\;\theta-\nu
\end{align}
Such that we are naturally led to define another of the COEs:  the argument of periapse, \eq{\upomega=\theta-\nu} (not related to angular velocity). From the above, we see that the three classic elements \eq{\Omega,i,} and \eq{\upomega} are the 3-1-3 Euler angles relating the perifocal basis to the inertial basis. 
From Eq.~\eqref{Rlmn_mat}, it is seen that the matrix representation of this tensor is given by \eq{ R_{313}(\Omega,i,\upomega) = M_3(\Omega)M_1(i)M_3(\upomega)}. 
The angular velocity of the perifocal basis, with respect to the inertial basis, is then given by
\begin{small}
\begin{align} \label{w_o/i}
    \gvgr{\txw}_{\hat{o}/\hat{\iota}} \,=\, \dot{\Omega}\,\ihatb_3  \,+\, \diff{i}{t}\uvecb{n} \,+\, \dot{\upomega}\,\ohatb_3
    &\,=\,
    (\diff{i}{t}\cos{\upomega}+\dot{\Omega}\sin{\upomega}\sin{i})\ohatb_1
    \,+\,
    (\dot{\Omega}\cos{\upomega}\sin{i} - \diff{i}{t}\sin{\upomega} )\ohatb_2
    \,+\,(\dot{\upomega}+\dot{\Omega}\cos{i})\ohatb_3\,
\\ \nonumber 
    &\,=\, \tfrac{r}{h}\cos{\nu}\, (\uhatb_3 \cdot\gvb{a}) \ohatb_1 \,+\, \tfrac{r}{h}\sin{\nu}\, (\uhatb_3\cdot\gvb{a})\ohatb_2 
    \,-\, \tfrac{r}{h e}\Big(\big[\tfrac{p}{r}\cos{\nu}\,\uhatb_1 - (\tfrac{p}{r}+1)\sin{\nu}\,\uhatb_2  \big] \cdot\gvb{a}\Big)\ohatb_3
\end{align}
\end{small}
where the second line again follows from substitution of \eq{\diff{i}{t}}, \eq{\dot{\Omega}} and \eq{\dot{\upomega}} from Eq.~\eqref{diW}.  
Letting \eq{\crd{\omega_i}{o} = \gvgr{\txw}_{\hat{o}/\hat{\iota}}\cdot \ohatb_i}, we note that the variational equations for the classic elements \eq{e}, \eq{\upomega} and \eq{\nu} given in Eq.~\eqref{COE_eom_0} and \eqref{d_angle} are equivalent to
\begin{small}
\begin{align}
    \dot{e} \;=\; \ohatb_1\cdot  \tfrac{1}{\mu} (  \wt{\dot{\mbf{r}}}\cdot\tvb{r} -\tvb{h} ) \cdot\gvb{a} 
    &&,&&
    \dot{\upomega} \;=\; \crd{\omega_3}{o} \,-\, \tfrac{r}{h}\tfrac{\sin\theta}{\tan{i}}\ohatb_3\cdot\gvb{a}
    &&,&&
    \dot{\nu} \;=\;  -\crd{\omega_3}{o}  \,+\, \tfrac{h^2}{r}
\end{align}
\end{small}
The orbiting body's instantaneous position and velocity may be expressed in the perifocal basis as
\begin{small}
\begin{align} \label{v_peri}
        \gvb{r} \,=\, r\cos{\nu} \ohatb_1 \,+\, r\sin{\nu}\ohatb_2  
 &&,&& 
       \dot{\gvb{r}} = \tfrac{\mu}{h^2}\tvb{h}\cdot(\gvb{e}+\uvecb{r}) \;=\; \tfrac{\mu}{h}\tvb{o}_3\cdot(e\ohatb_1 + \uhatb_1) 
\;=\; -\tfrac{\mu}{h}\sin \nu \ohatb_1 \,+\, \tfrac{\mu}{h} (e +\cos{\nu}) \ohatb_2
\end{align}
\end{small}
where we have used the definitions of \eq{\gvb{e}}, \eq{\gvb{k}}, and \eq{\gvb{h}} in Eq.~\eqref{LRL_e} to obtain the above expression for \eq{\dot{\gvb{r}}}. From this expression, we find the radial velocity component as \eq{\dot{r}=\dot{\gvb{r}} \cdot \uvecb{u}_1 = \tfrac{\mu}{h}e\sin{\nu}}.



\subsection{The Equinoctial Basis}
A less-often encountered basis is  the \textit{equinoctial basis}. This basis, which we denote  as \eq{\shatb_i}, is defined in relation to the inertial basis, \eq{\ihatb_i}, in terms of the classic elements \eq{\Omega} and \eq{i}, and the line of nodes, \eq{\nhatb},  as follows: 
\begin{align} \label{Shat_def}
    \shatb_j \;=\; \mbf{R}_{313}(\Omega,i ,-\Omega)\cdot \ihatb_j
    \;=\; \mbf{R}(\uvecb{n},i)\cdot\ihatb_j
   &&
   \fnsz{\text{note: }}
   \; \uvecb{s}_3=\uvecb{u}_3=\ohatb_3=\hhatb
\end{align}
The above gives two equivalent  definitions. The first says that  \eq{\shatb_i} is obtained from \eq{\ihatb_i} by a 3-1-3 simple rotation sequence through angles \eq{\Omega}-\eq{i}-(\eq{-\Omega}).  The second says that \eq{\shatb_i} is obtained from \eq{\ihatb_i} by a rotation about \eq{\uvecb{n}} through \eq{i}. It can be verified that this latter definition is equivalent to the former by finding the eigenvectors and eigenvalues of the \eq{\mbf{R}_{313}(\Omega,i,-\Omega)} rotation tensor. The result will show that the principal axis and angle are indeed \eq{\uvecb{n}} and \eq{i}, respectively. 
Since the components of the principal axis are the same in both bases, Eq.~\eqref{nodes_1} leads to
\begin{align} \label{nodes_is}
    \uvecb{n}\;=\; \mbf{R}_3(\Omega)\cdot\ihatb_1 \;=\; \cos\Omega\ihatb_1 \;+\; \sin\Omega\ihatb_2 \;=\; \cos\Omega\shatb_1 \;+\; \sin\Omega\shatb_2
     \;\;=\; \frac{\ihatb_3 \times \mbf{h}}{\Vert  \ihatb_3  \times \mbf{h} \Vert }
\end{align}
\sloppy From Eq.~\eqref{Rlmn_mat}  and \eqref{Shat_def}, it is seen that the matrix representation of the tensor relating \eq{\shatb_i} and \eq{\ihatb_i} is  given by \eq{R_{313}(\Omega,i,-\Omega)  = M_3(\Omega)M_1(i)\trn{M}_3(\Omega)}. 
From equations  \eqref{w_313} and \eqref{Shat_def}, it is seen that the angular velocity of the equinoctial basis is given in terms of \eq{i} and \eq{\Omega} as
 \begin{small}
 \begin{align} \label{w_shat}
      \gvgr{\txw}_{\hat{s}/\hat{\iota}} \,=\, \dot{{\Omega}} \,\ihatb_3 \,+\, \diff{i}{t}\uvecb{n} \,-\, \dot{{\Omega}}\,\shatb_3
       &\,=\, \tfrac{r}{h}(\uhatb_3\cdot\gvb{a}) \Big( ( \cos\theta\cos\Omega - \sin\theta\sin\Omega) \shatb_1
    \,+\,  ( \sin\theta\cos\Omega + \cos\theta\sin\Omega)  \shatb_2  
    \, - \,      \tan\tfrac{i}{2}\sin{\theta} \, \shatb_3  \Big)
\\   \nonumber 
    &\,=\, \tfrac{r}{h}(\uhatb_3\cdot\gvb{a}) 
   \big( \cos{l} \, \shatb_1 \,+\, \sin{l} \, \shatb_2 \,-\,   \tan\tfrac{i}{2}\sin{\theta} \,\shatb_3 \big)
 \end{align}
 \end{small}
 where the first line follows from substitution of \eq{\diff{i}{t}} and \eq{\dot{\Omega}} from Eq.~\eqref{diW}, along with the relation \eq{\tfrac{1-\cos{x}}{\sin{x}}= \tan\tfrac{x}{2}}. In the second line above, we have gotten slightly ahead of ourselves by introducing a new angle, \eq{l\;\dot{=}\; \Omega+\upw + \nu = \Omega + \theta}, whose only purpose at the moment is to simplify the above expression for \eq{ \gvgr{\txw}_{\hat{s}/\hat{\iota}} } via useful trigonometric relations given later in Eq.~\eqref{MEE_trig}.  However, as indicated in the introduction and discussed further in the next section, \eq{l} is known as the \textit{true longitude} and is one of the modified equinoctial elements introduced by Walker et al.  
 We will return our attention to the equinoctial basis when we examine  these elements in the next section. For now, note the definition of the equinoctial basis as a principal rotation given in Eq.~\eqref{Shat_def} is inherently singular at \eq{i=0} where the line of nodes, \eq{\nhatb}, becomes undefined.


\section{The Equinoctial Elements} \label{sec:MEE}

In this section, we present the modified equinoctial elements (MEEs) and their variational equations as they typically appear in the literature \cite{arya2021low}. We then show how these elements are connected to the reference frames and attitude representations discussed in previous sections. 


\subsection{The Modified Equinoctial Elements Defined In Terms of Classic Orbit Elements} 
As discussed in the introduction, the equinoctial elements are  a set of six orbit elements commonly used in celestial mechanics which avoid the singularity issues seen in the classical elements at \eq{i=0} and \eq{e=0}. We consider the \textit{modified equinoctial elements} (MEEs), given in the third column of Eq.~\eqref{EE_defs}, and employ the following labeling convention:
\footnote{What we have labeled \eq{e_1}, \eq{e_2}, \eq{q_1}, and \eq{q_2} above, other sources often label using some permutation of the letters \eq{f}, \eq{g}, \eq{h}, \eq{p}, and \eq{k}, with different sources using different labeling conventions.
However, these letters are already in common use for various other parameters in celestial mechanics (e.g., this work and many others use \eq{h} and \eq{p} for the specific angular momentum and semilatus rectum, respectively, and \eq{f} is often used for the true anomaly). } 
\begin{small}
\begin{align} \label{MEEdef}
    \begin{array}{cc}
          \text{modified equinoctial elements}
    \end{array} \left\{ \quad
    \begin{array}{ll}
         p \;=\; \tfrac{h^2}{\mu} \;=\; a(1-e^2)  
     \\
          e_1 \;=\;  e\cos (\Omega+\upomega)
     \\
        e_2 \;=\;  e\sin (\Omega+\upomega)
    \end{array} \right.
    &&
    \begin{array}{ll}
        &q_1 \;=\;  \tan(\tfrac{i}{2})\cos\Omega
    \\
        &q_2 \;=\;  \tan(\tfrac{i}{2})\sin\Omega
    \\
       & l \;=\; \Omega + \upomega + \nu \;=\; \Omega +\theta
    \end{array} 
    \qquad\qquad&&\qquad
\end{align}
\end{small}
Like the COEs, the first five MEEs are integrals of motion for the Kepler problem and osculating elements for the perturbed problem and the sixth MEE, \eq{l} (true longitude), is a fast phase angle variable. 
The governing equations for the MEEs under the influence of a non-Keplerian perturbing force (per unit reduced mass), \eq{\gvb{a}}, are given as follows \cite{walker1985set}:
\begin{small}
\begin{align} \label{MEE_eom_tru}
\begin{array}{lll}
    \dot{p}   =  \tfrac{r}{h}2p\,\uvecb{u}_2 \cdot\gvb{a} &,
\\[4pt]
    \dot{e}_1     =  \tfrac{r}{h} \big\{ w\sin{l}\,\uvecb{u}_1 + \big[(w+1)\cos{l}+  e_1 \big]\uvecb{u}_2 -   e_2(  q_1 \sin{l}-  q_2  \cos{l})\uvecb{u}_3 \big\}\cdot\gvb{a} &,
\\[4pt]
    \dot{e}_2   =  \tfrac{r}{h} \big\{ -w\cos{l}\,\uvecb{u}_1 + \big[(w+1)\sin{l}+  e_2 \big]\uvecb{u}_2 +   e_1(  q_1 \sin{l}-  q_2  \cos{l})\uvecb{u}_3 \big\}\cdot\gvb{a} &,
\end{array}
&&
\begin{array}{llll}
    \dot{q}_1     =   \tfrac{r}{h}\tfrac{1}{2}(1+  q_1^2+  q_2^2)\cos{l}\,\uvecb{u}_3\cdot\gvb{a}
\\[4pt]
    \dot{q}_2    =  \tfrac{r}{h}\tfrac{1}{2}(1+  q_1^2+  q_2^2)\sin{l}\,\uvecb{u}_3\cdot\gvb{a}
\\[4pt]
    \hdot{l}     =  \tfrac{r}{h}(  q_1 \sin{l}-  q_2  \cos{l})\uvecb{u}_3\cdot\gvb{a} + \tfrac{h}{r^2}
\end{array} 
\end{align}
\end{small}
Where \eq{\{\uvecb{u}_1,\uvecb{u}_2,\uvecb{u}_3\}} are the LVLH basis vectors defined in Eq.~\eqref{LVLHdef} and where the radial distance, \eq{r}, specific angular momentum magnitude, \eq{h}, and the unnamed  \eq{w}, and eccentricity terms are defined in terms of MEEs by
\begin{small}
\begin{align} \label{w_def}
     h  \,=\,  \sqrt{\mu p} 
  \quad, && 
    r \,=\, \tfrac{p}{w} \,=\, \tfrac{h^2}{\mu w}
\quad, && 
w \,=\, \tfrac{p}{r} \,=\,  1+ e\cos\nu  
  \quad, && 
  \begin{array}{llll}
       e\cos\nu  \,=\,    e_1\cos{l}+  e_2\sin{l} \\
        e\sin\nu  \,=\,    e_1\sin{l} - e_2\cos{l}
  \end{array}
\end{align}
\end{small}
Other useful trigonometric  relations involving  \eq{l}, \eq{\Omega}, \eq{\upw}, \eq{\nu}, and \eq{\theta} are given in the footnote.\footnote{The relation \eq{l = \Omega + \upomega + \nu = \Omega+\theta = \varpi +\nu }, where \eq{\theta= \upomega + \nu} and \eq{\varpi = \Omega + \upomega}, leads to the following trigonometric identities:
\begin{align} \label{MEE_trig}
     &\cos\theta\cos\Omega - \sin\theta\sin\Omega = \cos{l}
  &&,&& 
      \sin\theta\cos\Omega + \cos\theta\sin\Omega = \sin{l}
  &&,&& 
      \sin\theta = \cos\Omega\sin{l}-\sin\Omega\cos{l}
\\ \nonumber 
 &\cos\varpi \cos\nu - \sin\varpi \sin\nu = \cos{l}
  &&,&& 
      \cos\varpi \sin\nu + \sin\varpi \cos\nu  = \sin{l}
  &&,&& 
      \cos\theta = \cos\Omega\cos{l}-\sin\Omega\sin{l}
\end{align}
}

\paragraph{Other Versions.} 
As discussed in the introduction, the MEEs are not the only version of equinoctial elements in use. 
In place of the semilatus rectum,  \eq{p}, some sources instead use the mean motion, \eq{n}, the specific angular momentum magnitude, \eq{h}, the semi-major axis, \eq{a}, or the specific energy, \eq{\mcal{E}}. The governing equation for \eq{a} is already given in Eq.~\eqref{COE_eom_0}. The governing equations for \eq{h}, \eq{n}, and \eq{\mcal{E}}  may be found from straightforward differentiation and substitution: 
\begin{small}
\begin{align}
\begin{array}{ll}
       h   = \sqrt{\mu p}
 \\[4pt]
     n= \sqrt{\tfrac{\mu}{a^3}} 
 \\[4pt]
     \mcal{E} = -\tfrac{\mu}{2a}
\end{array}
\qquad\qquad
\rightarrow 
\qquad\qquad
\begin{array}{ll}
      \hdot{h} \;=\; \tfrac{1}{2}\sqrt{\tfrac{\mu}{p}}\,\dot{p} \;=\; r \,\uvecb{u}_2\cdot\gvb{a} 
 \\[4pt]
     \dot{n} \;=\; -\tfrac{3}{2}\tfrac{n}{a}\dot{a} \;=\; 
     -3 na \tfrac{r}{h}( e\sin \nu \uvecb{u}_1 + \tfrac{p}{r}\uvecb{u}_2) \cdot\gvb{a}
\\[4pt]
      \dot{\mcal{E}}  \;=\; \tfrac{\mu}{2a^2}\dot{a} \;=\; 
      \tfrac{\mu}{h}( e\sin \nu \uvecb{u}_1 + \tfrac{p}{r}\uvecb{u}_2) \cdot\gvb{a}
\end{array}
\end{align}
\end{small}
where we have substituted the expressions for \eq{\dot{a}} and \eq{\dot{p}} from Eq.~\eqref{COE_eom_0} and \eqref{MEE_eom_tru}.  
As is the case for the COEs, there is also variation in sixth equinoctial element. Although the MEEs classically use the true longitude, \eq{l=\Omega+\upomega+\nu}, some other sources use the mean longtidue, \eq{\lambda=\Omega+\upomega+M}, or the eccentric longitude,  $\upepsilon = \Omega+\upomega+E$ or their ``slow' counterparts,  \eq{\lambda_0=\Omega+\upomega+M_0} or \eq{\upepsilon_0 = \Omega+\upomega+E_0}. Kepler's equation then takesthe form \eq{\lambda = \upepsilon - e_1\sin\upepsilon + e_2\cos \upepsilon } and the radial distance may be  written \eq{r=a(1-e_1\cos \upepsilon - e_2\sin\upepsilon}) \cite{broucke1972equinoctial}.



\subsection{Geometric Meaning of The Modified Equinoctial Elements} \label{sec:MEEmeaning}

\paragraph{The Elements $q_1$ and $q_2$. }
Consider again the equinoctial basis, \eq{\shatb_i}, depicted in Figure \ref{fig:bases} and defined in Eq.~\eqref{Shat_def} where it is seen that the principal axis and angle relating \eq{\shatb_i} to the inertial basis, \eq{\ihatb_i}, are the line of nodes, \eq{\uvecb{n}}, and the inclination, \eq{i}, respectively.
From Eq.~\eqref{CRPdef}, it is seen that the Rodrigues vector, \eq{\gvb{q}},  relating the equinoctial and inertial bases is then given by
\begin{align} \label{crp_eq}
\shatb_j=\mbf{R}(\uvecb{n},i)\cdot\ihatb_j 
&&\rightarrow &&
    \gvb{q} \;=\; \tan\tfrac{i}{2}\uvecb{n} \;=\; \tan\tfrac{i}{2}(\cos\Omega\ihatb_1 \;+\; \sin\Omega\ihatb_2) \;=\; \tan\tfrac{i}{2}(\cos\Omega\shatb_1 \;+\; \sin\Omega\shatb_2)
\end{align}
where we have used Eq.~\eqref{nodes_is} for \eq{\uvecb{n}}. 
The components of the above \eq{\gvb{q}} take the same form in both the equinoctial and inertial bases. Letting  \eq{q_i\;\dot{=}\;\gvb{q}\cdot\ihatb_i=\gvb{q}\cdot\shatb_i}, the equinoctial basis CRPs are given in terms of \eq{i} and \eq{\Omega} by
\begin{align} \label{CRPdef_shat}
         q_1\;=\;  \tan\tfrac{i}{2}\cos\Omega  
     &&,&& 
          q_2\;=\;  \tan\tfrac{i}{2}\sin\Omega 
         &&,&& 
           q_3 \;=\; 0
\end{align}
Looking at the above, it should be apparent why we have chosen the labels \eq{q_1} and \eq{q_2} for fourth and fifth MEEs defined in Eq.~\eqref{MEEdef}. Note that we may express the rotation tensor \eq{\mbf{R}(\gvb{q})} that relates \eq{\shatb_i} and \eq{\ihatb_i} using Eq.~\eqref{CRPdef}. Or, with \eq{q_3=0}, Eq.~\eqref{Rij_crp} leads to the matrix representation of \eq{\mbf{R}(\gvb{q})} (in the \eq{\shatb_i} and \eq{\ihatb_i} bases) as
\begin{small}
\begin{align} \label{Rcrp_shat}
  \shatb_i=\mbf{R}(\gvb{q})\cdot\ihatb_i &&, &&
     R(q_1,q_2) = 
     \frac{1}{1+ q^2}
 \begin{pmatrix}  
    1 +   q_1^2 - q_2^2    &    2q_1  q_2 &   2 q_2  
 \\
   2 q_1  q_2    &  1+ q_2^2 -  q_1^2  &   -2q_1 
\\
    -2 q_2   &   2 q_1  &  1  -  q_1^2 -   q_2^2 
\end{pmatrix}  
 =\, \begin{pmatrix}
     \crd{[\shatb_1]}{\iota} &  \crd{[\shatb_2]}{\iota} &  \crd{[\shatb_3]}{\iota}
 \end{pmatrix}
\end{align}
\end{small}

\begin{remrm}[meaning of $q_1$ and $q_2$] 
The elements \eq{q_1} and \eq{q_2} defined in the  Eq.~\eqref{MEEdef} are the two non-zero components (in the \eq{\shatb_i} and \eq{\ihatb_i} bases) of the Rodrigues vector, \eq{\gvb{q}},
that relates the equinoctial and inertial bases by \eq{\shatb_i=\mbf{R}(\gvb{q})\cdot\ihatb_i}. 
The third component of this vector is zero in the  \eq{\shatb_i} and \eq{\ihatb_i} bases. 
\end{remrm}

\begin{remrm}[equatorial orbits]
The definition of the equinoctial basis as a principal rotation, \eq{\shatb_i=\mbf{R}(\nhatb,i)\cdot\ihatb_i}, is inherently singular at \eq{i=0} (equatorial orbit) where the line of nodes, \eq{\nhatb}, becomes undefined. 
Despite this, the Rodrigues vector, \eq{\gvb{q}=\tan\tfrac{i}{2}\nhatb}, is perfectly well defined at \eq{i=0}; it is simply the zero vector, \eq{\gvb{q}=\gvb{0}}. 
\end{remrm} \label{rem:crp_i0}

Informally, we may reconcile the above remark by noting that \eq{\shatb_i} is also defined by an Euler angle sequence as \eq{\shatb_i=\mbf{R}_{313}(\Omega,i,-\Omega)\cdot\ihatb_i} such that \eq{i=0} always reduces to the identity transformation,  \eq{\shatb_i=\mbf{R}_{3}(\Omega-\Omega)\cdot\ihatb_i = \mbf{I}\cdot\ihatb_i},  and the Rodrigues vector that generates an identity transformation (through Eq.~\eqref{MRPdef}) is the zero vector. Thus, in the case \eq{i=0},  \eq{\gvb{q}=\gvb{0}} and \eq{\shatb_i=\ihatb_i} is just the inertial reference frame. 
However, this reasoning is not fully satisfactory as the angle \eq{\Omega} itself is undefined when \eq{i=0}. 
For a more concrete verification of the above remark, note that Eq.~\eqref{Rcrp_shat} leads to an alternative definition of the two non-zero equinoctial CRPs:
\begin{small}
\begin{align} \label{crp_i0}
        \hhatb \,=\, \shatb_3 \,=\, \tfrac{1}{1+q^2} \big(  2q_2 \ihatb_1 - 2q_1\ihatb_2  + (1-q^2)\ihatb_3 \big)
        && \Rightarrow &&
        \begin{array}{lllll}
             q_1 \,=\, - \dfrac{\hhatb\cdot\ihatb_2}{1+\hhatb\cdot\ihatb_3} = - \dfrac{\hhatb\cdot\ihatb_2}{1+\cos{i}} 
             \\[6pt]
             q_2  \,=\, \dfrac{\hhatb\cdot\ihatb_1}{1+\hhatb\cdot\ihatb_3}
             = \dfrac{\hhatb\cdot\ihatb_1}{1+\cos{i}}
        \end{array}
\end{align}
\end{small}
 If  \eq{i=0}, then \eq{\hhatb = \ihatb_3} and \eq{\cos{i}=1} such that the above relations lead smoothly to \eq{q_1=q_2=0} for equatorial orbits.\footnote{Alternatively,  we may convince ourselves that, in the relation  \eq{\gvb{q}=\tan\tfrac{i}{2}\nhatb}, the factor \eq{\tan\tfrac{i}{2}} approaches zero more rapidly than the line of nodes becomes undefined as \eq{i\to 0}, effectively cancelling out the apparent singularity in the equinoctial basis at \eq{i=0}. From the definition of $q_1$ and $q_2$, we see that $\tan\tfrac{i}{2}= q_1\cos\Omega +q_2\sin\Omega$ and thus \eq{\tan{\Omega}=q_2/q_1} becomes undefined as $(q_1,q_2)\to(0,0)$. Since $\tan\tfrac{i}{2}= q_1\cos\Omega +q_2\sin\Omega$ (and \eq{\cos\Omega \leq 1} and \eq{\sin\Omega \leq 1}), we see that $\tan\tfrac{i}{2}\to 0$ faster than $(q_1,q_2)\to(0,0)$.    }  


\paragraph{The Elements $e_1$, $e_2$, and $l$. }
To see the meaning of \eq{e_1,\; e_2}, and \eq{l}  defined in Eq.~\eqref{MEEdef}, let us examine the relation of the equinoctial basis with the the more familiar LVLH basis, \eq{\uhatb_i}, and perifocal basis, \eq{\ohatb_i}, defined earlier.   
From Eqs. \eqref{nhat_def_R}, \eqref{o_i_rel}, and \eqref{Shat_def}, it is seen that the equinoctial basis is related to these  bases by a simple-3 rotation about the orbit normal direction, \eq{\hhatb}, such that all three bases share this axis and the relation between \eq{\uhatb_i} and \eq{\shatb_i} naturally leads to the true longitude:
\begin{small}
\begin{align} \label{R_os}  
      \ohatb_i \;=\; \mbf{R}_3(\Omega+\upomega)\cdot\shatb_i  
    &&\uhatb_i 
    \;=\; \mbf{R}_3(\Omega+\upomega+\nu)\cdot\shatb_i 
&&
    \shatb_3 \;=\; \ohatb_3 \;=\; \uhatb_3 \;=\; \hhatb  
\end{align}
\end{small}

\begin{remrm}[meaning of $l$]
 The true longitude, \eq{l},  is the principal rotation angle relating the equinoctial and LVLH bases by a simple 3-rotation about the orbit normal axis:
 \begin{align}
\uhatb_i \;=\; \mbf{R}(\hhatb, l )\shatb_i 
    \;=\; \mbf{R}_3(l)\cdot\shatb_i 
   &&,&&
    l = \Omega + \upomega + \nu \;=\; \Omega+\theta
\end{align}
\end{remrm}
Note that, even though \eq{\Omega}, \eq{\upomega}, or \eq{\nu} may become undefined at \eq{i=0} and/or \eq{e=0},  the true longitude, \eq{l =\Omega + \upomega + \nu}, is always uniquely defined. When \eq{e\neq 0} and \eq{i=0},  \eq{\nu} is defined but \eq{\Omega} and \eq{\upomega} are, individually, undefined. Yet, their sum, \eq{\Omega+\upomega}, is still defined (in this case it is the angle between the \eq{\ihatb_1} reference direction and direction of periapse, \eq{\ohatb_1}). Similarly, if \eq{i\neq 0} but \eq{e=0}, then \eq{\upomega} and \eq{\nu} are, individually, undefined yet their sum, \eq{\theta=\upomega+\nu}, is still defined (it is the angle between the line of nodes, \eq{\nhatb}, and instantaneous position, \eq{\uvecb{r}}). If both \eq{i=0} and \eq{e=0}, then  \eq{\Omega}, \eq{\upomega}, and \eq{\nu} are all, individually, undefined yet their sum is still defined (in this case it is the angle between \eq{\ihatb_1} and  \eq{\uvecb{r}}). In all cases, \eq{l=\Omega + \upomega +\nu} is well defined.  

Next, consider the eccentricity vector defined in Eq.~\eqref{LRL_e}. In the perifocal basis, \eq{\ohatb_i}, this vector is given simply as \eq{\gvb{e}=e\ohatb_1}. 
Using the the first relation in Eq.~\eqref{R_os} to expand \eq{\gvb{e}} in the equinoctial basis, it becomes clear why we have chosen to label the fourth and fifth MEEs as \eq{e_1} and \eq{e_2}:
\begin{align}
    \gvb{e} \;=\; e\ohatb_1 \;=\; e\mbf{R}_3(\Omega+\upomega)\cdot\shatb_1
    \;=\; e\cos(\Omega+\upomega)\shatb_1 \;+\; e\sin(\Omega+\upomega)\shatb_2
\end{align}

\begin{remrm}[meaning of $e_1$ and $e_2$]
    As was already noted in \cite{cefola1972equinoctial},  the elements \eq{e_1} and \eq{e_2} are the non-zero components of the LRL/eccentricity vector, \eq{\gvb{e}}, in the equinoctial basis, \eq{\shatb_i}  (the third component is zero in this basis): 
\begin{align}
      e_1 \;=\; \gvb{e}\cdot\shatb_1 \;=\; e\cos(\Omega+\upomega)
  &&,&&  
     e_2 \;=\; \gvb{e}\cdot\shatb_2 \;=\; e\sin(\Omega+\upomega)
     &&,&&
     \gvb{e} \;=\; e_1\shatb_1 \,+\, e_2 \shatb_2
\end{align}
\end{remrm}

\paragraph{Attitude Dynamics for the Equinoctial CRPs.} 
To verify that \eq{q_1} and \eq{q_2} are indeed the CRPs for the equinoctial basis, we may show that the CRP attitude dynamics developed in section \ref{sec:CRPdyn} can be used to recover the governing equations for \eq{\dot{q}_1} and \eq{\dot{q}_2} given in Eq.~\eqref{MEE_eom_tru} (which have, historically, been derived using variation of parameters or algebraic relations to the COEs). To that end, consider again the angular velocity of the equinoctial basis with respect to the inertial basis, \eq{\gvgr{\txw}_{\hat{s}/\hat{\iota}}}, given by Eq.~\eqref{w_shat} and let \eq{\crd{\omega_i}{s}=\gvgr{\txw}_{\hat{s}/\hat{\iota}}\cdot\shatb_i}. From the trigonometric identities in Eq.~\eqref{MEE_trig}, these  components may be written as
 \begin{small}
\begin{align} \label{w_s_crp}
 \begin{array}{lll}
          \crd{\omega_1}{s}
         = \tfrac{r}{h}( \cos\theta\cos\Omega - \sin\theta\sin\Omega) \hhatb \cdot\gvb{a}
         &=\; \tfrac{r}{h}\cos{l} \hhatb \cdot\gvb{a}
\\[4pt]
          \crd{\omega_2}{s}
         = \tfrac{r}{h}( \sin\theta\cos\Omega + \cos\theta\sin\Omega) \hhatb \cdot\gvb{a}
         &=\;  \tfrac{r}{h}\sin{l} \hhatb \cdot\gvb{a}
\\[4pt]
         \crd{\omega_3}{s} =
        -\tfrac{r}{h}\tan\tfrac{i}{2}\sin{\theta} \hhatb\cdot\gvb{a}
        &=\;  \tfrac{r}{h}
        (q_2\cos{l}-q_1\sin{l}) \hhatb \cdot\gvb{a} &=\;  q_2\crd{\omega_1}{s} \;-\; q_1\crd{\omega_2}{s}
\end{array} 
\end{align}
\end{small}
Now, with \eq{q_3=0} and \eq{\crd{\omega_i}{s}}  given above, Eq.~\eqref{w_crp} and \eqref{crp_ode_gen} for the CRP attitude kinematics lead to
\begin{small}
\begin{align} \label{CRPdot_scalar}
 \begin{array}{llll}
   \dot{q}_1 \,=\, \tfrac{1}{2} \big[  (1+q_1^2)\crd{\omega_1}{s} +  q_1 q_2\crd{\omega_2}{s} +  q_2\crd{\omega_3}{s}     \big]
   & \!=\; \frac{1}{2}(1+q^2)\crd{\omega_1}{s}
   &\!=\, 
   \frac{1}{2}\tfrac{r}{h}(1+q^2)\cos{l} \hhatb\cdot\gvb{a} 
\\[4pt]
    \dot{q}_2 \,=\,  \tfrac{1}{2} \big[  q_1 q_2\crd{\omega_1}{s}  +  (1+ q_2^2 )\crd{\omega_2}{s} - q_1\crd{\omega_3}{s}   \big]
    &\! =\,
    \frac{1}{2}(1+q^2)\crd{\omega_2}{s}
     & \!=\,  
   \frac{1}{2}\tfrac{r}{h}(1+q^2)\sin{l} \hhatb\cdot\gvb{a} 
\\[4pt]
    \dot{q}_3 \,=\, \tfrac{1}{2} \big[  -q_2 \crd{\omega_1}{s} +   q_1\crd{\omega_2}{s}  + \crd{\omega_3}{s} \big]
    &\!=\, 0
 \end{array} 
\end{align}
\end{small}
where \eq{q^2=q_1^2+q_2^2} on account of the fact that \eq{q_3=0} remains constant (as confirmed by the above). The far right-hand-side of the above are indeed in agreement with the standard variational equations given in Eq.~\eqref{MEE_eom_tru} (recall \eq{\hhatb=\uhatb_3=\shatb_3}).
We also note that the equation of motion for the true longitude, \eq{\hdot{l}}, given in Eq.~\eqref{MEE_eom_tru}, is equivalent to
\begin{small}
\begin{align} \label{ldot_ws}
    \hdot{l} \;=\;  
    -\crd{\omega_3}{s} \;+\;  \tfrac{h}{r^2}
\end{align}
\end{small}
which is precisely what we would expect given the relation \eq{\uhatb_i=\mbf{R}_3(l)\cdot \shatb_i} with \eq{l=\Omega+\upomega+\nu} and \eq{\dot{\nu}^{\text{kep}}=\tfrac{h}{r^2}}. Examining Eq.~\eqref{MEE_eom_tru}, we see that \eq{\crd{\omega_3}{s}} also appears in the equations for \eq{\dot{e}_1} and \eq{\dot{e}_2}.

\begin{remrm}
Although the equinoctial elements have seen significant use in the last decades, 
the connection we have drawn between these elements and the geometry of the 2BP does not appear widely appreciated, or even formally recognized,  in the literature. The rotation matrix seen in Eq.~\eqref{Rcrp_shat} has been noted by others \cite{broucke1972equinoctial,cefola1972equinoctial,battin1999introduction} and yet, to the best of the authors' knowledge, the  direct connection to Euler's rotation theorem and attitude kinematics (specifically, the Rodrigues parameters) has never been explicitly identified nor have the above equations of motion been derived using attitude dynamics.\footnote{We mention, however, that analogous variation of parameters equations for 3-1-3 Euler angles were derived from rotational kinematics in \cite{junkins1979analogy}.}   
\end{remrm}

\paragraph{Conversion to Position \& Velocity Coordinates.}
From \eq{\gvb{r}=r\uhatb_1 =  \mbf{R}_3(l)\cdot r\shatb_1},  
the instantaneous position vector is
\begin{align} \label{r_eq}
    \gvb{r}  \;=\; r\cos{l}\shatb_1 \;+\; r\sin{l}\shatb_2
&&
r = \tfrac{p}{w} = \tfrac{p}{1+e_1\cos{l}+e_2\sin{l}}
\end{align}
Using the kinematic transport equation from Eq.~\eqref{transport_0}, 
the velocity vector is then 
\begin{small}
\begin{align}
    \dot{\gvb{r}} \;=\; \diff{}{t}(\gvb{r}\cdot\shatb_i) \shatb_i  \,+\, \tvgr{\txw}_{\hat{s}/\hat{\iota}}\cdot\gvb{r}
     \;=\;  (\dot{r} \cos{l} - \tfrac{h}{r}\sin{l}) \shatb_1 \;+\; (\dot{r} \sin{l} + \tfrac{h}{r}\cos{l}) \shatb_2
\end{align}
\end{small}
where the second equality has made use of  Eq.~\eqref{w_s_crp} and \eqref{ldot_ws} and where \eq{\dot{r} = \tfrac{\mu e}{h}\sin\nu} with \eq{e=(e_1^2+e_2^2)^{\sfrac{1}{2}}} and \eq{\mu / h =(\mu/ p)^{\sfrac{1}{2}}}. 
The above velocity may be written more concisely in terms of the equinoctial elements using Eq.~\eqref{v_peri} along with the relation \eq{\ohatb_i = \mbf{R}_3(\Omega + \upomega)\cdot\shatb_i} to obtain 
\begin{small}
\begin{align} \label{v_eq}
   \dot{\gvb{r}} \;=\;   -\tfrac{\mu}{h}( e_2 + \sin{l}
    ) \shatb_1
    \;+\;
    \tfrac{\mu}{h}( e_1 + \cos{l}
    )\shatb_2
\end{align}
\end{small}

The coordinate vectors of \eq{\gvb{r}} and \eq{\dot{\gvb{r}}} in the inertial basis, \eq{\ihatb_i},  expressed in terms of the MEEs are found using \eq{\crd{r_{\!i}}{\iota}=R_{ij}\crd{r_{\!\!j}}{s} } and  \eq{\crd{v_{\!i}}{\iota}=R_{ij}\crd{v_{\!\!j}}{s} } with the matrix \eq{R=R(q_1,q_2)} given in Eq.  \eqref{Rcrp_shat}. 
The result is
\begin{small}
\begin{align}
    \crd{\bm{r}}{\iota}  \,=\,  \frac{r}{1+q^2} 
\fnsz{\begin{pmatrix}
   ( 1 +  q_1^2 -q_2^2 )\cos{l}  +  2q_1q_2\sin{l} 
   \\
   2q_1q_2\cos{l}   +  ( 1 - q_1^2 +q_2^2 )\sin{l}
   \\
   -2 (q_2 \cos{l} -  q_1 \sin{l}) 
\end{pmatrix} }
&&,&&
 \crd{\bm{v}}{\iota}  \,=\, \frac{\mu}{h}  \frac{1}{1+q^2}
 \fnsz{\begin{pmatrix}
   -( 1 +   q_1^2 - q_2^2 )(e_2+\sin{l})  +  2 q_1 q_2(e_1+\cos{l}) 
   \\
   -2 q_1 q_2(e_2+\sin{l})   +   ( 1 -   q_1^2 + q_2^2 )(e_1+\cos{l})
   \\
   2 \big[  q_2(e_2+\sin{l})    +   q_1 (e_1+\cos{l}) \big]
\end{pmatrix}  }
\end{align}
\end{small}

\section{A New Set of Equinoctial Elements Using Modified Rodrigues Parameters}  \label{sec:MRP MEE}

The equinoctial basis CRPs given in Eq.~\eqref{CRPdef_shat} are proportional to \eq{\tan\tfrac{i}{2}}. As such, they are singular at \eq{i=\pm\pi}. We are therefore motivated to develop a new version of the equinoctial elements based on the equinoctial basis MRPs, proportional to \eq{\tan\tfrac{i}{4}}, which will shift this singularity to \eq{\pm 2\pi}.   
The authors have creatively dubbed these new orbit elements the \textit{MRP Modified Equinoctial Elements} (MRP MEEs). Although we will not always do so, much of the following can be arrived at simply using the coordinate transformation between CRPs and MRPs:
\begin{small}
\begin{align} \label{crp2mrp}
      \tan\tfrac{x}{2} \;=\; \frac{2\tan{(x/4)}}{1-\tan^2{(x/4)}}
     &&\Rightarrow&&
     q_i \;=\; \frac{2\sigma_i}{1-\sigma^2}
     \qquad \Leftrightarrow \qquad 
     \sigma_i \;=\; \frac{q_i}{1+ \sqrt{1+q^2}}
\end{align}
\end{small}
along with the corresponding Jacobians:
\begin{align} \label{crp_mrp_J}
    &\pderiv{q_i}{\sigma_j}  \,=\, \tfrac{1}{k}( \delta_{ij} \,+\, \tfrac{1}{k} \sigma_i\sigma_j )
 &&,&&
     \pderiv{\sigma_i}{q_j}
     \,=\,
     k( \delta_{ij} \,-\, \tfrac{k^2}{1 -k } q_i q_j )
     &&,&&  
      k\,\,\dot{=}\,\, \tfrac{1}{2}(1-\sigma^2) \, \equiv\,  \tfrac{1}{1 + \sqrt{1+q^2}} 
\end{align}

\paragraph{The Equinoctial MRPs.}
The modified Rodrigues vector, \eq{\gvgr{\txsig}} is defined using the principal axis and angle, \eq{\uvecb{n}} and \eq{i}, of the equinoctial basis as in Eq.~\eqref{MRPdef}:
\begin{align} \label{MRP_def}
\shatb_j = \mbf{R}(\nhatb,i) \cdot\ihatb_j
&&,&&
    \gvgr{\txsig}  \;=\; \tan(\tfrac{i}{4})\uvecb{n} 
    \;=\; \tan\tfrac{i}{4}(\cos\Omega\ihatb_1 \;+\; \sin\Omega\ihatb_2) \;=\; \tan\tfrac{i}{4}(\cos\Omega\shatb_1 \;+\; \sin\Omega\shatb_2)
\end{align}
where we have used Eq.~\eqref{nodes_is} for the line of nodes, \eq{\uvecb{n}}. 
The components of \eq{\gvgr{\txsig}} are the same in the inertial and equinoctial bases and we define the equinoctial MRPs as  \eq{\sigma_i=\gvgr{\txsig}\cdot\ihatb_i=\gvgr{\txsig}\cdot\shatb_i}:
\begin{align}
     \sigma_1   \;=\; \tan\tfrac{i}{4}\cos\Omega
   &&,&& 
      \sigma_2  \;=\;\tan\tfrac{i}{4}\sin\Omega 
 &&,&& 
     \sigma_3  \;=\; 0
\end{align}
Note from the discussion surrounding Remark \ref{rem:crp_i0} and Eq.~\eqref{crp_i0}, along with Eq.~\eqref{crp2mrp}, that the above MRPs are perfectly well defined for equatorial orbits (\eq{i=0}) where the line of nodes and \eq{\Omega} become undefined; they are simply zero. 
We can obtain the governing equations for these  MRPs using the attitude dynamics given by Eq.~\eqref{MRPdyn_gen}:
\begin{small}
\begin{align} \label{MRP_eom_scal}
     \begin{array}{ll}
      \dot{\sigma}_1 \;=\; \tfrac{1}{4} \big[ (1+\sigma_1^2 - \sigma_2^2  )\crd{\omega_1}{s}  +2\sigma_1 \sigma_2\crd{\omega_2}{s}   + 2\sigma_2 \crd{\omega_3}{s} \big]
\\[4pt]
     \dot{\sigma}_2 \;=\; \tfrac{1}{4} \big[ 2\sigma_1 \sigma_2\crd{\omega_1}{s}  + (1 -\sigma_1^2 + \sigma_2^2 )\crd{\omega_2}{s}  - 2\sigma_1 \crd{\omega_3}{s} \big]
 \\[4pt]
    \dot{\sigma}_3 \;=\; \tfrac{1}{4} \big[ -2\sigma_2 \crd{\omega_1}{s} +
     2\sigma_1 \crd{\omega_2}{s}  + (1 -\sigma_1^2 - \sigma_2^2 )\crd{\omega_3}{s} \big]
\end{array} 
\end{align}
\end{small}
where \eq{\sigma_3=0} and \eq{\crd{\omega_i}{s}} are given in Eq.~\eqref{w_s_crp}. 
Using Eq.~\eqref{crp2mrp}, \eq{\crd{\omega_3}{s}} may be expressed in terms of \eq{\sigma_i} to obtain 
\begin{small}
\begin{align} \label{w3_mrp} 
          \crd{\omega_1}{s}  =  
         \tfrac{r}{h}\cos{l}\hhatb  \cdot\gvb{a} 
&&,&& 
          \crd{\omega_2}{s}  = 
         \tfrac{r}{h}\sin{l}\hhatb \cdot\gvb{a} 
&&,&& 
         \crd{\omega_3}{s}  = 
        \tfrac{r}{h}\tfrac{2}{1-\sigma^2}(\sigma_2\cos{l} - \sigma_1\sin{l} ) \hhatb \cdot\gvb{a} 
        \,=\, \tfrac{2}{1-\sigma^2}( \sigma_2\crd{\omega_1}{s} - \sigma_1\crd{\omega_2}{s})
\end{align}
\end{small}
Substituting the above into  Eq.~\eqref{MRP_eom_scal} we find, after some simplification, the governing equations:
\begin{small}
\begin{align} \label{mrp_dot_fin}
   \begin{array}{llll}
         \dot{\sigma}_1 \;=\; 
         \tfrac{1}{4}(1+\sigma^2)(\crd{\omega_1}{s} \,+\, \sigma_2\crd{\omega_3}{s}) 
    &=\;  \tfrac{1}{4} \tfrac{r}{h}\tfrac{1+\sigma^2}{1-\sigma^2} \big[ (1-\sigma_1^2+\sigma_2^2)\cos{l} \;-\; 
     2\sigma_1\sigma_2\sin{l}  
    \big]\hhatb \cdot\gvb{a}
\\
       \dot{\sigma}_2 \;=\;  \tfrac{1}{4}(1+\sigma^2)(\crd{\omega_2}{s} \,-\, \sigma_1\crd{\omega_3}{s}) 
     &=\;    \tfrac{1}{4} \tfrac{r}{h}\tfrac{1+\sigma^2}{1-\sigma^2} \big[ (1+\sigma_1^2-\sigma_2^2)\sin{l} \;-\;
     2\sigma_1\sigma_2\cos{l}  
    \big]\hhatb \cdot\gvb{a}
 \\
    \dot{\sigma}_3 \;=\; 0
    \end{array}  
\end{align}
\end{small}
where \eq{\sigma^2=\sigma_1^2+\sigma_2^2} and the third equation confirms that \eq{\sigma_3=0} remains constant, as expected.

\paragraph{The MRP Equinoctial Elements.}
Rather than the MEEs given in Eq.~\eqref{MEEdef}, an alternative set of equinoctial elements are now proposed in which the equinoctial CRPs, \eq{q_1} and \eq{q_2}, are replaced by the equinoctial MRPs, \eq{\sigma_1} and \eq{\sigma_2}. The algebraic relation between this set of MRP equinoctial elements and the classical elements is thus
\begin{small}
\begin{align} \label{MEEMRPdef}
    \text{MRP equinoctial elements}
    \quad
    \left\{ \quad
    \begin{array}{ll}
         p \;=\; \tfrac{h^2}{\mu} \;=\; a(1-e^2)  
     \\
          e_1 \;=\;  e\cos{(\Omega+\upomega)} \;=\; \gvb{e}\cdot\shatb_1
     \\
        e_2 \;=\;  e\sin{(\Omega+\upomega)} \;=\; \gvb{e}\cdot\shatb_2
    \end{array} \right.
    &&
    \begin{array}{ll}
        \sigma_1 \;=\;  \tan(\tfrac{i}{4})\cos\Omega \;=\;  \gvgr{\txsig}\cdot\shatb_1
    \\
        \sigma_2 \;=\;  \tan(\tfrac{i}{4})\sin\Omega  \;=\; \gvgr{\txsig}\cdot\shatb_2
    \\
        l \;=\; \Omega + \upomega + \nu
    \end{array} 
    \qquad
\end{align}
\end{small}
From equations \eqref{mrp_dot_fin} and \eqref{MEE_eom_tru}, the governing equations for the above MRP equinoctial elements are
\begin{small}
\begin{align} \label{MEEMRP_eom_tru}
\begin{array}{lll}
    &\,\dot{p}   \,=\, 2\tfrac{r}{h}p\,\uhatb_2 \cdot\gvb{a}
    &,
\\[4pt]
    &\dot{e}_1  \,=\, \tfrac{r}{h} \big\{ w\sin{l}\uhatb_1 + \big[(w+1)\cos{l}+e_1 \big]\uhatb_2 \big\} \cdot\gvb{a}  \,+\,  e_2 \crd{\omega_3}{s}  
    &,
\\[4pt]
    &\dot{e}_2  \,=\, \tfrac{r}{h} \big\{ -w\cos{l}\uhatb_1 + \big[(w+1)\sin{l}+e_2 \big]\uhatb_2 \big\} \cdot\gvb{a}  \,-\,  e_1 \crd{\omega_3}{s}   
    &,
\end{array}
&&
\begin{array}{llll}
    &\dot{\sigma}_ 1  \,=\,
         \tfrac{1}{4}(1+\sigma^2)(\crd{\omega_1}{s} \,+\, \sigma_2\crd{\omega_3}{s}) 
\\[4pt]
     &\dot{\sigma}_2   \,=\, 
         \tfrac{1}{4}(1+\sigma^2)(\crd{\omega_2}{s} \,-\, \sigma_1\crd{\omega_3}{s}) 
\\[4pt]
    &\;\;\;\hdot{l}  \,=\, 
    -\crd{\omega_3}{s} \,+\,  \tfrac{h}{r^2}
\end{array} 
\end{align}
\end{small}
where the \eq{\crd{\omega_i}{s}} are given by Eq.~\eqref{w3_mrp} which, upon substituting into the above, leads to 
\begin{align} \label{MEEMRP_eom_2}
\begin{array}{lll}
    &\,\dot{p}   \,=\, 2\tfrac{r}{h}p\,\uhatb_2 \cdot\gvb{a}
\\[4pt]
    &\dot{e}_1  \,=\, \tfrac{r}{h} \big\{ w\sin{l}\uhatb_1 + \big[(w+1)\cos{l}+e_1 \big]\uhatb_2 
    \,+\,  e_2\tfrac{2}{1-\sigma^2}(\sigma_2\cos{l} - \sigma_1\sin{l} ) \uvecb{u}_3 \big\} \cdot\gvb{a} 
\\[4pt]
    &\dot{e}_2  \,=\, \tfrac{r}{h} \big\{ -w\cos{l}\uhatb_1 + \big[(w+1)\sin{l}+e_2 \big]\uhatb_2 
     \,-\, e_1\tfrac{2}{1-\sigma^2}(\sigma_2\cos{l} - \sigma_1\sin{l} ) \uvecb{u}_3 \big\} \cdot\gvb{a}
\\[4pt]
    &\dot{\sigma}_ 1  \,=\,
         \tfrac{1}{4}\tfrac{r}{h}\tfrac{1+\sigma^2}{1-\sigma^2} \big\{ (1-\sigma_1^2+\sigma_2^2)\cos{l} \,-\, 
      2\sigma_1\sigma_2\sin{l}  
    \big\} \uhatb_3\cdot\gvb{a}
\\[4pt]
     &\dot{\sigma}_2   \,=\, 
         \tfrac{1}{4}\tfrac{r}{h}\tfrac{1+\sigma^2}{1-\sigma^2} \big\{ (1+\sigma_1^2-\sigma_2^2)\sin{l} \,-\, 
     2\sigma_1\sigma_2\cos{l}  
    \big\} \uhatb_3\cdot\gvb{a}
\\[4pt]
    &\;\;\;\hdot{l}  \,=\, 
    \tfrac{r}{h}\tfrac{2}{1-\sigma^2}(\sigma_1 \sin{l}-\sigma_2  \cos{l})\uhatb_3\cdot\gvb{a} \,+\,  \tfrac{h}{r^2}
\end{array} 
\end{align}
where \eq{h=\sqrt{\mu p}}, and \eq{r=\tfrac{p}{w}},  and \eq{w=1+e\cos\nu = 1+e_1\cos{l}+ e_2\sin{l}}, as given by Eq.~\eqref{w_def}. 

\begin{remrm}
Exchanging the CRPs for MRPs shifts the \emph{coordinate singularity} from  \eq{i=\pm \pi} to \eq{i=\pm 2 \pi}. However, the \emph{dynamics} for the new elements still contain a singularity at \eq{i=\pm \pi} (corresponding to \eq{\sigma^2 =1}). This, perhaps unexpected, result arises from the fact that the equinoctial elements are tied to the attitude of the equinoctial basis,   whose angular velocity component  along the orbit normal,  \eq{\shatb_3=\hhatb}, is itself singular at \eq{i=\pm \pi} (corresponding to \eq{q_i=\pm \infty} and \eq{\sigma^2 =1})
\begin{small}
\begin{align}
    \crd{\omega_3}{s} \,\,\dot{=}\,\, {\textbf{\textomega}}_{\hat{s}/\hat{\iota}}
    \cdot\shatb_3 \,=\,   -\tfrac{r}{h}\tan\tfrac{i}{2}\sin{\theta}\hhatb\cdot\gvb{a}
    \;=\;  \tfrac{r}{h}
        (q_2\cos{l}-q_1\sin{l}) \hhatb \cdot\gvb{a}
        \;=\; \tfrac{r}{h}\tfrac{2}{1-\sigma^2}(\sigma_2\cos{l} - \sigma_1\sin{l} ) \hhatb \cdot\gvb{a} 
\end{align}
\end{small}
That is, the well-known singularity in the CRP MEEs at \eq{i=\pm \pi} is not merely a \emph{coordinate} singularity but is, in fact, a \emph{kinematic} singularity inherent to the equinoctial basis itself. 
\end{remrm}

\paragraph{Conversion to Position \& Velocity Coordinates. }
From Eq.~\eqref{r_eq} and \eqref{v_eq}, 
\begin{small}
\begin{align} \label{v_eq2}
    \gvb{r}  \;=\; r\cos{l}\shatb_1 \;+\; r\sin{l}\shatb_2
 &&,&&
\dot{\gvb{r}} \;=\;   -\tfrac{\mu}{h}( e_2 + \sin{l}
    ) \shatb_1
    \;+\;
    \tfrac{\mu}{h}( e_1 + \cos{l}
    )\shatb_2
\end{align}
\end{small}
are the position and velocity vectors expanded in the equinoctial basis. 
The components in the inertial basis, \eq{\ihatb_i},  are then found using \eq{\crd{r_{\!i}}{\iota}=R_{ij}\crd{r_{\!\!j}}{s} } and  \eq{\crd{v_{\!i}}{\iota}=R_{ij}\crd{v_{\!\!j}}{s} } with the matrix \eq{R(\sigma_1,\sigma_2)} given by Eq.~\eqref{Rij_mrp} with \eq{\sigma_3=0}:
\begin{small}
\begin{align} \label{Rij_s}
 &R \;= \tfrac{4}{(1+\sigma^2)^2}
    \fnsz{
    \begin{pmatrix}
     k^2 +  \sigma_1^2 -\sigma_2^2    &    2 \sigma_1 \sigma_2 &   2 k\sigma_2 
 \\
  2\sigma_1 \sigma_2   &  k^2+\sigma_2^2 - \sigma_1^2     &   -2 k\sigma_1 
\\
    -2 k\sigma_2   &   2 k\sigma_1  &  k^2  - \sigma_1^2 -  \sigma_2^2 
    \end{pmatrix}
    }
    \;\;,&&
    k\;\dot{=}\; \tfrac{1}{2}(1-\sigma^2)
\end{align}
\end{small}
leading to
\begin{small}
\begin{align}
    \crd{\bm{r}}{\iota}  \,=\, r  \tfrac{4}{(1+\sigma^2)^2} 
\fnsz{\begin{pmatrix}
   ( k^2 +  \sigma_1^2 -\sigma_2^2 )\cos{l}  +  2\sigma_1\sigma_2\sin{l} 
   \\
   2\sigma_1\sigma_2\cos{l}   +  ( k^2 -  \sigma_1^2 +\sigma_2^2 )\sin{l}
   \\
   -2k (\sigma_2 \cos{l} -  \sigma_1 \sin{l}) 
\end{pmatrix} }
&&,&&
 \crd{\bm{v}}{\iota}  \,=\, \tfrac{\mu}{h}  \tfrac{4}{(1+\sigma^2)^2} 
 \fnsz{\begin{pmatrix}
   -( k^2 +  \sigma_1^2 -\sigma_2^2 )(e_2+\sin{l})  +  2\sigma_1\sigma_2(e_1+\cos{l}) 
   \\
   -2\sigma_1\sigma_2(e_2+\sin{l})   +   ( k^2 -  \sigma_1^2 +\sigma_2^2 )(e_1+\cos{l})
   \\
   2k \big[ \sigma_2(e_2+\sin{l})    +   \sigma_1 (e_1+\cos{l}) \big]
\end{pmatrix}  }
\end{align}
\end{small}


\section{Trajectory Optimization} \label{sec:traj}
Trajectory optimization is one of the direct applications which benefits greatly from the selection of an appropriate coordinate set as per the optimal control problem at hand. Therefore, we will now solve a low thrust rendezvous problem using the introduced MRP MEE coordinate set in order to validate it in this regard and establish its practical utility. Regularization (elimination of troublesome singularities) is therefore coupled to a well-behaved trajectory optimization formulation.  

\subsection{Optimal Control Formulation}
In this section, we will formulate the optimal control problem to evaluate the necessary conditions for a fuel-optimal interplanetary transfer. We start by first defining the dynamics of the spacecraft using the introduced MRP MEEs coordinate set.

\paragraph{State Dynamics.}
We have assumed a two body dynamics model here as it is generally adopted for preliminary mission design and analysis. Also, only the heliocentric phase of the trajectory where the motion is predominantly governed by the Sun’s gravitational force will be assumed. As a consequence,  weakly perturbed Keplerian motion is preserved. 
Using MRP MEEs defined in  Eq.~\eqref{MEEMRPdef},  \eq{\bm{x}= (p,e_1,e_2,\sigma_1,\sigma_2,l)\in\mathbb{R}^6} as our state vector, along with the spacecraft mass, \eq{m}, the state dynamics given in Eqs.~\eqref{MEEMRP_eom_tru} and \eqref{MEEMRP_eom_2}
may be split into a term linear in the thrust acceleration, \eq{\bm{a}\in\mbb{R}^3}, and a Keplerian term, \eq{\bm{k}\in\mbb{R}^6}, as
\begin{small}
\begin{align} \label{xdot_rh}
    \dot{\bm{x}} \;=\; \tfrac{r}{h} B \cdot\bm{a} \,+\, \bm{k} 
    \qquad\quad\text{or, }\quad
    \dot{x}_i \,=\, \tfrac{r}{h}\bm{b}_i\cdot\bm{a}  \,+\, k_i
    &&,&&
    \dot{m} = -\tfrac{m\mag{\bm{a}}}{c}
\end{align}
\end{small}
 where \eq{\bm{a}} is the LVLH thrust \textit{acceleration} components (\eq{a_i=\gvb{a}\cdot\uhatb_i}),  where \eq{k_i=k_6 \delta_{i6}} with \eq{k_6=\tfrac{h}{r^2}} being the only non-zero component of \eq{\bm{k}}, and where \eq{c} is the exhaust velocity (constant). We have factored out the \eq{\tfrac{r}{h}} term that appears in all of the equations. The six \eq{\bm{b}_i\in\mathbb{R}^3}, which make up \eq{B\in\mathbb{R}^{6\times 3}}, are given by
\begin{small}
\begin{align} \label{bvecs}
     B =  \begin{pmatrix}
  \text{— }  \bm{b}_1 \text{ —}
\\
    \vdots
\\
  \text{— }  \bm{b}_6 \text{ —}
\end{pmatrix}
&&
\begin{array}{lllllllll} 
      \bm{b}_1 = (0\;,\; 2p\;,\; 0)
\\[4pt]
    \bm{b}_2 = 
        ( w\sin{l} \;,\;
         (w+1)\cos{l} + e_1 \;,\; 
         -e_2 b_6 )
\\[4pt]
    \bm{b}_3 = 
       ( -w\cos{l} \;,\;
         (w+1)\sin{l} + e_2 \;,\;
         e_1 b_6 )
\end{array}
&&
\begin{array}{lll}
    \bm{b}_4   = 
            ( 0 \;,\;  0 \;,\; b_4 )
\\[4pt]
    \bm{b}_5 = 
            ( 0 \;,\;  0 \;,\; b_5 )
\\[4pt]
     \bm{b}_6  = 
             (0 \;,\;  0 \;,\; b_6 )
\end{array} 
\end{align}
\end{small}
where \eq{b_4,} \eq{b_5}, and \eq{b_6} are functions of \eq{\sigma_1}, \eq{\sigma_2}, and \eq{l} given by
\begin{small}
\begin{align} \label{b456}
   \begin{array}{lll}
        b_4 =  \tfrac{1}{4}\tfrac{1+\sigma^2}{1-\sigma^2} \big\{ (1-\sigma_1^2+\sigma_2^2)\cos{l} -      2\sigma_1\sigma_2\sin{l}      \big\} 
    &=\;   \tfrac{1}{4}(1 + \sigma^2)( \cos{l} -  \sigma_2 b_6 )
\\[4pt] 
    b_5 =  \tfrac{1}{4}\tfrac{1+\sigma^2}{1-\sigma^2} \big\{ (1+\sigma_1^2-\sigma_2^2)\sin{l} \;-\;      2\sigma_1\sigma_2\cos{l}       \big\} 
    &=\;  \tfrac{1}{4}(1 + \sigma^2)( \sin{l} +  \sigma_1 b_6 )
\\[4pt]
         b_6= \tfrac{2}{1-\sigma^2}( \sigma_1 \sin{l} -\sigma_2  \cos{l} ) 
   \end{array}
\end{align}
\end{small}
For setting up a minimum fuel orbit transfer problem where low-thrust propulsion is the only non-Keplerian force, we have parameterized the control acceleration vector as
$ \bm{a} = \frac{T \delta}{m} \hat{\bm{\alpha}} $. Here, $\hat{\bm{\alpha}}$ denotes the thrust steering unit vector, $\delta$ denotes the throttle input that determines if the thruster is ``on'' or ``off'' and \textit{m} is the mass of the spacecraft.

\paragraph{Cost Functional and the Hamiltonian.}
Letting $c = I_\text{sp} g_0$ denote the exhaust velocity (constant), the Lagrangian cost function for the minimum-fuel trajectory optimization problem is
\begin{small}
\begin{align}
    J \;=\; \int_{t_0}^{t_f} L \d t \;=\;   \int_{t_0}^{t_f} \big[ H(\bm{x},\bm{\lambda},\bm{a},t) \,-\, \bm{\lambda}\cdot \dot{\bm{x}} \,-\, \lambda_m \dot{m}  \big] \d t
    &&,&&
      L = -\dot{m} = \tfrac{m\mag{\bm{a}}}{c} 
      &&,&&
      H = L + \bm{\lambda}\cdot\dot{\bm{x}} + \lambda_m\dot{m}
\end{align}
\end{small} Letting \eq{\bm{a}= a \uvec{\alpha}} with \eq{a=\mag{\bm{a}}}, 
the Hamiltonian, \eq{H = L + \bm{\lambda}\cdot\dot{\bm{x}} + \lambda_m\dot{m}}, is given by
\begin{small}
\begin{align} \label{ham}
\begin{array}{rllllll}
     H(\bm{x},\bm{\lambda},m,\lambda_m,\bm{a}) &\!\!\!=\;  
    \bm{\lambda}\cdot \big( \tfrac{r}{h}B\cdot\bm{a} \,+\, \bm{k} \big)  \,-\, \tfrac{1}{c}(\lambda_m-1) m a
    &=\; 
    \bm{\lambda} \cdot\bm{k} \;+\; \Big( \bm{\lambda}\cdot \tfrac{r}{h}B\cdot \uvec{\alpha} \,-\, \tfrac{m}{c}(\lambda_m-1) \Big)  a
\\
     &\!\!\!=\;  \lambda_i( \tfrac{r}{h} \bm{b}_i\cdot\bm{a} \,+\, k_i \big)  \,-\, \tfrac{1}{c}(\lambda_m-1) m a
      &=\; 
    \lambda_6 k_6 \;+\;   \Big(  \lambda_i  \tfrac{r}{h} \bm{b}_i\cdot\uvec{\alpha}    \,-\, \tfrac{m}{c}(\lambda_m-1)  \Big) a
\end{array}
\end{align}
\end{small}

\paragraph{Costate Dynamics.}
The dynamics of the co-states are derived using the Euler-Lagrange equations, \eq{\dot{\lambda}_k = -\pderiv{H}{x_k}}, as per the first order necessary conditions of optimality. These equations are most easily calculated when expressed in terms of gradients of the individual \eq{\bm{b}_i} (rather than the matrix \eq{B}):
\begin{small}
\begin{align} \label{ydot}
\begin{array}{rllll}
     &\dot{\bm{\lambda}} \;=\; -\pderiv{H}{\bm{x}} \,=\, - \lambda_i \trn{[\pderiv{}{\bm{x}}( \tfrac{r}{h}\bm{b}_i )  ]} \cdot \bm{a} \,-\, \lambda_6 \pderiv{k_6}{\bm{x}}
    \;\;=\;\;
     - \lambda_i \Big( \bm{b}_i\otimes (\pderiv{}{\bm{x}}\tfrac{r}{h}) + \tfrac{r}{h}\pderiv{\bm{b}_i}{\bm{x}}        \Big)^{\!\!\top} \cdot\bm{a}
     \,-\, \lambda_6 \pderiv{k_6}{\bm{x}} 
 \\[4pt]
  \tx{or, }   &  \dot{\lambda}_k  \;=\; -\pderiv{H}{x_k}\,=\, - \lambda_i \big(  \bm{b}_i \pderiv{}{x_k}\tfrac{r}{h} \,+\, \tfrac{r}{h}\pderiv{\bm{b}_i}{x_k} \big) \cdot\bm{a} 
\,- \,  \lambda_6\pderiv{k_6}{x_k}
\end{array}
\end{align}
\end{small}
And the governing equation for the mass costate, \eq{\lambda_m}:\footnote{The equation for \eq{\dot{\lambda}_m} given in Eq.~\eqref{ym_dot} is for the case that the thrust \textit{acceleration}, \eq{\bm{a}}, is the control. If we were to instead use the actual thrust \textit{force}, \eq{\bm{f}=m\bm{a}}, as the control, then the Hamiltonian given in equation \eqref{ham} would be changed using \eq{\bm{a}=\tfrac{1}{m}\bm{f}} and \eq{\dot{\lambda}_m} would instead be given by
\begin{align} \nonumber
     \dot{\lambda}_m \,=\, \tfrac{1}{m^2} \bm{\lambda}\cdot B \cdot \bm{f} \;=\; \mag{ \trn{B}\cdot\bm{\lambda}} \tfrac{f}{m^2}
\end{align}
 }
\begin{small}
\begin{align} \label{ym_dot}
    \dot{\lambda}_m \,=\, -\pderiv{H}{m} \,=\, \tfrac{1}{c}(\lambda_m -1 )a
\end{align}
\end{small}
The gradients  of \eq{\tfrac{r}{h}} and of \eq{k_6=\tfrac{h}{r^2}}  appearing in Eq.~\eqref{ydot} are given by
\begin{small}
\begin{align}
\begin{array}{lllll}
 \tfrac{r}{h} = \tfrac{h}{\mu w} =  \sqrt{\tfrac{p}{\mu}}\tfrac{1}{w}
     &,\qquad&\qquad
      \pderiv{}{\bm{x}}\tfrac{r}{h} \;=\; 
    \tfrac{r}{h w}  
      \big(
    \tfrac{1}{2}h
    \;,\;
  -\cos{l} 
    \;,\;
  -\sin{l} 
    \;,\;
    0 
    \;,\;
    0 
    \;,\;
   e\sin\nu
    \big)
\\[5pt]
     k_6 = \tfrac{h}{r^2} = \sqrt{\tfrac{\mu}{p^3}}w^2  
     &,\qquad&\qquad
     \pderiv{}{\bm{x}} k_6 \;=\;
    \tfrac{h}{r^2 w}
    \big( -\tfrac{3}{2} \tfrac{\mu w}{h^2 }
   \;,\;
      2\cos{l} 
    \;,\;
      2\sin{l} 
    \;,\;
    0 
    \;,\;
    0 
    \;,\;
      -2 e\sin\nu \big) 
\end{array}
\end{align}
\end{small}
where \eq{e\sin\nu =  e_1\sin{l} - e_2\cos{l} }. The remaining \eq{\pderiv{\bm{b}_i}{\bm{x}}} in  Eq.~\eqref{ydot} are given explicitly in Appendix \ref{app:costate}.

\paragraph{Costate Transformations.}  The CRP MEEs, \eq{(p,e_1,e_2,q_1,q_2,l)}, given in the introduction and discussed in section \ref{sec:MEE}, have been a popular choice of coordinates for trajectory optimization over the past decades \cite{arya2021low}. As such, it is worthwhile to have a map between the CRP MEE costates and the new MRP MEE costates. For general optimal control problems of the type considered here, it can be shown that, given state-space coordinates, \eq{\bm{x}\in \mbb{R}^m}, with costates \eq{\bm{\lambda}\in \mbb{R}^m}, a bijective differentiable coordinate transformation,  \eq{\bm{x}\mapsto\bm{y}(\bm{x})\in \mbb{R}^m}, induces a costate transformation, \eq{ \bm{\lambda}\mapsto \bm{\gamma}(\bm{\lambda})\in \mbb{R}^m}, given by \cite{arya_costates}:
\begin{align}
    \bm{\gamma} \;=\; \invtrn{\pderiv{\bm{y}}{\bm{x}}} \cdot \bm{\lambda} \; \equiv \; \trn{\pderiv{\bm{x}}{\bm{y}}} \cdot \bm{\lambda}
    &&
    \bm{\lambda} \;=\; \trn{\pderiv{\bm{y}}{\bm{x}}} \cdot \bm{\gamma} \;\equiv\;
    \invtrn{\pderiv{\bm{x}}{\bm{y}}} \cdot \bm{\gamma}
\end{align}
where the right side of the above is the inverse costate transformation (the requirement that the coordinate transformation is bijective and differentiable guarantees the Jacobians are square non-degenerate matrices \cite{arya_costates}).  For the case at hand, let \eq{\bm{x}=(p,e_1,e_2,\sigma_1,\sigma_2,l)} be the MRP MEEs and  \eq{\bm{y}=(p,e_1,e_2,q_1,q_2,l)} be the CRP MEEs. Since, only the fourth and fifth elements differ, it is only the fourth and fifth costates that will be different, with the transformation given by
\begin{align}
     \bm{\gamma}_q \,=\,  \invtrn{\pderiv{\bm{q}}{\bm{\sigma}}} \cdot \bm{\lambda}_\sigma \, = \, \trn{\pderiv{\bm{\sigma}}{\bm{q}}} \cdot \bm{\lambda}_\sigma
&& \Leftrightarrow &&
    \bm{\lambda}_\sigma \,=\, \invtrn{\pderiv{\bm{\sigma}}{\bm{q}}} \cdot \bm{\lambda}_\sigma \, = \, \trn{\pderiv{\bm{q}}{\bm{\sigma}}} \cdot \bm{\gamma}_q
\end{align}
where \eq{\bm{\gamma}_q ,\bm{\lambda}_\sigma\in\mbb{R}^2} (\eq{q_3=\sigma_3=0}) and where the above Jacobians are given by Eq.~\eqref{crp_mrp_J} from which we see that, if $\sigma_1$ and $\sigma_2$ are small enough (close to zero), then $\pderiv{\bm{q}}{\bm{\sigma}}\approx 2 I\in\mbb{R}^{2\times 2}$.
Given the range of inclination that are effectively used in space applications and the range space of tangent, sine and cosine functions , the values of $\sigma_1$ and $\sigma_2$ tend to be rather small.
\paragraph{Optimal Control and Transversality Conditions.}
 Pontryagin's minimum principle (PMP) requires that the Hamiltonian be minimized over the set of all admissible controls. Since the Hamiltonian is bi-linear in $\delta$ and $\hat{\bm{\alpha}}$ and since $\delta \geq 0 $, by applying PMP, the optimal unit thrust direction vector, $\hat{\alpha}$, as well as the throttle setting, $\delta^*$ are given by: 
\begin{small}
\begin{align} 
\hat{\bm{\alpha}}^*= -\frac{B^{\top} \bm{\lambda}}{\mag{B^{\top} \bm{\lambda}}},
&&,&&
\delta^* = \arg \min_{\delta \in [0,1]} H (\bm{x}^*(t),\delta(t),\bm{\lambda}^*(t)),
\label{eq:primer}
\end{align}
\end{small}
According to PMP, the optimal value of $\delta$ depends on the sign of the switching function ($\tilde{S}$) (see Eq.~\eqref{SF}):
\begin{small}
\begin{equation} \label{SF}
\tilde{S}= \frac{r}{h}\frac{\mag{B^{\top} \bm{\lambda}}c}{m} +\lambda_m -1.
\end{equation}
\end{small}
We make use of the hyperbolic tangent smoothing (HTS) method \cite{taheri2018generic} for regularizing the control input as
\begin{small}
\begin{equation} \label{eq:smoothedcontrolp1}
\delta^*(\tilde{S}) = \frac{1}{2} \left [ 1 + \text{sign}(\tilde{S}) \right] \cong  \delta^*(\tilde{S},\rho) = \frac{1}{2} \left [ 1 + \tanh \left( \frac{\tilde{S}}{\rho} \right ) \right],
\end{equation}
\end{small}
where $\rho$ is the smoothing parameter (also adopted as a continuation parameter in the numerical continuation procedure that controls the sharpness of the switches. Notice that the approximation of Eq.\eqref{eq:smoothedcontrolp1} becomes exact as $\rho \xrightarrow{\Delta} 0$). There is only one co-state transversality condition that applies to the co-state associated with the free final mass, i.e., its value at the final time has to be zero, $\lambda_m(t_f) = 0$. 
\paragraph{The Two Point Boundary Value Problem.} 
 All together, the state and costate dynamics are given by Eqs. \eqref{xdot_rh}, \eqref{ydot}, and \eqref{ym_dot}. For fixed time, fixed initial and final states, the required boundary conditions  are:
\begin{small}
\begin{align} \label{eq:BCs}
\begin{array}{rllllll}
    t_0 = \text{fixed} 
\\[4pt]
     t_f = \text{fixed} 
\end{array}
 && \left| \quad  
\begin{array}{rllllll}
     \bm{x}(t_0) = \bm{x}_0 = \text{fixed} & \to &   \bm{\lambda}(t_0) = \tx{free} 
\\[4pt]
    \bm{x}(t_f) = \bm{x}_f = \text{fixed}  & \to &  \bm{\lambda}(t_f) = \text{free} 
\end{array}
\quad \right| 
&&
\begin{array}{rllllll}
     m(t_0) = m_0 = \text{fixed} & \to &  \lambda_m(t_0) = \tx{free}  
\\[4pt]
     m(t_f) = \text{free} & \to &  \lambda_m(t_f) = 0
\end{array}
\end{align}
\end{small}
The optimal control problem now amounts to 
finding the initial costate values, \eq{\bm{\lambda}(t_0)} and \eq{\lambda_m(t_0)},  such that, when Eqs. \eqref{xdot_rh}, \eqref{ydot}, and \eqref{ym_dot} are integrated from \eq{t_0} to \eq{t_f} with controls obtained from Eqs. \eqref{eq:primer} \& \eqref{eq:smoothedcontrolp1}, the final conditions given in Eq. \eqref{eq:BCs} are satisfied.

\subsection{Numerical Example}
The formulation for optimal control problem defined using the MRP MEEs coordinates is applied to solve an interplanetary rendezvous problem from the Earth to a significantly inclined orbit of a near-Earth asteroid, 138925 (2001 AU43). The multi-year transfer involves two revolutions around the Sun to reach the asteroid in 1720 days while departing from Earth on  February, 5 2018. The procedure for finding the boundary epochs for a trajectory optimization problem involves evaluating a reasonable time of flight for a maneuver between the selected bodies. In our case, after selecting an arbitrary starting epoch, time of flight for a minimum time maneuver is solved for between the two bodies keeping the final true longitude free as to not fix the number of revolutions around the Sun and the intercept point on the final orbit. The resulting time of flight is then reasonably and carefully increased to allow for some coast arcs in the thrust profile as per any minimum fuel maneuver without resulting in a change in number of revolutions. The final epoch is then realized to get the destination state vector (using JPL's Horizons database) for the final trajectory maneuver.
The boundary conditions hence obtained from JPL's Horizons database are given by [a (km), e, i(rad.), $\Omega $ (rad.), $\omega $(rad.), $\nu$(rad.)]: 
\begin{small}
\begin{align}
\Psi_{\text{E}} &= [1.497251E+08,0.0173,7.6438E-05,2.8152,5.2940,0.7221] \\
\Psi_{\text{NEA}} &= [2.83738E+08,0.3765,	1.2593,	2.2567, 2.60614, 0.634857] 
\end{align}
\end{small}

where $\Psi_{\text{E}}$ denotes the departure state and $\Psi_{\text{NEA}}$ denotes the arrival state.
A single thruster engine with initial mass, $m_0$ = 2800 kg delivering thrust, T = 0.45N, $I_\text{sp}$ = 3000 seconds is assumed.  The states are normalized using canonical units such that one distance unit (DU) is equal to one astronomical unit(AU), and the Time Units (TU) is equal to 1/2π year. Scaling of the state variables and time gives rise to
corresponding scaling of the co-state variables. The TPBVP is solved using \textit{MATLAB}'s standard solver, \textit{fsolve} with gradients evaluated using finite difference method.

\begin{figure}
\centering    
\begin{subfigure}[b]{0.48\textwidth}
\centering    
    \includegraphics[width = \textwidth]{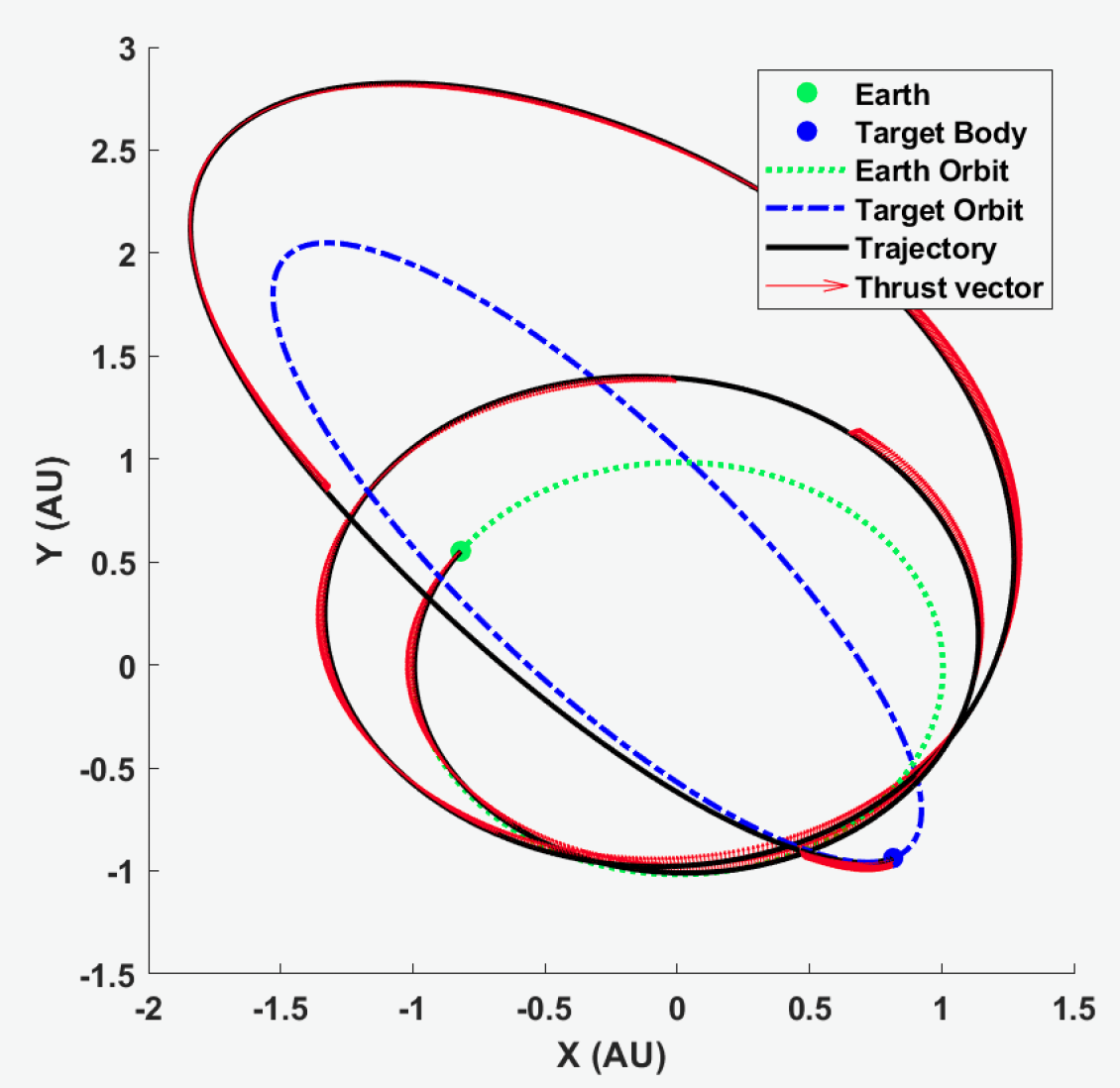}
   \caption{3D trajectory with thrust arcs for the given rendezvous problem.}
    \label{fig:traj} 
    \end{subfigure}
    \hfill
    \begin{subfigure}[b]{0.48\textwidth}
    \centering
    \includegraphics[width =\textwidth]{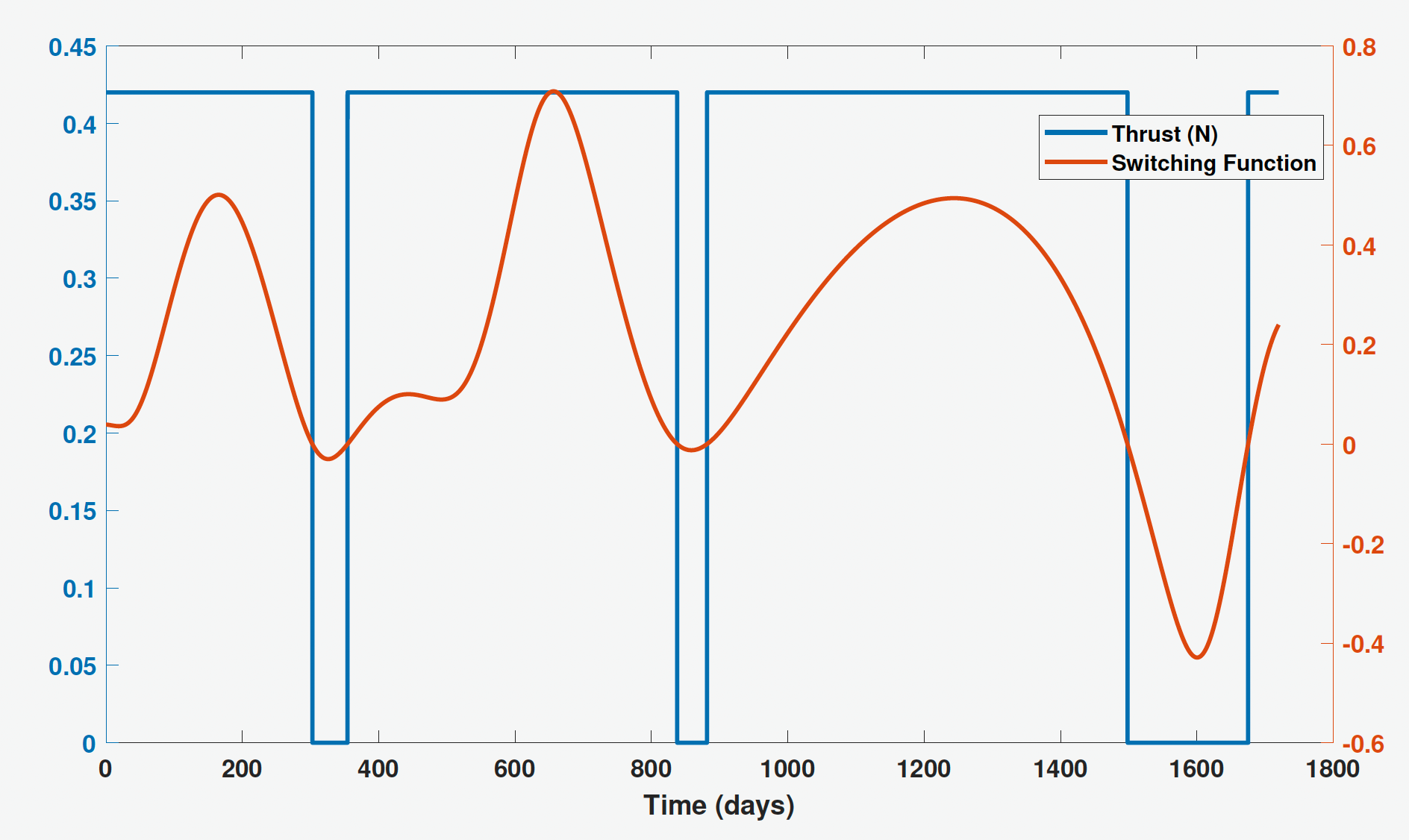}
    \caption{ Thrust and switching profile for the optimal maneuver to the asteroid.} 
    \label{fig:thrust}
    \end{subfigure}
\end{figure}

\begin{figure}
    \centering    
    \includegraphics[width = 0.65\linewidth]{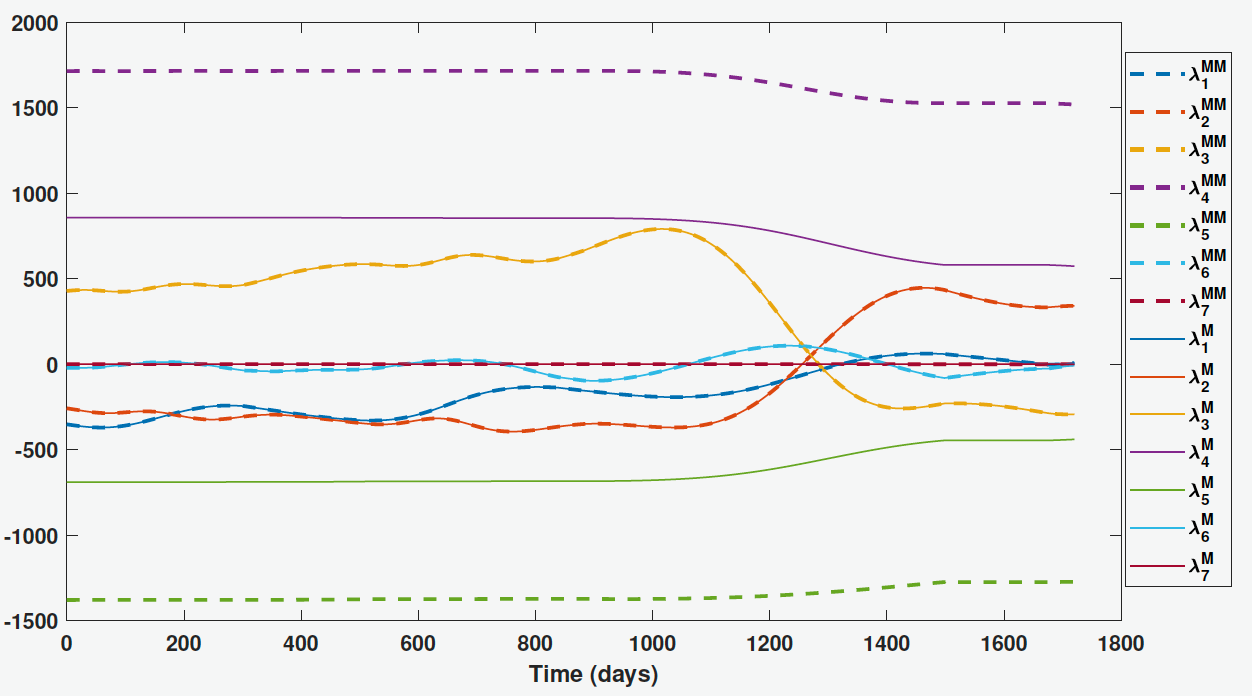}
    \caption{Time history of co-states for both MRP MEEs (denoted by $\lambda^{MM}$) and MEEs (denoted by $\lambda^{M}$).}
    \label{fig:Comp_lambda}
\end{figure}

The obtained 3D trajectory and the fuel-optimal thrust profile are given in Fig.~\ref{fig:traj} and Fig.~\ref{fig:thrust} respectively. The majority of the inclination change happens during the third thrust arc given in Fig.~\ref{fig:thrust}.  The history of costates associated with both MEEs and MRP MEEs is provided in Fig.~\ref{fig:Comp_lambda} for comparison and verification of the previously mentioned relationship between the costates of the two sets.
The new coordinates perform at par with MEEs with a slight inferiority in numerical convergence even for a transfer involving fairly high inclination change. It could be due to more complex expressions in states co-states differential equations in case of MRP MEEs as almost all co-states feature same values for both coordinate sets (see Fig.~\ref{fig:Comp_lambda}). Table \ref{tab:Comp_Stats} provides the statistics establishing the utility of the introduced coordinate set in solving TPBVPs with a large domain of convergence. The Table \ref{tab:Comp_Stats} summarize percentage of converged cases (success),  mean value for the number of iterations ($\mu_\text{iter}$) and function evaluations ($\mu_\text{fun}$) and mean simulation time ($\mu_\text{time}$) for both MEEs and MRP MEEs. These values are evaluated using the standard outputs of \textit{MATLAB}'s \textit{fsolve} solver. The mean is evaluated over 50 randomly initialized initial costates for solving the TPBVP \cite{Vishala2019}. 
 Obviously, convergence statistics are a function of the particular random sample and 50 trials (for solving a TPBVP with 14 coupled nonlinear ODEs) is not conclusive evidence with regard to efficiency. So we conclude only that the new formulation with the MRP MEEs is validated. Code optimization and relative computational efficiency is left for future studies.

\begin{table}[]
\caption{Averaged numerical convergence performance for (CRP) MEEs and MRP MEEs}
\centering
\begin{tabular}{lllll}
\hline
     & $\mu_\text{fun}$                & $\mu_\text{iter}$               & $\mu_\text{time}$(sec)              & success                   \\
     \hline
(CRP) MEEs  & 982                      & 118                     & 17.898                 & 88 \%                     \\
MRP MEEs & \multicolumn{1}{c}{1167} & \multicolumn{1}{c}{144} & \multicolumn{1}{c}{25} & \multicolumn{1}{c}{82 \%}    \\     
\hline
\end{tabular}

\label{tab:Comp_Stats}
\end{table}


\section{Conclusion} 

Fundamentals of proper orthogonal rotations and attitude kinematics in Euclidean 3-space were used to
connect the equinoctial elements to the geometry of the two-body problem, thereby filling a gap in the current understanding of these elements.
It was shown that the often-seen algebraic relations between the equinoctial elements and the classic elements follow from the geometric relations between the equinoctial basis and the LVLH, perifocal, and inertial bases. 
Most importantly, it was shown for the first time, to the authors'  knowledge, that two of the equinoctial elements are attitude coordinates — the classic Rodrigues parameters (CRPs) — for the equinoctial basis. This fact was further verified through a new derivation of the governing equations for these two elements using attitude dynamics. A new set of equinoctial elements was then developed in which the two CRPs are exchanged for the corresponding modified Rodrigues parameters (MRPs).  Principles of attitude dynamic were again used to derive the governing equations for these new \textit{MRP modified equinoctial elements} (MRP MEEs).  The new elements have a \textit{coordinate} singularity only at inclinations of \eq{i=\pm 2\pi}, as opposed to \eq{i=\pm \pi} for the the usual CRP MEEs. 
 However, it was shown that the well-known singularity at \eq{i=\pm \pi} for the CRP MEEs is not merely a coordinate singularity but is, in fact, a  \textit{kinematic} singularity inherent to the equinoctial basis itself and, as such, remains present in the \textit{dynamics} for the new MRP MEEs. This feature of the equinoctial frame  motivates the question of whether a new reference frame may be defined which dispenses with the kinematic singularity at  \eq{i=\pm \pi} while maintaining the salient features of the equinoctial frame. This question is left for future work to address.


Finally, to validate the new MRP MEEs in a significant application,  a ready set of equations to formulate a low-thrust trajectory optimization problem using these new elements is presented and verified numerically for an interplanetary minimum-fuel orbit transfer. 
Interestingly, the costates of the new MRP MEEs and the usual CRP MEEs are found to carry a near- scalar relationship with each other.
In terms of numerical convergence for solving two point boundary value problems of the type found in trajectory optimization, the new MRP MEEs do not appear to provide any clear advantage over the standard CRP MEEs for the moderately-high inclination example considered in this work. This may change for orbit transfers closer to the CRPs singularity at \eq{i=\pm \pi} or for any other application which has not yet been fully realized by the authors. Code optimization and more extensive numerical study are needed to establish relative merits. 




\appendix

\section{Appendix: Orthogonal Rotations \& Angular Velocity} \label{app:rot}
In this appendix , we review  orthogonal rotations and basis transformations in Euclidean 3-space, \eq{\mathbb{E}^3}. Einstein summation convention will be used (summation over repeated indices). However, for general \eq{\mbb{E}^n}, we will consider only \textit{orthonormal} bases 
such that the standard Euclidean metric (positive-definite inner product) has components
\eq{g_{ij}=g^{ij}=\delta^i_j} in any such basis.
Therefore, no distinction will be made between \eq{\mathbb{E}^n} and its dual space,  between vectors and 1-forms, nor between contravariant and covariant tensor components; all indices will appear as 
subscripts.
By adopting this convention, the Euclidean metric will be treated as equivalent to the identity tensor, \eq{\mbf{I}}, such that it, along with the associated metric volume form (the Levi-Civita pseudotensor) on \eq{\mbb{E}^n}, 
are isotropic tensors with components given in every \textit{orthonormal} basis by the Kronecker and Levi-Civita symbols 
(see footnote\footnote{The \eq{n}-dimensional Kronecker symbol, \eq{\delta_{ij}}, may be thought of as the components of the \eq{n\times n} identity matrix; \eq{\delta_{ij}=+1} if \eq{i=j} and \eq{\delta_{ij}=0} if \eq{i\neq j}, with trace \eq{\delta_{ii}=n}.
The \eq{n}-diemensional Levi-Civita symbol, \eq{\epsilon_{i_1\dots i_n}}, is totally antisymmetric and thus completely defined by \eq{\epsilon_{1\dots n} = 1}. If the indices are any even (odd) permutation of  \eq{1\dots n} then it is equal to \eq{+1} (\eq{-1}), and if any index is repeated then it is equal to zero. 
For the case \eq{n=3}, this means the only nonzero values of \eq{\delta_{ij}} and \eq{\epsilon_{ijk}} are the following:
\begin{align} \label{LC_3d}
    \delta_{11}=\delta_{22} = \delta_{33}=1 
    &&,&&
    \epsilon_{123} = \epsilon_{312} =\epsilon_{231} = -\epsilon_{321} = -\epsilon_{132} = -\epsilon_{213} = 1
\end{align} 
A basis for \eq{\mbb{E}^n} is \textit{orthonormal} (with respect to the Euclidean metric) if and only if the components of the Euclidean metric in the basis are equal to \eq{\delta_{ij}}. Such an orthonormal basis is \textit{positively-oriented} if and only if the components of the metric volume form (Levi-Civita pseudotensor) are equal to \eq{\epsilon_{i_1\dots i_n}} in the basis (and negatively oriented if equal to \eq{-\epsilon_{i_1\dots i_n}}). 
Alternatively, one could  \textit{define} the Kronecker symbol, \eq{\delta_{ij}}, as the components of the Euclidean metric in any orthonormal basis, and one could \textit{define} the Levi-Civita symbol, \eq{\epsilon_{i_1\dots i_n}}, as the components of the Euclidean metric volume form (Levi-Civita pseudotensor) in any positively-oriented orthonormal basis.    
}), 
respectively: 
\begin{small}
\begin{align} \label{LCdel}
    \begin{array}{llll}
    \mbf{I} \;=\; \delta_{ij}\ehatb_i\otimes\ehatb_j \;=\; \ehatb_i\otimes\ehatb_i
    \;\; \in \mbb{E}^n\otimes\mbb{E}^n
\\[4pt]
      \bm{\epsilon} \;=\;  \epsilon_{i_1\dots i_n} \ehatb_{i_1}\otimes \dots \otimes\ehatb_{i_n} \;=\; \tfrac{1}{n!}\epsilon_{i_1\dots i_n}\ehatb_{i_1}\wedge\dots \wedge\ehatb_{i_n} \;=\; \ehatb_1\wedge\dots \wedge\ehatb_n
     \;\;\in \bwedge{n} \mathbb{E}^n
     &,&\qquad \epsilon_{1\dots n} =  1  
\end{array}
\end{align}
\end{small}
where, in addition to orthonormality, the above expression for the Levi-Civita tensor, \eq{\bm{\ep}}, further assumes that the \eq{\ehatb_i\in\mbb{E}^n} constitute a \textit{positively-oriented} (right-handed)  basis. It would be scaled by \eq{-1} for negatively-oriented (left-handed) orthonormal bases such that the above \eq{\bm{\ep}} is technically a pseudotensor.  
Although it is not necessary to relegate our analysis to orthonormal bases and cartesian tensors, doing so is common practice when dealing with orthogonal transformations in Euclidean space and allows us to dispense with the distinction between \eq{\mbb{E}^n} and its dual space. As such, we refer to all tensors of order \eq{0\leq k\leq n}  as \textit{k}-tensors (scalars for \eq{k=0} and vectors for \eq{k=1}) and consider them as elements of the \eq{k^{\tx{th}}} tensor power, \eq{\botimes{k}\mbb{E}^n}.  Likewise, all alternating (skew-symmetric) \textit{k}-tensors may be regarded as elements of the \eq{k^{\text{th}}} exterior power, \eq{\bwedge{k}\mbb{E}^n\subset \botimes{k}\mbb{E}^n}.  



\subsection{Transformation of Orthonormal Bases} \label{app:orth}
Consider two arbitrary sets of basis vectors in \eq{\mbb{E}^3}: \eq{\{\ehatb_1,\ehatb_2,\ehatb_3\}} and \eq{\{\ehatb'_1,\ehatb'_2,\ehatb'_3\}}.
We will refer to these simply as the \eq{\ehatb_i} and \eq{\ehatb'_i} bases and it is implied that \eq{i} ranges from \eq{1} to \eq{3} (although,  more generally,  Eqs.~\eqref{ib_1} - \eqref{hodge} are also valid on \eq{\mbb{E}^n} for finite \eq{n \geq 2}).
As true for any two bases, there exists some non-degenerate 2-tensor, \eq{\mbf{R}\in\mathbb{E}^3\otimes\mathbb{E}^3}, such that they are related by a linear transformation:
\begin{align} \label{ib_1}
    \ehatb'_i \;=\; \mbf{R}\cdot\ehatb_i 
    &&,&&
    \ehatb_i \;=\; \inv{\mbf{R}}\cdot\ehatb'_i
   &&,&&
   |\mbf{R}| \;\dot{=}\; \text{det}(\mbf{R})  \,\neq\,  0
\end{align}
However, we will impose two conditions on all bases considered in this work; (1) the bases are orthonormal and (2) they share the same ordering (right-handed). These two conditions further require for \eq{\mbf{R}} to be a member of the the group of special orthogonal tenors on \eq{\mbb{E}^3}, \eq{\mrm{SO}(\mbb{E}^3)}:
\begin{align} \label{Rprops}
        \mrm{SO}(\mbb{E}^n) \;\;\dot{=}\;\; \{ \mbf{Q}\in \mbb{E}^n\otimes\mbb{E}^n \;|\;  \mbf{Q}\cdot \trn{\mbf{Q}} = \trn{\mbf{Q}}\cdot\mbf{Q} = \mbf{I} \;,\; |\mbf{Q}|=1 \}
        &&,&&
       \dim (\mrm{SO}(\mbb{E}^n)) \,=\, \tfrac{n}{2}(n-1)
\end{align}
where  \eq{\mbf{I}=\delta_{ij}\ehatb_i\otimes\ehatb_j=\ehatb_i\otimes\ehatb_i} is the isotropic identity tensor (and inner product) and  \eq{\trn{(\;)}} denotes the transpose of any 2-tensor.  
The first property of the above ensures that, if \eq{\ehatb_i} is an orthonormal basis and \eq{\mbf{R}\in\mrm{SO}(\mbb{E}^3)}, then  \eq{\mbf{R}\cdot\ehatb_i} is also an orthonormal basis. More generally, it preserves the standard inner product on \eq{\mbb{E}^3}; \eq{(\mbf{R}\cdot\gvb{u})\cdot(\mbf{R}\cdot\gvb{v}) = \gvb{u}\cdot\gvb{v}} for all \eq{\gvb{u},\gvb{v}\in\mbb{E}^3}. The second property above, \eq{|\mbf{R}|=1}, ensures that any bases \eq{\ehatb} and \eq{\ehatb'_i} related as in \eqref{ib_1} share the same ordering or ``handedness''. This less-obvious relation follows from the fact that \eq{|\mbf{R}|=1} ensures that  the components of the Levi-Civita pseudotensor, \eq{\epsilon_{ijk}}, are unchanged under a basis transformation \eq{\ehatb'_i=\mbf{R}\cdot\ehatb_i}.
Members of \eq{\mrm{SO}(\mbb{E}^3)} are often referred to as \textit{proper} orthogonal tensors or  (rigid) \textit{rotation tensors}. 
For any bases related as in \eqref{ib_1},  it is easy to show that  components of \eq{\mbf{R}} in these bases — denoted \eq{R_{ij}} and \eq{R'_{ij}} — are equivalent and  given by the projections of the bases  onto one another:
\begin{align} \label{R_ij}
    R_{ij} \;=\; \ehatb_i\cdot\mbf{R}\cdot\ehatb_j \;=\; \ehatb_i\cdot\ehatb'_j
      \;=\;  (\trn{\mbf{R}}\cdot\ehatb'_i)\cdot\ehatb'_j \;=\; \ehatb'_i \cdot \mbf{R}  \cdot\ehatb'_j \;=\; R'_{ij}
      &&,&&
      R_{ij}\,=\, R'_{ij} \,=\,  \ehatb_i\cdot\ehatb'_j
\end{align}
The basis transformation may then be expressed as 
\begin{align}  \label{basis_ij}
     \ehatb'_i \;=\; \mbf{R}\cdot\ehatb_i \;=\; R_{ji}\ehatb_j
&&,&&
    \ehatb_i \;=\; \trn{\mbf{R}}\cdot\ehatb'_i \;=\; R_{ij}\ehatb'_j
 &&,&&
    \mbf{R} \,=\, \ehatb'_i\otimes\ehatb_i
\end{align}
The components \eq{R_{ij}} may be collected into a matrix, \eq{R\in \mrm{SO}(3)\subset \mathrm{M}_{3,3}(\mathbb{R})}, as
\begin{align}
    R \;=\; \Big( \begin{matrix}
    [\uvec{e}'_1]^{\hat{e}} &  [\uvec{e}'_2]^{\hat{e}} &  [\uvec{e}'_3]^{\hat{e}}
    \end{matrix} \Big)
    &&
    \inv{R}\;=\;\trn{R} \;=\; \Big( \begin{matrix}
     [\uvec{e}_1]^{\hat{e}'} &  [\uvec{e}_2]^{\hat{e}'} &  [\uvec{e}_3]^{\hat{e}'}
    \end{matrix} \Big)
\end{align}
where \eq{\crd{[\uvec{e}'_k]}{e}\in \mathrm{M}_{3,1}(\mathbb{R})} is the coordinate vector of \eq{\ehatb'_k} in the \eq{\ehatb_i} basis  and \eq{[\uvec{e}_k]^{\hat{e}'}\in \mathrm{M}_{3,1}(\mathbb{R})} is the coordinate vector of \eq{\ehatb_k} in the \eq{\ehatb'_i} basis. 

\paragraph{Transformation of (Cartesian) Tensor Components. }
We can expand any vector, \eq{\gvb{u}\in\mbb{E}^3}, in any basis as \eq{\gvb{u} =  u_j\ehatb_j  =  u'_i\ehatb'_i} where  \eq{u_i=\gvb{u}\cdot\ehatb_i} and \eq{u'_i=\gvb{u}\cdot\ehatb'_i} are the components of \eq{\gvb{u}} in the \eq{\ehatb_i} and \eq{\ehatb'_i} basis, respectively. If \eq{\ehatb'_i=\mbf{R}\cdot\ehatb_i} such that \eq{R_{ij}=\ehatb_i\cdot\ehatb'_j}, then these components must be related by \eq{ u'_i = \ehatb'_i \cdot \gvb{u} =  (\ehatb'_i\cdot\ehatb_j)u_j = R_{ji}u_j}. 
The extension of this for some arbitrary \textit{p}-tensor,  \eq{\mbf{T}\in \bigotimes^p \mathbb{E}^3}, is\footnote{ \textit{Pseudotensors:} There are tensor-like objects whose components transform as in Eq.~\eqref{Xform_p-tensor}, but with an additional scaling by \eq{\text{sgn}( | R |)}. Such objects are called \textit{pseudotensors}.
In \eq{\mathbb{E}^3}, pseudovectors (\eq{p=1}) are sometimes called \emph{axial vectors} and include things such as the angular velocity (pseudo)vector, the vector/cross product of vectors, and magnetic field.
Although we will deal with pseudotensors in this work, we limit consideration to \eq{\big| R \big|=1} and the distinction is therefore of little consequence; we will often refer to them simply as tensors.  }
\begin{align} \label{Xform_p-tensor}
\begin{array}{rl}
   \fnsz{\text{ for \textit{p}-tensor:}}  &
   \mbf{T} \;=\; T_{ i_1 \dots i_p}\ehatb_{ i_1}\otimes \dots \otimes\ehatb_{ i_p}
   \;\;=\;\;
   T'_{ i_1  \dots i_p}\ehatb'_{ i_1}\otimes \dots\otimes \ehatb'_{ i_p}
   \\[4pt]
    \fnsz{\text{components transform as:}}  &  T'_{j_1 \dots j_p} \;=\;   \big(
     R_{ i_1j_1} \dots R_{ i_pj_p}
     \big) T_{ i_1  \dots i_p}
\end{array}
\end{align}
For some vector, \eq{\gvb{u}}, and some 2-tensor, \eq{\mbf{T}}, the above leads to\footnote{It is important to note that \eq{\ehatb_i} and \eq{\ehatb'_i} are two different sets of basis vectors, while  \eq{[\bm{u}]=\trn{(
u_1\;\; u_2 \;\; u_3)}} and \eq{[\bm{u}]'=\trn{(
u'_1\;\; u'_2 \;\; u'_3)}}  are merely different coordinate representations of the \textit{same} vector, \eq{\gvb{u}}, in these bases.}  
\begin{align} \label{basisXform}
&\begin{array}{rrclll}
    & u'_i \;=\;{R}_{ji}\,u_j 
     &\leftrightarrow &
      u_i \;=\; R_{ij}\,u'_j 
      &\quad,
\\[4pt] 
 \fnsz{\text{matrix form: }} \quad 
  &[\bm{u}]' \;=\;\trn{R}\,[\bm{u}]
     &\leftrightarrow &
      [\bm{u}] \;=\; R\,[\bm{u}]'  
       &\quad,
\end{array} 
&&
\begin{array}{lcl}
       T'_{ij}\;=\; R_{ki}R_{sj}  T_{ks}
     &\leftrightarrow &
      T_{ij}\;=\;R_{ik} R_{js} T'_{ks}
\\[4pt]
     T'\;=\;\trn{R}\, TR 
     &\leftrightarrow &
      T\;=\;R\,T'\trn{R} 
\end{array}
\end{align}
\sloppy where the second line is simply the matrix form of the first line; we have defined the column vectors \eq{[\bm{u}],[\bm{u}]'\in \mathrm{M}_{3,1}(\mathbb{R})} as the coordinate vectors of \eq{\gvb{u}} in the \eq{\ehatb_i} and \eq{\ehatb'_i} bases, respectively.
Similarly, \eq{T,T'\in\mathrm{M}_{3,3}(\mathbb{R})} are the matrix representations of a 2-tensor, \eq{\mbf{T}}, in the \eq{\ehatb_i} and \eq{\ehatb'_i} bases, respectively, and
\eq{R\in SO(3)\subset \mathrm{M}_{3,3}(\mathbb{R})} is the matrix representation of the rotation tensor, \eq{\mbf{R}}  (same in both bases, as seen from Eq.~\eqref{R_ij}).



\subsection{The Axial Operator} \label{app:axial}
Here we define the \textit{axial operator}, used throughout this work. In \eq{\mathbb{E}^3}, this operator gives an ismorphism between vectors, \eq{\mathbb{E}^3}, and bivectors (antisymmetric 2-tensors), \eq{\bwedge{2} \mathbb{E}^3}.

\begin{defn}{\textbf{Axial Operator.}\footnote{What we are defining as the axial operator, \eq{\wt{(\cdot)}},  is actually the (negative of) the Hodge star operator, \eq{\hodge(\cdot)},  simplified for the special case of orthonormal bases in \eq{\mathbb{E}^n}.} } \label{def:ax}
We define the axial operator, \eq{\,\wt{\;\;}} , for  \eq{\mathbb{E}^n} as a linear map, \eq{\,\wt{\;\;}:\bwedge{p} \mathbb{E}^n \to \bwedge{n-p} \mathbb{E}^n} (\eq{p\leq n}), defined such that, for any \eq{\bm{\Omega}\in \bwedge{p} \mathbb{E}^n} ,  \eq{\tv{\Omega}\in \bwedge{n-p} \mathbb{E}^n} is given by\footnote{The symbol \eq{\wedge} denotes the exterior product (``wedge'' product). For vectors, it may be interpreted as the atisymmetrized tensor product:  \eq{\gvb{u}\wedge \gvb{v}=\gvb{u} \otimes \gvb{v} -\gvb{v}\otimes\gvb{u} = -\gvb{v} \wedge\gvb{u}}. }
\begin{align} \label{hodge}
   \text{for: }\quad  &\bm{\Omega} \;=\; \Omega_{ i_1 \dots  i_p}\ehatb_{ i_1} \otimes \dots\otimes\ehatb_{ i_p}
 \;=\; \tfrac{1}{p!} \Omega_{ i_1 \dots  i_p}\ehatb_{ i_1}\wedge\dots\wedge\ehatb_{ i_p}
\\[5pt]  \nonumber
     &\tv{\Omega} \;\,\dot{=}\; -\tfrac{1}{p!}\epsilon_{ i_1 \dots  i_{n-p}\,k_1\dots k_p} \, \Omega_{k_1\dots k_p} (\ehatb_{ i_1}\otimes\dots\otimes\ehatb_{ i_{n-p}}) 
    \;=\; \tfrac{1}{p!(n-p)!}\epsilon_{ i_1 \dots  i_{n-p}\, k_1\dots k_p} \, \Omega_{k_1\dots k_p} (\ehatb_{ i_1}\wedge\dots\wedge\ehatb_{ i_{n-p}})
\end{align}
where \eq{\epsilon_{i_1\dots i_n}} are the components of the  \eq{n}-dimensional Levi-Civita (pseudo)tensor, \eq{\bm{\ep} \in \bwedge{n}\mbb{E}^n}, given in Eq.~\eqref{LCdel}.
\end{defn}
Note the domain and codomain of the map \eq{\,\wt{\;\;}:\bwedge{p} \mathbb{E}^n \to \bwedge{n-p} \mathbb{E}^n} both have the same dimension; \eq{ {n\choose p} = {n\choose n-p}=\tfrac{n!}{p!(n-p)!}}.
It can be shown that applying the axial operator twice is an automorphism, \eq{\wt{\wt{\;\;}}: \bwedge{p} \mathbb{E}^n \to\bwedge{p} \mathbb{E}^n }, given by \eq{\wt{\tv{\Omega}}=(-1)^{p(n-p)}\bm{\Omega}}. 
 Thus, for the particular case \eq{n=3}, it is an involutive (self-inverse) operator satisfying   \eq{\wt{\tv{\Omega}}=\bm{\Omega}} for any \eq{\bm{\Omega}\in\bwedge{p} \mathbb{E}^3}  (\eq{p \leq 3}). 
In particular, any \eq{\bm{\Omega}\in\bwedge{2} \mathbb{E}^3} has an axial vector, \eq{\gvgr{\txw}= \tv{\Omega} \in\mathbb{E}^3}, such that \eq{\tvgr{\txw}=\bm{\Omega}}:
\begin{align} \label{hodge_3d}
    \mathbb{E}^3  \ni 
   \;\;
     \gvgr{\txw}  \,=\, \tv{\Omega} \,=\, -\tfrac{1}{2}\epsilon_{ijk}(\ehatb_j\cdot\bm{\Omega}\cdot\ehatb_k)\ehatb_i
\quad\;\;\leftrightarrow \quad\;\;
       \bm{\Omega}  \,=\, \tvgr{\txw} \,=\,  
      -\tfrac{1}{2}\epsilon_{ijk}(\gvgr{\txw}\cdot\ehatb_k)\ehatb_i\wedge\ehatb_j  
     \;\;  \in  \bwedge{2} \mathbb{E}^3 
\end{align}
The above can be read in either direction; it does not matter if we ``start with'' the vector or the antisymmetric 2-tensor;\footnote{There is actually a subtle distinction: if \eq{\bm{\Omega}} is an antsymmetric 2-tensor, then \eq{\tv{\Omega}} is technically a pseudovector. Likewise, if \eq{\gvb{u}} is a vector then \eq{\tvb{u}} is a pseudotensor. However, as mentioned, the distinction is inconsequential if we limit consideration to orthonormal bases with the same ``handedness''.  }  the axial operator maps between them and is self-inverse. 
Letting \eq{\wt{\omega}_{ij}=\ehatb_i\cdot\tvgr{\txw}\cdot\ehatb_j} and  \eq{\wt{\omega}\in \mathrm{M}_{3,3}(\mathbb{R})} denote the components/matrix representation in some basis,
\begin{align} \label{wx_matrix}
    \omega_i \;=\;     -\tfrac{1}{2} \epsilon_{ijk} \wt{\omega}_{jk}
     \qquad \leftrightarrow \qquad 
    \wt{\omega}_{ij} \;=\; -\epsilon_{ijk}w_k
    &&,&&
     \wt{\omega} \;=\; 
     { \begin{pmatrix} 0 & -\omega_3 & \omega_2 \\
    \omega_3 & 0 & -\omega_1 \\
    -\omega_2 & \omega_1 & 0
    \end{pmatrix}  }
\end{align}
The above is likely familiar to many readers. For any \eq{\gvb{u},\gvb{v},\gvb{w}\in\mathbb{E}^3},   we note the following useful relations:
\begin{align} \label{ux_props}
\begin{array}{lll}
       \tvb{u}\cdot\gvb{v} \,=\, \gvb{u}\times\gvb{v} &,
      \\[4pt]
        \wt{\gvb{u}\times\gvb{v}} \,=\, -\gvb{u}\wedge\gvb{v} &,
      \\[4pt]%
       \text{tr}(\tvb{u}\cdot\tvb{v}) = -2\gvb{u}\cdot\gvb{v} &,
 \end{array}
 &&
 \begin{array}{lll}
    \tvb{u}\cdot\tvb{v} \,=\, \gvb{v}\otimes\gvb{u} -(\gvb{u}\cdot\gvb{v})\mbf{I} &,
  \\[4pt]
      \tvb{u}\cdot\tvb{u}\cdot\tvb{u} 
  \,=\,  -u^2\tvb{u} &,
  \\[4pt]%
  \wt{ \gvb{u}\otimes\gvb{v} } \,=\,
      \wt{ \tvb{v}\cdot\tvb{u} } \,=\,
      -\tfrac{1}{2}\gvb{u}\times\gvb{v}
 \end{array}
 &&
 \begin{array}{lll}
      \tvb{u}\cdot \tvb{v} \cdot\gvb{w} 
     \,=\, \gvb{u}\times(\gvb{v} \times\gvb{w})
 \,=\, ( \gvb{v}\wedge \gvb{w})\cdot \gvb{u} 
 \\[4pt]
       \gvb{u}\cdot \tvb{v} \cdot\gvb{w} 
     \,=\, \gvb{u}\cdot(\gvb{v} \times\gvb{w}) 
     \\[4pt]%
    \;\;
 \end{array}
 \end{align}
A final note regarding the axial operator: 
In \eq{\mbb{E}^3}, it is typically used only for vectors and antisymmetric 2-tensors, as detailed above. However, we may also apply it to \textit{any} \eq{2}-tensor, \eq{\mbf{T}\in \mathbb{E}^3\otimes \mathbb{E}^3 }. It is straightforward to show that doing so is equivalent to applying it to the antisymmetric part of \eq{\mbf{T}}: 
\begin{align} \label{axial_T}
     \forall \; \mbf{T}\in \mathbb{E}^3\otimes \mathbb{E}^3  \;\;:
 &&
     \tvb{T} = \wt{\tfrac{1}{2} ( \mbf{T}-\trn{\mbf{T}}) } 
    \qquad,\qquad 
     \wt{\tvb{T}}=\tfrac{1}{2} ( \mbf{T}-\trn{\mbf{T}})
\end{align}

\subsection{The Angular Velocity Tensor \& Axial Vector} \label{sec:ang_vel}


There is a strong analogy between rigid body kinematics and basis transformations in \eq{\mathbb{E}^3}. A basis embedded in a rigid body can be viewed as a representation of that body's orientation in space (attitude) with respect to some set of reference directions (i.e., a basis).
If the bodies — represented as orthonormal bases — are rotating in some manner then the instantaneous relative orientation at some time \eq{t} is  described by  \eq{\ehatb'_i(t)=\mbf{R}(t)\cdot\ehatb_i(t)}, for some  \eq{\mbf{R}(t)\in\mrm{SO}(\mbb{E}^3)} (we again consider only orthonormal bases for reasons of convenience, not necessity).  The relative rotational motion is described by the relative angular velocity of \eq{\ehatb'_i} with respect to \eq{\ehatb_i} (or vice versa).

\paragraph{Corotational Derivative. } 
To make the following developments generalizable to any two rotating orthonormal bases, we preemptively define the \textit{corotational} time-derivative of any \textit{p}-tensor, \eq{\mbf{T}=T_{i_1 \dots i_p}(\ehatb_{i_1}\otimes \dots\otimes \ehatb_{i_p})}, with respect to an arbitrarily-rotating basis, \eq{\ehatb_i}, as follows: 
\begin{align} \label{frame_diff}
\fnsz{\begin{array}{cc}
     \text{\eq{\ehatb_i}-corotational}  \\
     \text{derivtive}
\end{array}} :
\qquad 
    \fdiff{\hat{e}} \mbf{T}  \;\;\dot{=}\;\; (\diff{}{t}{T}_{i_1 \dots i_p}) \ehatb_{i_1}\otimes \dots\otimes \ehatb_{i_p}
  \quad =\quad \diff{}{t} \mbf{T} \;-\; T_{i_1 \dots i_p} \diff{}{t}(\ehatb_{i_1}\otimes \dots\otimes \ehatb_{i_p})
\end{align}
That is, \eq{\fdiff{\hat{e}}} — called the ``\eq{\ehatb_i}-corotational derivative'' for short — is  a \textit{basis-dependent} time-derivative that is blind to the motion of the \eq{\ehatb_i} basis: \eq{\fdiff{\hat{e}} \ehatb_i = \gvb{0}}.  This distinction is only applicable for \textit{p}-tensors of order  \eq{p\geq 1} (i.e., \eq{\fdiff{\hat{e}} f = \diff{}{t} f} for any scalar-valued function, \eq{f}). 
If \eq{\ehatb_i} happens to be an inertial basis (in the Newtonian sense) then  \eq{\fdiff{\hat{e}}\equiv \diff{}{t}} is equivalent to the total/inertial time-derivative. 


\paragraph{Angular Velocity. }
From here forward, we consider two arbitrarily-rotating, right-handed, orthonormal bases related by  \eq{\ehatb'_i(t)=\mbf{R}(t)\cdot\ehatb_i(t)},  with \eq{\mbf{R}(t)\in\mrm{SO}(\mbb{E}^3)}, 
and adopt the following notation:
\begin{align}
        \hdot{(\cdot)} \;\;\dot{=}\;\; \diff{}{t}(\cdot)
        &&,&&
        \ring{(\cdot)} \;\;\dot{=}\;\; \fdiff{\hat{e}} (\cdot)
        \qquad \Rightarrow \qquad 
        \ring{\ehatb}_i\,=\, \gvb{0}
\end{align} 
By definition, \eq{\mbf{R}} must, at all times,  satisfy
\eq{\mbf{R}\cdot\trn{\mbf{R}}=\trn{\mbf{R}}\cdot\mbf{R}=\mbf{I}} (we will suppress the argument \eq{t}). Taking the inertial time-derivative  yields  \eq{ \diff{}{t} (\mbf{R}\cdot\trn{\mbf{R}}) = \diff{}{t} (\trn{\mbf{R}}\cdot\mbf{R}) =\mbf{0} }. 
However, it can be shown that this requirement holds not only for the inertial derivative but also for the corotational derivative with respect to 
\textit{any} arbitrary orthonormal basis:
\begin{align} \label{Rdot_SO3}
\forall \; \mbf{R} \in \mrm{SO}(\mbb{E}^3) \;\;\;
\left\{
\begin{array}{llll}
 \diff{}{t}(\mbf{R}\cdot\trn{\mbf{R}}) 
      \;=\; \mbf{0} \;=\;
      \fdiff{\hat{n}}(\mbf{R}\cdot\trn{\mbf{R}}) \;=\;
  \fdiff{\hat{n}}{\mbf{R}}\cdot\trn{\mbf{R}} \;+\;  \mbf{R}\cdot \fdiff{\hat{n}}\trn{\mbf{R}} 
\\[4pt]
    \diff{}{t}(\trn{\mbf{R}}\cdot\mbf{R}) \;=\; \mbf{0} \;=\;
      \fdiff{\hat{n}}(\trn{\mbf{R}}\cdot\mbf{R}) \;=\; \trn{\mbf{R}}\cdot \fdiff{\hat{n}}{\mbf{R}}  \;+\;
    \fdiff{\hat{n}}\trn{\mbf{R}}\cdot\mbf{R} 
\end{array}\right.
 &&
   { \text{(any \eq{\uvecb{n}_i})}}
\end{align}
such that, for any choice of  \eq{\nhatb_i} (inertial or otherwise), the above yields two equations of the form \eq{\mbf{A}+\trn{\mbf{A}} = \mbf{0}}. For the case that \eq{\ehatb'_i=\mbf{R}\cdot\ehatb_i},  it is useful to express the above using the \eq{\ehatb_i}-corotational derivative (denoted \eq{\ring{(\cdot)}}) and to define the two resulting antisymmetric 2-tensors
in terms of \eq{\mbf{R}=\ehatb'_i\otimes\ehatb_i} and \eq{\ring{\mbf{R}}=\ring{\ehatb}'_i\otimes\ehatb_i} as
follows:\footnote{The definitions of \eq{\tvgr{\txw}} and \eq{\tvgr{\txy}} in Eq.~\eqref{peepee} are equivalently expressed in terms of the \eq{\ehatb'_i}-corotational derivative as  \eq{\tvgr{\txw}= \trn{\mbf{R}} \cdot \fdiff{\hat{e}'}\mbf{R}} and \eq{\tvgr{\txy}= \fdiff{\hat{e}'}\mbf{R} \cdot\trn{\mbf{R}} }. }
\begin{align} \label{peepee}
\begin{array}{llll}
     \tvgr{\txw} \;\; \dot{=} \;\; \ring{\mbf{R}}\cdot\trn{\mbf{R}}  \,=\, -\mbf{R}\cdot\trn{\ring{\mbf{R}}}
    & =\; 
   (\ring{\ehatb}'_j\cdot\ehatb'_i)\ehatb'_i\otimes\ehatb'_j
   \,=\,
   \tfrac{1}{2}(\ring{\ehatb}'_j\cdot\ehatb'_i)\ehatb'_i\wedge\ehatb'_j
   & =\; \ring{\ehatb}'_j \otimes\ehatb'_j 
   \;\;\equiv\; \ehatb_i\otimes \fdiff{\hat{e}'\!\!}\ehatb_i
\\
     \tvgr{\txy} \;\; {\dot{=}} \;\; \trn{\mbf{R}}\cdot\ring{\mbf{R}}  \,=\, -\trn{\ring{\mbf{R}}}\cdot\mbf{R} 
     & =\; (\ring{\ehatb}'_j\cdot\ehatb'_i)\ehatb_i\otimes\ehatb_j
     \,=\,
   \tfrac{1}{2}(\ring{\ehatb}'_j\cdot\ehatb'_i)\ehatb_i\wedge\ehatb_j
\end{array}
\end{align}
where, in light of  Eq.~\eqref{hodge_3d}, we have preemptively written \eq{\tvgr{\txw},\tvgr{\txy} \in \bwedge{2}\mathbb{E}^3}
 as the ``axial dual'' of some (pseudo)vectors, \eq{\gvgr{\txw},\gvgr{\txy}\in\mathbb{E}^3}, which themselves may be obtained from the above using Eqs. \eqref{hodge_3d} and \eqref{ux_props}. In doing so, it is most natural to use the \eq{\ehatb'_i} basis to obtain  \eq{\gvgr{\txw}=-\tfrac{1}{2}\epsilon_{ijk} (\ehatb'_j \cdot \tvgr{\txw}\cdot\ehatb'_k)}, and the \eq{\ehatb_i} basis to obtain \eq{\gvgr{\txy}=-\tfrac{1}{2}\epsilon_{ijk}(\ehatb_j \cdot \tvgr{\txy}\cdot\ehatb_k)\ehatb_i}: 
\begin{align} \label{wvec_1}
\begin{array}{ll}
     \gvgr{\txw}  \;=\; \wt{\tvgr{\txw}} \;=\; 
   \tfrac{1}{2}\epsilon_{ijk} (\ring{\ehatb}'_j\cdot\ehatb'_k)\ehatb'_i
    \;=\;
    \tfrac{1}{2}(\ring{\ehatb}'_i\cdot\ehatb'_j) \ehatb'_i\times \ehatb'_j
    \;=\;
    \tfrac{1}{2}  \ehatb'_i \times \ring{\ehatb}'_i
\\[4pt]
      \gvgr{\txy} \;=\;   \wt{\tvgr{\txy}} 
       \;=\;  \tfrac{1}{2}\epsilon_{ijk} (\ring{\ehatb}'_j\cdot\ehatb'_k)\ehatb_i
       \;=\;
    \tfrac{1}{2}(\ring{\ehatb}'_i\cdot\ehatb'_j) \ehatb_i\times \ehatb_j
\end{array}
\end{align}
By inspection of Eqs. \eqref{peepee} and \eqref{wvec_1}, it is seen  that the components of \eq{\tvgr{\txy}} and \eq{\gvgr{\txy}} in the \eq{\ehatb_i} basis (denoted \eq{\wt{\lambda}_{ij}}  and \eq{\lambda_i}) are equal to those of \eq{\tvgr{\txw}} and \eq{\gvgr{\txw}} in the \eq{\ehatb'_i} basis (denoted \eq{\wt{\omega}'_{ij}}  and \eq{\omega'_i}):
\begin{align} \label{Omega_xform}
    \wt{\lambda}_{ij} \;=\;  \wt{\omega}'_{ij} \;=\; \ehatb'_i \cdot \ring{\ehatb}'_j
    \qquad\quad, \qquad\quad \lambda_i \;=\; \omega'_i \;=\; -\tfrac{1}{2}\epsilon_{ijk}\ehatb'_j \cdot \ring{\ehatb}'_k 
\end{align}
The above relations are further confirmed by noting that Eqs. \eqref{peepee} and \eqref{wvec_1} lead to the following:
\begin{align} \label{Rdot_tens}
    \ring{\mbf{R}} \;=\; \tvgr{\txw}\cdot\mbf{R}  \;=\;  \mbf{R}\cdot\tvgr{\txy} 
    \;=\; \hdot{R}_{ij}\ehatb_i\otimes\ehatb_j
&&,&&
\begin{array}{ll}
    \tvgr{\txw} \;=\; \mbf{R}\cdot\tvgr{\txy}\cdot \trn{\mbf{R}}
\\[4pt]
 \gvgr{\txw} \,=\, \mbf{R}\cdot\gvgr{\txy}
\end{array}
&&,&&
\begin{array}{lll}
       \tvgr{\txy} \;=\; \trn{\mbf{R}}\cdot\tvgr{\txw}\cdot\mbf{R} 
\\[4pt]
     \gvgr{\txy} \,=\, \trn{\mbf{R}}\cdot\gvgr{\txw}
\end{array}
\end{align}
As one may have inferred, \eq{\gvgr{\txw}} is the familiar angular velocity (pseudo)vector (of \eq{\ehatb'_i} relative to \eq{\ehatb_i}). It is in fact the  axial vector of the angular velocity tensor, \eq{\tvgr{\txw}}.\footnote{There is some ambiguity in whether to call \eq{\tvgr{\txw}} or \eq{\tvgr{\txy}} the angular velocity. We use \eq{\tvgr{\txw}} to maintain consistency with most formulations.} 
Eqs. \eqref{peepee} and \eqref{wvec_1} give a coordinate-free definition of this angular velocity in terms of \eq{\mbf{R}}, \eq{\ehatb_i}, \eq{\ehatb'_i}, and their \eq{\ehatb_i}-corotational derivatives. 
For computational purposes, we are typically interested in the components of these tensor relations in either the \eq{\ehatb_i} or \eq{\ehatb'_i} basis. To that end, note that \eq{\ring{\mbf{R}}} is simply \eq{\ring{\mbf{R}}=\hdot{R}_{ij}\ehatb_i\otimes\ehatb_j}. 
The coordinate representation of Eqs. \eqref{peepee} and \eqref{Rdot_tens}  in the \eq{\ehatb_i} basis then leads to 
\begin{align} \label{wx_R}  
\begin{array}{rllll}
       &\qquad    \wt{\omega}_{ij} \;=\;  \hdot{R}_{ik}R_{jk} 
      \\[4pt]
        \fnsz{\text{matrix form: }} &\qquad   \wt{\omega} \;=\; 
      \hdot{R}\trn{R}
\end{array}
&&,&&
\begin{array}{lllll}
    \wt{\omega}'_{ij} \;=\; \wt{\lambda}_{ij} \;=\;  R_{ki}\hdot{R}_{kj} 
\\[4pt]
       \wt{\omega}' \;=\; 
    \trn{R}\hdot{R}
\end{array}
&&,&&
\begin{array}{llll}
       \hdot{R}_{ij} \;=\; \wt{\omega}_{ik}R_{kj}  \;=\; R_{ik}\wt{\omega}'_{kj}
   \\[4pt]
       \hdot{R} \;=\; \wt{\omega}R \;=\; R\wt{\omega}'
\end{array}
\end{align}
where we have used the fact that \eq{\wt{\lambda}_{ij}=\wt{\omega}_{ij}'}, where \eq{\wt{\omega},\wt{\omega}' \in \mathrm{M}_{3,3}(\mathbb{R})} are the matrix representations of the tensor \eq{\tvgr{\txw}} in \eq{\ehatb_i} and \eq{\ehatb'_i}, and    
where \eq{\hdot{R}} is simply the time-derivative of the matrix \eq{R}. The components of  \eq{\gvgr{\txw}} in the two bases are easily read off from the matrices \eq{\wt{\omega}} and \eq{\wt{\omega}'} using Eq.~\eqref{wx_matrix}. 
Alternatively,  Eq.~\eqref{wx_R} into Eq.~\eqref{wvec_1} leads to
\begin{align} \label{wx_R2}
     \omega_k \;=\;
     -\tfrac{1}{2}\epsilon_{ijk}\wt{\omega}_{ij}
     \;=\; -\tfrac{1}{2}\epsilon_{ijk}\hdot{R}_{is}R_{js} 
&&,&& 
      \omega'_k \;=\;
     -\tfrac{1}{2}\epsilon_{ijk}\wt{\omega}'_{ij}
     \;=\; -\tfrac{1}{2}\epsilon_{ijk}R_{si}\hdot{R}_{sj}
\end{align}


\paragraph{Kinematic Transport.}
The \eq{\ehatb_i}-corotational derivative, \eq{\ring{(\cdot)}\equiv \fdiff{\hat{e}}(\cdot)}, of some arbitrary vector, \eq{\gvb{u}}, and  2-tensor, \eq{\mbf{T}}, may be expressed using the \eq{\ehatb_i} and  \eq{\ehatb'_i} bases (both of which may be rotating) as
\footnote{\eq{\dot{u}'_i} and  \eq{\dot{T}'_{ij}} are the derivitives of the components of \eq{\gvb{u}} and \eq{\mbf{T}} in the \eq{\ehatb'_i} basis, \textit{not} the components of the derivatives.}
\begin{align} \label{udot_I}
    &\ring{\gvb{u}} \;=\;  \dot{u}_i\ehatb_i 
    \;=\;   \dot{u}'_i\ehatb'_i \;+\; u'_i\ring{\ehatb}'_i
    && 
     \ring{\mbf{T}} \;=\;  \dot{T}_{ij} \ehatb_i\otimes\ehatb_j
     \;=\;  \dot{T}'_{ij}\ehatb'_i\otimes\ehatb'_j \;+\; T'_{ij}\fdiff{\hat{e}} (\ehatb'_i\otimes\ehatb'_j)
\end{align}
\sloppy
The \eq{\ehatb'_i} basis has angular velocity (with respect to \eq{\ehatb_i}) defined by Eq.~\eqref{peepee} as \eq{\tvgr{\txw}=\ring{\mbf{R}}\cdot\trn{\mbf{R}}= \ring{\ehatb}'_k\otimes\ehatb'_k}. Contracting with \eq{\ehatb'_i}, we obtain the \eq{\ehatb_i}-corotational derivatives of the \eq{\ehatb'_i} basis:
\begin{align} \label{ehat_dot_2}
    \ring{\ehatb}'_i \;=\; \tvgr{\txw} \cdot\ehatb'_i
    \;=\; \gvgr{\txw} \times\ehatb'_i
    &&,&&
    \fdiff{\hat{e}} (\ehatb'_i\otimes\ehatb'_j) \,=\,  \tvgr{\txw}\cdot(\ehatb'_i\otimes\ehatb'_j) \,-\, (\ehatb'_i\otimes\ehatb'_j)\cdot\tvgr{\txw}
\end{align}
where the relation on the right follows from that on the left. Using the above,
Eq.~\eqref{udot_I} then
leads to the familiar kinematic transport equation for a vector and 2-tensor:
 \begin{align} 
\begin{array}{lllll}
\ring{\gvb{u}} \; =\; 
     \dot{u}_i\ehatb_i \;=\;
     \dot{u}'_i\ehatb'_i \,+\, \tvgr{\txw}\cdot\gvb{u}
    &,
\\[4pt]
    \ring{\mbf{T}} \;=\;  
    \dot{T}_{ij} \ehatb_i\otimes\ehatb_j
    \;=\;
    \dot{T}'_{ij}\ehatb'_i\otimes\ehatb'_j \,+\, \tvgr{\txw}\cdot\mbf{T} \,-\, \mbf{T}\cdot\tvgr{\txw}
&,
\end{array}
&&
\begin{array}{lllll}
     \dot{u}_i\ehatb_i \,=\,
        (\dot{u}'_i \,+\, \wt{\omega}'_{ij}u'_j )\ehatb'_i
 \\[4pt]
    \dot{T}_{ij} \ehatb_i\otimes\ehatb_j \,=\,
      (\dot{T}'_{ij} \,+\, \wt{\omega}'_{ik}T'_{kj} \,-\, T'_{ik}\wt{\omega}'_{kj} ) \ehatb'_i\otimes\ehatb'_j
\end{array}
\end{align}
The left-hand-side of these relations is \eq{\ehatb_i}-corotational derivative and the
first term on the right-hand-side is the \eq{\ehatb'_i}-corotational derivative. Thus, the following holds for any \eq{\ehatb'_i=\mbf{R}\cdot\ehatb_i}, 
 \begin{align} \label{transport_0}
     \fdiff{\hat{e}}\gvb{u}
    \,=\,
     \fdiff{\hat{e}'\!\!}\gvb{u} \,+\, \tvgr{\txw}\cdot\gvb{u}  
 &&,&& 
   \fdiff{\hat{e}}\mbf{T} 
   \,=\,  \fdiff{\hat{e}'\!\!}\mbf{T}  \,+\, \tvgr{\txw}\cdot\mbf{T} \,-\, \mbf{T}\cdot\tvgr{\txw}
&&,&&
    \tvgr{\txw}  \;\dot{=}\;  \fdiff{\hat{e}}\mbf{R} \cdot\trn{\mbf{R}}  \;=\; \trn{\mbf{R}}\cdot\fdiff{\hat{e}'\!\!}\mbf{R}
\end{align}

\section{Appendix: Costate Gradients} \label{app:costate}

The costate equations require the gradients, \eq{\pderiv{\bm{b}_i}{\bm{x}}} of the six \eq{\bm{b}_i\in\mbb{R}^3} given in Eq.~\eqref{bvecs}.  These matrices are:
\begin{small}
\begin{align}
\begin{array}{llllllll}
     \pderiv{\bm{b}_1}{\bm{x}} = 
    \begin{pmatrix}
     0 & 0 & 0 & 0 & 0 & 0 \\
     2 & 0 & 0 & 0 & 0 & 0 \\
      0 & 0 & 0 & 0 & 0 & 0 \\
    \end{pmatrix}
\\
\\
      \pderiv{\bm{b}_2}{\bm{x}} = 
    \begin{pmatrix}
     0 & \cos{l}\sin{l} & \sin^2{l} & 0 & 0 & -e\sin\nu \sin{l} + w\cos{l} 
     \\
    0 & 1+\cos^2{l} & \cos{l}\sin{l} & 0 & 0 & -[e\sin\nu \cos{l} + (w+1)\sin{l}]
     \\
      0 & 0 & -b_6 & -e_2\pderiv{b_6}{\sigma_1} & -e_2\pderiv{b_6}{\sigma_2} & -e_2\pderiv{b_6}{l} 
    \end{pmatrix}
\\
\\
     \pderiv{\bm{b}_3}{\bm{x}} = 
    \begin{pmatrix}
     0 & -\cos^2{l} & -\cos{l}\sin{l} & 0 & 0 & e\sin\nu \cos{l} + w\sin{l} 
     \\
    0 & \cos{l}\sin{l} & 1 + \sin^2{l}  & 0 & 0 & -e\sin\nu \sin{l} + (w+1)\cos{l}]
     \\
      0 & b_6 & 0 & e_1\pderiv{b_6}{\sigma_1} & e_1\pderiv{b_6}{\sigma_2} & e_1\pderiv{b_6}{l} 
    \end{pmatrix}
\end{array}
\end{align}
\end{small}

\begin{small}
\begin{align} \nonumber
\begin{array}{lllll}
    \pderiv{\bm{b}_4}{\bm{x}} = 
    \begin{pmatrix}
         0 & 0 & 0 & 0 & 0 & 0
         \\
        0 & 0 & 0 & 0 & 0 & 0
         \\
          0 & 0 & 0 
          & \pderiv{b_4}{\sigma_1}
          & \pderiv{b_4}{\sigma_2}
          & \pderiv{b_4}{l}
    \end{pmatrix}
     \qquad,\qquad\qquad
     \begin{array}{llll}
           \pderiv{b_4}{\sigma_1} 
        \;=\; \tfrac{4 \sigma_1}{1-\sigma^4}b_4 - \tfrac{1}{2}\tfrac{1+\sigma^2}{1-\sigma^2}( \sigma_1\cos{l} + \sigma_2\sin{l} )
        \\
        \pderiv{b_4}{\sigma_2}
        \;=\; \tfrac{4 \sigma_2}{1-\sigma^4}b_4 + \tfrac{1}{2}\tfrac{1+\sigma^2}{1-\sigma^2}( \sigma_2\cos{l} - \sigma_1\sin{l} )
    \\
        \pderiv{b_4}{l} = b_5 
    \end{array}
\\
\\
    \pderiv{\bm{b}_5}{\bm{x}} = 
    \begin{pmatrix}
         0 & 0 & 0 & 0 & 0 & 0
         \\
        0 & 0 & 0 & 0 & 0 & 0
         \\
          0 & 0 & 0 
          & \pderiv{b_5}{\sigma_1}
          & \pderiv{b_5}{\sigma_2}
          & \pderiv{b_5}{l}
    \end{pmatrix}
    \qquad,\qquad\qquad
    \begin{array}{lll}
          \pderiv{b_5}{\sigma_1} 
        \;=\; \tfrac{4 \sigma_1}{1-\sigma^4}b_5 + \tfrac{1}{2}\tfrac{1+\sigma^2}{1-\sigma^2}( \sigma_1\sin{l} - \sigma_2\cos{l} )
    \\
        \pderiv{b_5}{\sigma_2} 
        \;=\; \tfrac{4 \sigma_2}{1-\sigma^4}b_5  - \tfrac{1}{2}\tfrac{1+\sigma^2}{1-\sigma^2}( \sigma_2\sin{l} + \sigma_1\cos{l} )
    \\
          \pderiv{b_5}{l} = -b_4
    \end{array}
\\
\\
    \pderiv{\bm{b}_6}{\bm{x}} = 
      \begin{pmatrix}
        0 & 0 & 0 & 0 & 0 & 0
        \\  0 & 0 & 0 & 0 & 0 & 0
        \\ 0 & 0 & 0 & \pderiv{b_6}{\sigma_1} 
        & \pderiv{b_6}{\sigma_2}
        & \pderiv{b_6}{l}
    \end{pmatrix}
    \qquad,\qquad\qquad
        \begin{array}{lllll}
         \pderiv{b_6}{\sigma_1} 
        \;=\; 
       \tfrac{2\sigma_1}{1-\sigma^2} b_6 + \tfrac{2}{1-\sigma^2}\sin{l}
    \\
         \pderiv{b_6}{\sigma_2} 
        \;=\; 
       \tfrac{2\sigma_2}{1-\sigma^2} b_6  - \tfrac{2}{1-\sigma^2}\cos{l}
    \\
        \pderiv{b_6}{l} \;=\; \tfrac{2}{1-\sigma^2}( \sigma_1 \cos{l}  +  \sigma_2  \sin{l}  )
    \end{array}
\end{array}   
\end{align}
\end{small}
where \eq{b_4}, \eq{b_5}, and \eq{b_6} are given in terms of the MRP MEEs by Eq.~\eqref{b456}.   

\end{document}